\documentclass[11pt, letterpaper]{article}

\usepackage{cite}
\usepackage{graphicx}
\usepackage{amsmath}
\usepackage{amssymb}
\usepackage{setspace}
\usepackage{epstopdf}
\usepackage{color}
\usepackage[letterpaper,left=1in,right=1in,top=1in,bottom=1in]{geometry}
\usepackage[linesnumbered,ruled,vlined,longend]{algorithm2e}
\usepackage{multirow}
\usepackage{longtable}
\usepackage{rotating}
\usepackage{threeparttable}
\usepackage{colortbl}
\usepackage{amsthm}
\usepackage[modulo]{lineno}
\usepackage{enumerate}
\usepackage{bm}
\usepackage{algorithmic}

\newtheorem{remark}{Remark}

\theoremstyle{definition}

\newcommand{\rev}{\textcolor{black}}

\title{\Large \bf Smart energy management: process structure-based hybrid neural networks for optimal scheduling and economic predictive control in integrated systems}

\author{\centerline{\normalsize Long Wu$^{a,b}$, Xunyuan Yin$^{c}$, Lei Pan$^{a,\ast}$, Jinfeng Liu$^{b,}$\thanks{Corresponding authors: L. Pan. Email: panlei@seu.edu.cn; J. Liu. Email: jinfeng@ualberta.ca}}\vspace{5mm}\\
	\centerline{\small $^{a}$ National Engineering Research Center of Power Generation Control and Safety,}\\
	\centerline{\small School of Energy and Environment, Southeast University, Nanjing, 210096, China}\\
	\centerline{\small $^{b}$ Department of Chemical \& Materials Engineering, University of Alberta,}\\
	\centerline{\small Edmonton, Alberta, Canada, T6G 1H9}\\
	\centerline{\small $^{c}$ School of Chemistry, Chemical Engineering and Biotechnology,}\\
	\centerline{\small Nanyang Technological University, 62 Nanyang Drive, Singapore, 637459}}

\allowdisplaybreaks

\begin{document}
	
	\date{}
	
	\maketitle
	\setstretch{1.39}
	
	\begin{abstract}
		Integrated energy systems (IESs) are complex prosumers consisting of diverse operating units spanning multiple domains. The \rev{tight} integration of these units results in varied dynamic characteristics and intricate nonlinear process interactions, making detailed dynamic modeling and successful operational optimization challenging. To address these \rev{concerns}, we propose a \rev{physics-informed} hybrid time-series neural network (NN) surrogate to predict the dynamic performance of IESs across multiple time scales. This neural network-based modeling approach develops time-series multi-layer perceptrons (MLPs) for the operating units and integrates them with prior process knowledge about system structure and fundamental dynamics. This integration forms three hybrid NNs—long-term, slow, and fast MLPs—that predict the entire system dynamics across multiple time scales. Leveraging these MLPs, we design an NN-based scheduler and an NN-based economic model predictive control (NEMPC) framework to meet global operational requirements: rapid electrical power responsiveness to operators' requests, adequate cooling supply to customers, and increased system profitability, while addressing the dynamic time-scale multiplicity present in IESs. The proposed day-ahead scheduler is formulated using the ReLU network-based MLP, which effectively represents IES performance under a broad range of conditions from a long-term perspective. The scheduler is then exactly recast into a mixed-integer linear programming problem for efficient evaluation. The real-time NEMPC, based on slow and fast MLPs, comprises two sequential distributed control agents: a slow NEMPC for the cooling-dominant subsystem with slower transient responses and a fast NEMPC for the power-dominant subsystem with faster responses. These agents collaborate in the decision-making process to achieve dynamic synergy in real time while reducing computational costs. Extensive simulations demonstrate that the developed scheduler and NEMPC schemes outperform their respective benchmark scheduler and controller by about 25\% and 40\%. Together, they enhance overall system performance by over 70\% compared to benchmark approaches.
	\end{abstract}
	
	
	\section{Introduction}
	
	The rapid advancement of sustainable energy utilization and cutting-edge energy network technologies has fueled the ambitions of industry, academia, and policymakers toward an energy transition and a green revolution \cite{arent2021multi}. In pursuit of enhanced environmental, economic, and social benefits, the composition and management of energy systems have undergone substantial transformations \cite{berjawi2021evaluation}. An increasing number of energy producers and consumers are being integrated into large-scale energy systems, commonly referred to as integrated energy systems (IESs) or multi-energy systems \cite{ramsebner2021single}. A typical IES spans multiple sectors, encompassing a diverse range of operating units, including clean renewable energy sources, controllable power generation units, cooling/heating facilities, energy storage systems, and multi-energy end-customers. While theoretically, the intensive integration within an IES promises greater overall energy efficiency and operational flexibility, in practice, the diverse nonlinear process interactions within the system introduce complex multi-time-scale dynamic characteristics \cite{wu2019large}. Effectively describing, coordinating, and managing these operating units, while unlocking the dynamic synergy potential across various sectors, has become a pressing challenge \cite{zheng2024systematic}.
	
	As multi-energy prosumers, IESs amalgamate various energy operating units across both supply and demand sides, showcasing their potentiality to promptly adjust power generation, assist utility grids in balancing electricity supply and demand, and address the growing intermittency of renewable energy generation \cite{yi2024low, jalving2023beyond}. Beyond securing additional profits by responding to unscheduled power regulation requests from utility grids, IESs offer flexible support to maintain grid reliability through grid response programs \cite{dowling2017multi, jalving2023beyond}. Notably, existing studies indicate that failing to adequately consider the system dynamics and time-scale multiplicity inherent in IESs constrains their capability in power response during daily operations \cite{paiva2013controllable, wu2022economic}. Moreover, this inadequacy can diminish the overall energy supply levels within the system and negatively impact its profitability \cite{lei2022research, wu2023distributed}.
	
	Due to the unique real-time nature of electricity, power systems typically divide operational management into two stages: day-ahead scheduling and real-time precise control, ensuring alignment with electricity markets \cite{gao2022multiscale}. To manage computational complexity, models used in IES scheduling, such as those in robust and multi-stage scheduling \cite{lu2021robust, liu2024multi}, \rev{typically} rely on linear economic models based on the system's nominal conditions. Since the accuracy of a single linear model is limited, \rev{there can be a significant mismatch between the dynamic behaviors of the actual plant and the linear model} \cite{bindlish2016power}. \rev{To address this limitation, researchers} have begun exploring other preferable models for scheduling strategies, such as piecewise linear (PWL) models at several operating points \cite{urbanucci2018limits}. K{\"a}mper et al. proposed an automated data-driven model generation method employing piecewise linear regression for multi-energy systems as their optimization basis \cite{kamper2021automog}. Han et al. embedded PWL models into a robust scheduling design for the optimal operation of a hydro-wind generation system \cite{han2023overcoming}, while Cui et al. adopted PWL models in the scheduler of building virtual power plants \cite{cui2024data}. Various PWL models have also been utilized in optimal scheduling for hybrid systems with water distribution and multiple storage \cite{moazeni2020optimal, cheng2021multi}. Recently, Wu et al. examined a two-layer scheduling approach using multi-linearization procedures to boost the energy and environmental efficiency of IESs \cite{wu2024two}. While these models provide a better approximation of local behavior compared to a single linear model, they still fall short in accurately representing the complex nonlinear dynamics of the system across a wide range of conditions and may thus diverge from the global optimum \cite{bemporad2000performance, igarashi2020mpc}. These shortcomings hinder improvements in system energy efficiency and overall performance in the long term, ultimately weakening the foundation for subsequent coordinated control implementation to achieve dynamic synergy.
	
	In real-time coordinated control, some existing approaches continue to rely on classical hierarchical optimization concepts, employing linear steady-state models for real-time adjustment and dispatch. \rev{These approaches overlook} system dynamics and therefore lead to degraded dynamic performance \cite{ellis2017economic, baldea2014integrated}. Nevertheless, such steady-state models are widely applied to intraday operational optimization, particularly in popular multi-level rolling optimization approaches \cite{yao2023multi, li2021multi}. Meanwhile, other researchers have integrated the dynamic properties of certain operating units \rev{as} an attempt to enhance system responsiveness. Qin et al. presented a multi-timescale hierarchical scheduling framework for a power-heating IES, taking into account the thermal inertia of combined heat and power (CHP) units and the heat network \cite{qin2022multi}. Similarly, Yang et al. integrated the inertia of buildings and the heat network into a two-timescale predictive control scheme to enhance the dynamic flexibility of CHP units \cite{yang2023dynamic}. Additionally, the dynamic responses of a microturbine and gas boiler were incorporated into a bi-level predictive control for a community IES to boost its dynamic response performance \cite{liu2022bi}. Although accounting for the response characteristics of some units can improve system dynamic performance \rev{to some extent}, it still falls short of leveraging the diverse and complementary dynamic behaviors between operating units for precise and rapid regulation of transient processes \cite{zhang2022rapid, jin2022power}. Consequently, this oversight in real-time control significantly limits the system’s capacity for dynamic synergistic enhancement, undermining the system's rapid power responsiveness to utility grid requests, energy supply quality to microgrids, and economic benefits \cite{wu2023distributed, jin2022power}.
	
	However, real-time operational methods that \rev{characterize} complete system dynamics of IESs have rarely been discussed and have only recently started to attract the attention of researchers. For a heat-power station with renewable sources, Lei et al. analyzed the system's dynamic responses in detail and designed a hierarchical control scheme using first-principles models, where the upper layer controls the heating sector and the lower layer controls the power sector \cite{lei2022research}. Similarly, Jin et al. studied a power-heat IES with dynamic response tools and introduced a distributed model predictive controller based on identified linear state-space models, enabling real-time dynamic coordination between power and heating sectors with different sampling intervals \cite{jin2022power}. Wu et al. examined an illustrative IES for cooling production and power generation, developing a time-scale separation method based on first-principles models to identify dynamic response differences and implement model reduction across multiple time scales; they then proposed a composite economic model predictive control approach to address the multi-time-scale properties exhibited in IES dynamics \cite{wu2022economic}. Building on this, Wu et al. developed a systematic subsystem decomposition guideline with a directed graph representation of IESs to reveal the underlying topology of their dynamics; they subsequently designed a distributed economic model predictive control, also based on first-principles models, for the modular management and dynamic coordination of IESs \cite{wu2023distributed}. These studies emphasize the crucial importance of detailed representations and careful consideration of dynamic characteristics in real-time coordination and electrical power responses. Nonetheless, a significant gap persists between existing real-time dispatch or control strategies and the expected system-wide dynamic synergy, which remains largely unaddressed.
	
	Among wide-ranging control frameworks, model predictive control (MPC) stands out \rev{for its ability to provide optimal real-time system operation performance while handling various constraints} \cite{rawlings2017model}. \rev{We have witnessed extensive applications of MPC in real-time operational management of energy systems.} In addition to the previously mentioned studies, supervisory MPC and distributed MPCs have been applied to wind-solar-battery systems based on nonlinear first-principles models \cite{qi2011supervisory, kong2019hierarchical}. Similar supervisory and distributed MPC approaches have also been proposed for grid-interactive buildings \cite{tang2019model, bay2022distributed}, though these employ linearization models for simpler controller implementation. To improve the accuracy of linear models, linear parameter varying (LPV) models have been introduced into MPC design for waste heat recovery systems and IESs \cite{shi2023data, li2024optimal}. With the increasing prevalence of MPC, researchers are turning to a flexible optimal control and decision-making tool known as economic MPC (EMPC) \cite{rawlings2012fundamentals}. EMPC incorporates global economic and dynamic design, providing much greater flexibility in optimizing dynamic processes and achieving system-wide synergy, distinguishing it from conventional tracking MPC \cite{ellis2017economic, baldea2014integrated}. Apart from the aforementioned studies \cite{wu2022economic, wu2023distributed}, Clarke et al. presented an EMPC framework for controlling solar-battery-generator systems \cite{clarke2020control}, while Wang et al. and Zhu et al. investigated the applicability of distributed EMPC for wind farms and hybrid solar-hydrogen power generation \cite{wang2024wind, zhu2024pv}. Furthermore, Borja-Conde et al., Rawlings et al., and Zhang et al. developed EMPC strategies for heating, ventilation, and air-conditioning (HVAC) systems \cite{borja2024efficient, rawlings2018economic} and traditional power plants \cite{zhang2020zone}, respectively. All these EMPC approaches require first-principles models in the controller design procedure, except for \cite{rawlings2018economic}, which utilizes PWL models. Despite the success of MPC in process control, the linearization models used in MPC generally sacrifice dynamic performance for simpler and more efficient control implementation \cite{henson1998nonlinear}. In contrast, while creating elaborate first-principles models for MPC can provide superior descriptions of nonlinear dynamics \cite{morari1999model}, it remains a demanding task due to the diverse operating units and intricate dynamic interactions within IESs, and it also comes with high computational costs \cite{morari1999model}. Additionally, to avoid unmanageable computational complexity and ill-conditioned optimization problems, first-principles models often require further model reduction to facilitate the application of control schemes \cite{decardi2023toward, wu2023decomposition, yin2017distributed}, raising extra barriers between advanced control approaches and engineering practice. As the number of integrated operating units increases, the complexity and interactivity of the system’s processes and dynamics also rise, making the pursuit of alternative dynamic models for system optimization increasingly urgent and necessary.
	
	In this context, powerful machine learning \rev{approaches}, particularly neural network (NN) modeling tools capable of predicting dynamics close to those of first-principles models \cite{wu2019machine, ren2022tutorial, chen2020machine}, are emerging as effective surrogates for conventional models and demonstrate significant appeal \cite{venkatasubramanian2019promise}. In energy and process system optimization, methods based on neural networks are gaining traction as promising solutions, and researchers have made considerable efforts and explorations in this area. Xiao et al. proposed a novel energy dispatching plan for grid-interactive communities that combines a recurrent neural network (RNN) and long short-term memory (LSTM) integrated into MPC, preventing indoor comfort constraint violations while improving grid stability \cite{xiao2024physically}. Jalving et al. presented a feedforward neural network (FNN) surrogate for IESs participating in energy markets, which shows proactive performance surpassing traditional algebraic models \cite{jalving2023beyond}. Ghafariasl et al. compared the performance of several neural network types, including LSTM, artificial neural network (ANN), and convolutional neural network (CNN), in multi-objective optimization of multi-generation systems, offering practical recommendations for related studies \cite{ghafariasl2024neural}. Elnour et al. evaluated the performance of ANN-based MPC for building climate control in sports facilities \cite{elnour2022neural}, while Wang et al. used an FNN-based controller to enhance HVAC systems for frequency regulation service \cite{wang2022machine}. Ellis et al. embedded encoder-decoder LSTM models into EMPC for superior control of building HVAC systems \cite{ellis2020encoder}, with similar findings reported by Li et al., who used encoder-decoder RNN models \cite{li2021model}. Neural networks have also been widely employed to predict uncertain disturbances or cooling demands, facilitating the application of MPC in multi-energy and building energy systems \cite{coccia2021artificial, cox2019real, kumar2023grey}. Furthermore, neural network-based MPC and optimization have been effectively applied to the real-time control of solar collector fields \cite{katz2020integration}, irrigation process systems \cite{huang2023model}, and biomass fluidized bed reactors \cite{wang2023predictive}, as well as in state estimation of carbon capture units \cite{liu2023state} and the prediction of the remaining useful life of batteries \cite{bhadriraju2023adaptive}. While neural network-based operational frameworks have shown preliminary success and broad application potentiality in studies, several issues remain unresolved for large-scale IESs.
	
	First, for IESs with intricate \rev{dynamics and inter-/intra-unit} interactions, accurately describing dynamic behaviors is \rev{essential} for successful scheduling and control schemes. However, how to develop suitable neural network-driven dynamic models that accurately capture the system's behavior across multiple time scales while facilitating the design of operational strategies for synergistic enhancement is still unclear. Second, accurately predicting the dynamic behavior of complex, nonlinear, multi-input, multi-output process systems, such as IESs, often requires large-scale neural network models to capture the underlying temporal evolution of data points \cite{gizaw2023artificial}. This not only complicates model training but also increases the computational complexity of control problems arising from numerous decision variables \cite{ren2022tutorial, alhajeri2022process}. Existing studies have demonstrated that appropriately incorporating process structural knowledge helps reduce model size and enhance prediction accuracy \cite{wu2020process, alhajeri2022process}. However, neural network methods in the field of complex energy systems management are not yet sufficiently explored, particularly regarding leveraging prior knowledge of system structures and fundamental dynamics during neural network modeling. Finally, there remains a lack of research on designing successful scheduling and distributed control schemes based on developed neural networks to deal with time-scale multiplicity and meet operational requirements, such as fulfilling energy supply demands, increasing operational revenue, improving grid responsiveness, and mitigating computational burdens.
	
	These pressing issues \rev{motivate us to explore fresh approaches to} control-oriented dynamic modeling and energy management schemes for IESs. Taking a grid-connected IES for power generation and cooling production as an illustration, we propose a hybrid time-series neural network-based dynamic modeling approach incorporating limited prior process knowledge and develop a neural network-based optimal scheduling and economic model predictive control framework for achieving dynamic synergy and economic optimization of IESs. Specifically, we first introduce a generic time-series multi-layer perceptron (MLP) to describe the dynamic behavior of various operating units across multiple time scales. By integrating commonly used input-output structures, unit interconnections, and fundamental dynamics, we construct three hybrid MLPs: a long-term MLP based on ReLU networks, a slow MLP, and a fast MLP, which accurately reflect the entire system's dynamic characteristics at hourly, minute, and second levels while avoiding unnecessary modeling efforts. Subsequently, we design a ReLU network-based day-ahead scheduler that considers global optimization objectives and exploits the long-term potential to enhance economic and dynamic performance, providing a preferable optimization foundation for subsequent real-time control. The scheduling optimization problem is then equivalently converted into a mixed-integer linear programming (MILP) problem for efficient solving. Finally, utilizing the developed slow and fast MLPs, we propose a neural network-based economic model predictive control (NEMPC) scheme aimed at improving rapid power responsiveness during transient processes, meeting customers' cooling demands, and maximizing operational revenue. The NEMPC framework, with a global optimization focus, employs a distributed approach featuring two local sequential NEMPC agents: a slow NEMPC based on the slow MLP for regulating the cooling-dominant subsystem with a relatively slower dynamic response, and a high-frequency fast NEMPC based on the fast MLP for controlling the power-dominant subsystem exhibiting relatively rapid transient response. These control agents collaborate through distributed design and information exchange, effectively tackling multi-time-scale coordination in real-time control while enhancing overall system performance and reducing computational costs. Extensive simulation results confirm the effectiveness and applicability of the proposed framework.
	
	The main contributions of this study are:
	\begin{enumerate}
		
		\item We \rev{develop} a hybrid time-series neural network model architecture for complex energy systems that incorporates process structural knowledge, providing an accurate representation of dynamic performance across various conditions and time scales.
		
		\item We \rev{propose} a neural network-based optimal scheduling scheme for complex energy systems, demonstrating how long-term optimization can significantly enhance both dynamic and economic performance.
		
		\item We \rev{present} a sequential distributed neural network-based economic model predictive control framework that achieves real-time dynamic synergy, thereby improving grid responsiveness, cooling satisfaction, and system profitability, while minimizing computational effort.
		
	\end{enumerate}
	
	The remainder of this paper is organized as follows: Section 2 explains the studied IESs; Section 3 provides the proposed hybrid time-series neural networks; Section 4 describes the neural network-based scheduling and control strategy; Section 5 presents the simulations and comparisons; and Section 6 concludes the paper.
	
	\section{System description and control problem statement}
	
	\subsection{System description}
	
	\begin{figure}[!ht]
		\centering
		\includegraphics[width=0.95\linewidth]{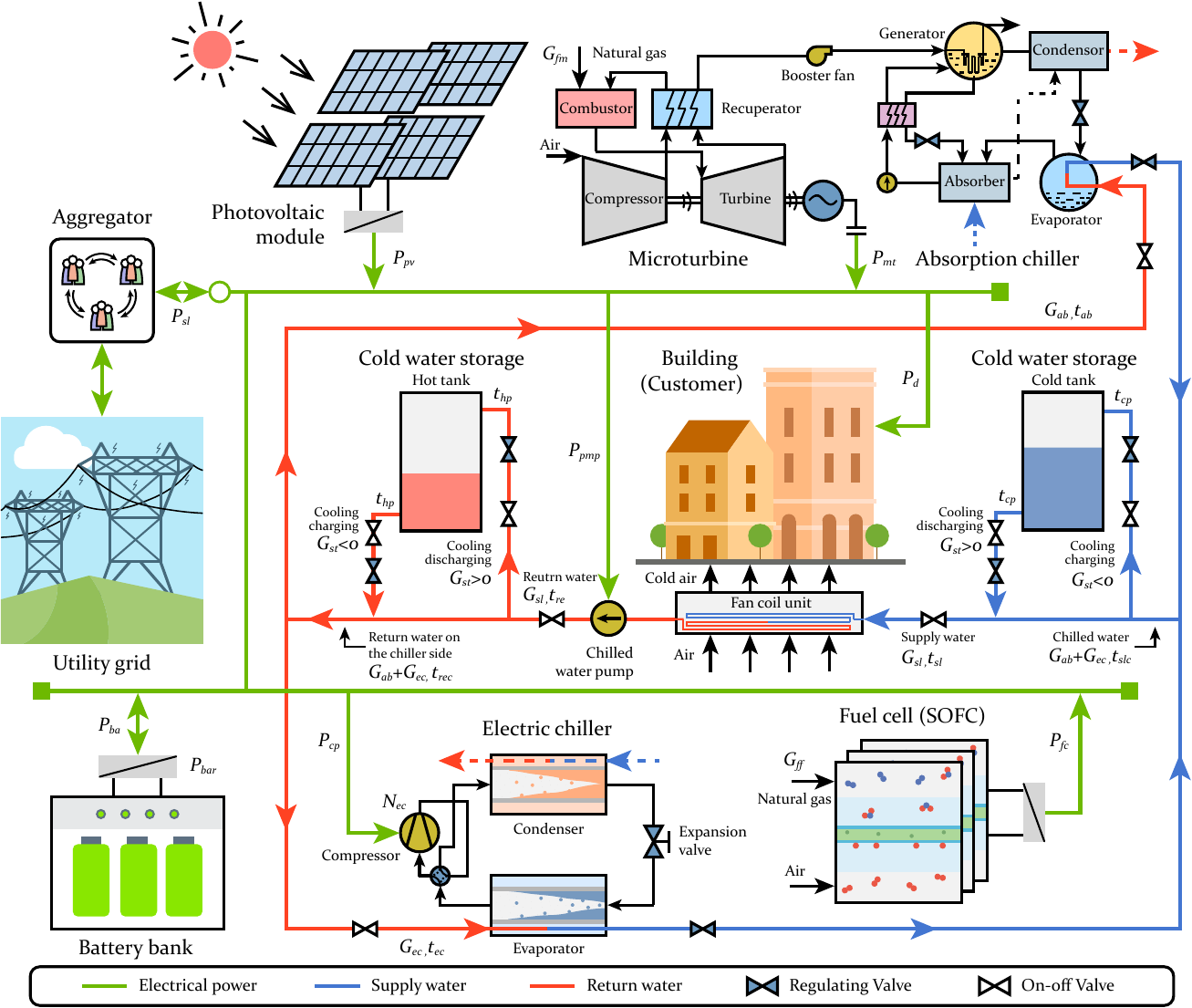}
		\caption{\rev{The considered grid-connected IES for power and cooling generation.}}
		\label{F1}
	\end{figure}
	
	This study examines a grid-connected integrated energy system (IES) that generates both power and cooling, as illustrated in Figure~\ref{F1}. The IES operates with three primary objectives: (i) to meet local customers' demands for electricity and cooling (microgrid demands); (ii) to supply electrical power to the utility grid through tie-lines, either according to the day-ahead scheduled baseline when not participating in real-time grid response, or based on the utility grid operator's real-time instructions when engaged in grid response (grid demands) \cite{dowling2017multi, kaundinya2009grid}; and (iii) to maximize operating earnings. Conflicts may arise between the electrical power requests of intra-microgrid customers and the requirements of the utility grid due to uncertainties associated with renewable energy sources, variable loads, and responsive instructions. In such situations, the IES prioritizes meeting local customers' electrical demands to ensure the safety and reliability of the microgrid, while the utility grid can rely on other dispatchable generation resources to address any electrical shortfalls \cite{gao2022multiscale}.
	
	The integrated energy system (IES) under consideration consists of several interconnected operating units: a photovoltaic system (PV), a battery bank (BA), a microturbine combined with an absorption chiller (MA), a fuel cell (FC), an electric chiller (EC), a cold water storage tank (CS), a building (BD), and auxiliary components such as water pumps (PMP). These units are connected via a microgrid and pipelines, forming an integrated power-cooling nexus. In operation, the PV system, fuel cell, and microturbine generate electricity from solar energy and natural gas, respectively. The battery bank manages power flow by discharging or absorbing electricity to smooth output fluctuations and facilitate long-term load shifting. Most of the generated power is supplied to local customers and the utility grid, while a portion is used by the electric chiller and pumps for cooling. The electric chiller converts this power into chilled water, while the absorption chiller produces cooling by utilizing waste heat from the upstream microturbine. The chilled water mixes with stored cold water from the storage tank when in cooling discharge mode, creating the final supply water that is piped into the building's fan coil unit for indoor temperature regulation. After heat exchange, the supply water becomes return water. Part of this return water is circulated back to the chillers for the next cooling cycle, while the remainder is stored in the hot water tank.
	
	Constructing a conventional first-principles dynamic model for such systems is often time-consuming and labor-intensive. For a comprehensive nonlinear dynamic model of the IES, along with detailed explanations, readers are referred to our previous studies \cite{wu2023distributed, wu2022economic}. This subsection will provide essential information about the process structure, including the input-output configurations of each operating unit, their physical interconnections, and several fundamental dynamic formulations. These elements are crucial for developing the desired hybrid time-series neural networks for the IES.
	
	Specifically, the output power of the photovoltaics $P_{pv}$, which operates under maximum power point tracking (MPPT) to minimize the cost of power generation, can be expressed by a set of nonlinear algebraic equations:
	\begin{equation}
		P_{pv}(\tau) = g^{PV}_{fp}(t_{a}(\tau), S_{ra}(\tau))
	\end{equation}
	where $\tau$ denotes the time variable; $g^{PV}_{fp}$ represents a nonlinear function that maps the ambient temperature $t_{a}$ and solar radiation $S_{ra}$ to the photovoltaic power output $P_{pv}$. Under MPPT, the photovoltaic power output depends on the uncontrollable environmental conditions $t_{a}$ and $S_{ra}$, which act as disturbances for the IES.
	
	In the case of the fuel cell, the electrical power $P_{fc}$ generated is determined by:
	\begin{equation}
		P_{fc}(\tau) = g^{FC}_{fp}(x^{FC}(\tau), G_{ff}(\tau), z_{fc}(\tau))
	\end{equation}
	where $x^{FC} \in \mathbb{R}^{5}$ represents the state variables vector including parameters such as current and pressure within the fuel cell, which characterize its dynamic response. These variables $x^{FC}$ are governed by the five nonlinear ordinary differential equations, $\dot{x}^{FC}(\tau) = f^{FC}_{fp}(x^{FC}(\tau), G_{ff}(\tau), z_{fc}(\tau))$. Here, $G_{ff}$ represents the mass flow rate of natural gas supplied to the fuel cell, serving as a continuous manipulated input; $z_{fc}$ is a binary input that controls the on/off state of the fuel cell. Both $G_{ff}$ and $z_{fc}$ are integral parts of the manipulated inputs for the IES.
	
	Next, the power produced by the microturbine, $P_{mt}$, and the supplied chilled water temperature from the downstream absorption chiller, $t_{ab}$, can be computed as:
	\begin{equation}
		P_{mt}(\tau) = g^{MA}_{fp,1}(x^{MA}(\tau), G_{fm}(\tau), z_{ma}(\tau))
	\end{equation}
	\begin{equation}
		t_{ab}(\tau) = g^{MA}_{fp,2}(x^{MA}(\tau), G_{fm}(\tau), G_{ab}(\tau), z_{ma}(\tau), t_{rec}(\tau))
	\end{equation}
	where $x^{MA} \in \mathbb{R}^{4}$ represents the state variables vector depicting the transient response behavior of the microturbine integrated with the absorption chiller. The governing equations for these states are given by the set of four state equations, $\dot{x}^{MA}(\tau) = f^{MA}_{fp}(x^{MA}(\tau), G_{fm}(\tau), G_{ab}(\tau), z_{ma}(\tau))$. $G_{fm}(\tau)$ denotes the natural gas feed rate to the microturbine, $G_{ab}(\tau)$ indicates the chilled water flow rate within the absorption chiller, and $z_{ma}$ is a binary input controlling whether the unit is on or off. These inputs are part of the IES's control inputs. $t_{rec}$ is the return water temperature on the chiller side, which stands for an input for the absorption chiller but cannot be directly controlled by the current unit since $t_{rec}$ is affected by other operating units in the pipeline. Such variables, which are inputs to the current unit but depend on other units, are termed linked variables and are typically intermediate variables in the IES. Additionally, the cooling power of the chiller, $Q_{ab}$, can be calculated as $Q_{ab} = G_{ab}C_{w}(t_{rec} - t_{ab})$, where $C_{w}$ denotes the specific heat capacity of water.
	
	For the electric chiller, the power consumed by the compressor, $P_{cp}$, and the resulting chilled water temperature, $t_{ec}$, can be evaluated similarly:
	\begin{equation}
		P_{cp}(\tau) = g^{EC}_{fp,1}(x^{EC}(\tau), N_{ec}(\tau), G_{ec}(\tau), z_{ec}(\tau), t_{rec}(\tau))
	\end{equation}
	\begin{equation}
		t_{ec}(\tau) = g^{EC}_{fp,2}(x^{EC}(\tau), N_{ec}(\tau), G_{ec}(\tau), z_{ec}(\tau), t_{rec}(\tau))
	\end{equation}
	where $x^{EC} \in \mathbb{R}^{6}$ is the state vector reflecting the dynamics of the electric chiller, given by the differential equation $\dot{x}^{EC}(\tau) = f^{EC}_{fp}(x^{EC}(\tau), N_{ec}(\tau), G_{ec}(\tau), z_{ec}(\tau), t_{rec}(\tau))$. $N_{ec}(\tau)$, $G_{ec}(\tau)$, and $z_{ec}(\tau)$ stand for the manipulated inputs: the compressor speed, the chilled water flow rate, and the on-off switch, respectively. These inputs are part of the IES's control variables. $t_{rec}$ serves as the linked input, similar to its role in the absorption chiller, and is a intermediate variable in the IES. Thus, the cooling capacity of the chiller is given by $Q_{ec} = G_{ec}C_{w}(t_{rec} - t_{ec})$.
	
	In the battery bank, the electrical power $P_{ba}$ and the current $I_{ba}$ are described as follows:
	\begin{equation}
		P_{ba}(\tau) = g^{BA}_{fp,1}(x^{BA}(\tau), P_{bar}(\tau))
	\end{equation}
	\begin{equation}
		I_{ba}(\tau) = g^{BA}_{fp,2}(x^{BA}(\tau), P_{bar}(\tau))
	\end{equation}
	where $x^{BA} \in \mathbb{R}^{2}$ accounts for the voltage and current transient states in the battery, depicted by the state equation $\dot{x}^{BA}(\tau) = f^{BA}_{fp}(x^{BA}(\tau), P_{bar}(\tau))$. $P_{bar}$ is the power command, serving as a manipulated input for the battery and a controllable input within the IES. The battery bank discharges when $P_{ba} > 0$ and $I_{ba} > 0$; otherwise, it charges. The state of charge $C_{soc}$, which quantifies the energy currently available in the battery, can be obtained using the following straightforward differential equation:
	\begin{equation} \label{E12}
		\dot{C}_{soc}(\tau) = - \frac{I_{ba}(\tau)}{3600 C_{eb} \eta_{pb}}
	\end{equation}
	where $C_{eb}$ is the charge per battery cell and $\eta_{pb}$ is the number of cells connected in parallel.
	
	To proceed, a cold water tank and a hot water tank make up the cold storage module in the IES. The cold water tank stores the chilled water produced by the chiller, while the hot water tank holds the return water from the building (i.e., customers). Compared to other units, the dynamics of cold water storage are relatively straightforward and can be readily expressed. The discharged/charged cooling power of the cold storage, $Q_{st}$, is given by:
	\begin{equation} \label{E15}
		Q_{st} = G_{st} C_w (t_{hp} - t_{cp})
	\end{equation}
	where $G_{st} = z_{st} G_{stu} + (z_{st} - 1) G_{stu}$ represents the circulating water flow rate in the cold storage. Here, $t_{hp} = z_{st} t_{re} + (1 - z_{st}) t_{sth}$ is the water temperature flowing between the cold tank and the pipes, and $t_{cp} = z_{st} t_{stc} + (1 - z_{st}) t_{slc}$ is the water temperature flowing between the hot tank and the pipes. In these expressions, $z_{st}$ and $G_{stu}$ are the two manipulated inputs of the cold storage. $z_{st}$ indicates whether cooling is being discharged or charged, and $G_{stu}$ denotes the absolute value of the circulating water flow rate. $t_{re}$ and $t_{slc}$ are the return water temperature from the building and the chiller-supplied chilled water temperature, which are linked inputs for the cold storage. $t_{sth}$ and $t_{stc}$ stand for the temperatures of the water stored in the hot and cold tanks, respectively. They are calculated as $t_{sth} = C_{sth}/((1 - C_{sot}) M_{st})$ and $t_{stc} = C_{stc}/((1 - C_{sot}) M_{st})$. $C_{sth}$ and $C_{stc}$ are the heat capacities of the water in the cold and hot tanks, respectively. The dynamics of $C_{sth}$ and $C_{stc}$ are described by the differential equations:
	\begin{equation} \label{E16}
		\dot{C}_{sth}(\tau) = G_{st}(\tau) t_{hp}(\tau)
	\end{equation}
	\begin{equation} \label{E17}
		\dot{C}_{stc}(\tau) = -G_{st}(\tau) t_{cp}(\tau)
	\end{equation}
	$C_{sot}$, the capacity state of the cold storage, indicates the remaining cold water available and is calculated by the differential equation:
	\begin{equation} \label{E13}
		\begin{aligned}
			& \dot{C}_{sot}(\tau) = -\frac{G_{st}(\tau)}{M_{st}}
		\end{aligned}
	\end{equation}
	where $M_{st}$ is the maximum amount of water the cold storage can hold.
	
	Regarding the building equipped with a suite of fan coil units, the room temperature of concern to the customer, $t_{br}$, and the return water temperature, $t_{re}$, are determined by the following formulations:
	\begin{equation}
		t_{br}(\tau) = g^{BD}_{fp,1}(x^{BD}(\tau), t_{sl}(\tau), G_{sl}(\tau), t_{a}(\tau), Q_{o}(\tau))
	\end{equation}
	\begin{equation}
		t_{re}(\tau) = g^{BD}_{fp,2}(x^{BD}(\tau), t_{sl}(\tau), G_{sl}(\tau), t_{a}(\tau), Q_{o}(\tau))
	\end{equation}
	where $x^{BD} \in \mathbb{R}^{2}$ represents the state variables in the building that depict the dynamic behavior of the relevant temperatures. The state variables $x^{BD}$ follow the ordinary differential equation $\dot{x}^{BD}(\tau) = f^{BD}_{fp}(x^{BD}(\tau), t_{sl}(\tau), G_{sl}(\tau), t_{a}(\tau), Q_{o}(\tau))$. $t_{sl}$ and $G_{sl}$ are the final supplied water temperature, mixed from the chillers and cold storage, and the final supplied water flow rate, respectively. These are the linked inputs for the building, with their mathematical expressions provided later. $t_{a}$ indicates the ambient temperature, the same variable as in the photovoltaics, while $Q_{o}$ stands for the other cooling loads in the building unit. Both $t_{a}$ and $Q_{o}$ are uncontrollable conditions, acting as disturbances for the building and the entire IES. The customer-concerned $t_{br}$ is also one of the controlled outputs of the entire IES. Moreover, the final cooling power offered to the building is given by:
	\begin{equation} \label{E18}
		Q_{sl} = G_{sl}C_{w}(t_{re} - t_{sl})
	\end{equation}
	
	Furthermore, based on the fundamental laws of energy and mass conservation, the aforementioned relevant water temperature and flow rate in the pipeline network, as illustrated in Figure~\ref{F1}, can be determined as follows:
	\begin{equation} \label{E1}
		G_{sl} = G_{ab} + G_{ec} + G_{st}
	\end{equation}
	\begin{equation}
		t_{rec} = \frac{G_{sl}t_{re} - G_{st}t_{hp}}{G_{ab} + G_{ec}}
	\end{equation}
	\begin{equation}
		t_{slc} = \frac{G_{ab}t_{ab} + G_{ec}t_{ec}}{G_{ab} + G_{ec}}
	\end{equation}
	\begin{equation} \label{E2}
		t_{sl} = \frac{G_{ab}t_{ab} + G_{ec}t_{ec} + G_{st}t_{cp}}{G_{sl}}
	\end{equation}
	All the variables in the above equations have been previously defined and will not be repeated here. Moreover, the consumed electric power of the water pump straightly depends on the total amount of circulated water, which is expressed as follows:
	\begin{equation}
		P_{pmp}(\tau) = g^{PMP}_{fp}(G_{all}(\tau))
	\end{equation}
	where $P_{pmp}$ accounts for the pump power load; $g^{PMP}_{fp}$ is the nonlinear mapping from the total circulated water flow rate $G_{all}$ to $P_{pmp}$, and $G_{all}$ can be obtained as follows:
	\begin{equation} \label{E3}
		G_{all} = G_{ab} + G_{ec} + G_{stu}
	\end{equation}
	It should be noted that while $G_{ab}$, $G_{ec}$, and $G_{stu}$ are the manipulated inputs for the absorption chiller, electric chiller, and cold storage, respectively, once they are determined within their respective operating units, $G_{all}$ will serve as a linked input for the water pump rather than a manipulated input for the pump.
	
	Eventually, the electrical power delivered by the entire IES to the utility grid can be computed under the assumption that the local microgrid power loads $P_d$ are met first, and then any surplus or deficit is either supplied to or received from the utility grid. This can be expressed as follows, according to the energy flow balance:
	\begin{equation} \label{E4}
		P_{sl} = P_{pv} + P_{fc} + P_{mt} + P_{ba} - P_{cp} - P_{pmp} - P_{d}
	\end{equation}
	where $P_{sl}$ is the electrical power transported by the IES to the utility grid, which is another crucial controlled output of the entire IES; $P_d$ is the power demand of the local customers, a disturbance for the IES. Typically, $P_{sl}$ is greater than zero, indicating that power is being delivered from the IES to the utility grid. However, there are instances when the IES may need to receive power from the utility grid to cover power shortages. This occurs when $P_{sl}$ is less than zero, as the microgrid generation may not be able to meet the local power demands during very short transient periods.
	
	To conclude, the IES is composed of eight operating units (PV, FC, BA, MA, EC, PMP, CS, BD) that are integrated through linked inputs based on their cooling- and electricity-related interconnectivity (Eqs.\eqref{E1}-\eqref{E2}, Eq.\eqref{E3}, and Eq.\eqref{E4}). These units interact with each other, sharing disturbances, inputs, or outputs. The key variable values of these units under nominal working conditions are listed in Table~\ref{T1}. In addition, all operating units can be presented using a generic nonlinear dynamic state-space formulation. For an operating unit $j$ ($j \in \{PV, FC, BA, MA, EC, PMP, CS, BD\}$), we have:
	\begin{equation} \label{E5}
		\begin{aligned}
			&\dot{x}^j(\tau)=f^j(x^j(\tau), u^j(\tau), z^j(\tau), w^j(\tau), v^j(\tau)) \\
			&o^j(\tau)=g^j(x^j(\tau), u^j(\tau), z^j(\tau), w^j(\tau), v^j(\tau))
		\end{aligned}
	\end{equation}
	where $x^j$ denotes the dynamic state vector of unit $j$; $u^j$ and $z^j$ are the continuous and binary manipulated input vectors of unit $j$; $w^j$ represents the disturbance vector; $v^j$ is the linked input vector; and $o^j$ is the controlled output vector. Table~\ref{T2} provides a summary of all the elements of these vectors except for $x^j$, as this work focuses on developing hybrid time-series neural networks with input-output structures. Detailed explanations of $x^j$ can be found in \cite{wu2023distributed}.
	
	\begin{table}[!ht] \small
		\centering
		\caption{Nominal condition values for key variables}
		\label{T1}
		\renewcommand{\arraystretch}{1.3}
		\tabcolsep 2pt
		\begin{tabular}{p{2.3cm}p{2.3cm}p{2.3cm}p{2.3cm}p{2.3cm}p{2.3cm}} \hline
			\textbf{Variable} & \textbf{Value} & \textbf{Unit} & \textbf{Variable} & \textbf{Value} & \textbf{Unit} \\ \hline
			$P_{pv}$ & 44 & kW & $Q_{ab}$ & 75 & kW \\
			$P_{fc}$ & 40 & kW & $Q_{ec}$ & 50 & kW \\
			$P_{mt}$ & 80 & kW & $Q_{st}$ & 0 ($-21 \sim 21$) & kW \\
			$P_{ba}$ & 0 ($-40 \sim 40$) & kW & $Q_{sl}$ & 125 & kW \\
			$P_{cp}$ & 12.6 & kW & $t_{ab}$ & 7 & $^{\circ}$C \\
			$P_{pmp}$ & 13.9 & kW & $t_{ec}$ & 7 & $^{\circ}$C \\
			$t_{sl}$ & 7 & $^{\circ}$C & $t_{cp}$ & 7 & $^{\circ}$C \\
			$t_{re}$ & 12 & $^{\circ}$C & $t_{hp}$ & 12 & $^{\circ}$C \\
			\hline
		\end{tabular}
	\end{table}
	
	\begin{table}[!ht] \small
		\centering
		\caption{Summary of input and output variables for operating units}
		\label{T2}
		\renewcommand{\arraystretch}{1.3}
		\tabcolsep 2pt
		\begin{tabular}{p{2.5cm}p{2.5cm}p{2.5cm}p{2.5cm}p{2.5cm}p{2.5cm}} \hline
			\textbf{Operating unit} & \textbf{Continuous input} & \textbf{Binary input} & \textbf{Disturbance} & \textbf{Linked input} & \textbf{Output} \\ \hline
			PV & / & / & $t_{a}$, $S_{ra}$ & / & $P_{pv}$ \\
			FC & $G_{ff}$ & $z_{fc}$ & / & / & $P_{fc}$ \\
			BA & $P_{bar}$ &/ & / & / & $P_{ba}$, $I_{ba}$, $C_{soc}$ \\
			MA & $G_{fm}$, $G_{ab}$ & $z_{ma}$ & / & $t_{rec}$ & $P_{mt}$, $t_{ab}$ \\
			EC & $N_{ec}$, $G_{ec}$ & $z_{ec}$ & / & $t_{rec}$ & $P_{cp}$, $t_{ec}$ \\
			PMP & / & / & / & $G_{all}$ & $P_{pmp}$ \\
			BD & / & / & $t_{a}$, $Q_{o}$ & $t_{sl}$, $G_{sl}$ & $t_{re}$, $t_{br}$ \\
			CS & $G_{stu}$ & $z_{st}$ & / & $t_{slc}$, $t_{re}$ & $t_{cp}$, $t_{hp}$, $C_{sot}$ \\
			\hline
		\end{tabular}
	\end{table}
	
	\subsection{Control problem statement}
	
	From the perspective of the entire system, the IES comprises four uncontrollable external disturbances, four binary manipulated inputs for switching operating units on/off or charging/discharging storage, seven continuous manipulated inputs for regulating operating units, and two key controlled outputs. The dynamic characteristics of the entire system are represented by twenty-three states across the operating units within the IES. Define the disturbance vector as $w = [t_a, S_{ra}, P_d, Q_o]^T$, the binary input vector as $z = [z_{fc}, z_{ma}, z_{ec}, z_{st}]^T$, the continuous input vector as $u = [G_{ff}, G_{fm}, G_{ab}, N_{ec}, G_{ec}, G_{stu}, P_{bar}]^T$, the output vector as $y = [P_{sl}, t_{br}]^T$. Let $x$ denote the state vector consisting of $x^j$ ($j \in \{PV, FC, BA, MA, EC, PMP, CS, BD\}$). The entire grid-connected IES can then be described using a standard nonlinear dynamic state-space expression as follows:
	\begin{equation} \label{E6}
		\begin{aligned}
			& \dot{x}(\tau) = f(x(\tau), u(\tau), z(\tau), w(\tau)) \\
			& y(\tau) = g(x(\tau), u(\tau), z(\tau), w(\tau))
		\end{aligned}
	\end{equation}
	where $x \in \mathbb{R}^{23}$, $u \in \mathbb{R}^{7}$, $z \in \{0,1\}^{4}$, $w \in \mathbb{R}^{4}$, and $y \in \mathbb{R}^{2}$. It is worth noting that, in comparison with Eq.\eqref{E5} for a single operating unit, Eq.\eqref{E6} for the entire IES does not include the linked input $v$ since these variables are integrated into the connections between the operating units within the IES, instead of being treated as inputs to the IES. Comprehensive explanations of $x$ are also available in \cite{wu2023distributed}.
	
	Managing a complex grid-connected IES presents a primary challenge: accurately modeling its nonlinear dynamic behavior under various conditions. This modeling is essential for achieving successful operational optimization with high prediction accuracy. This challenge is further compounded by the diverse material transfers and energy flows among the interconnected operating units, as well as their dynamic interactions within the IES when using first-principles modeling. Standard scheduling and model predictive control schemes often struggle to manage the dynamic time-scale multiplicity exhibited in the IES and the numerous decision variables involved. This multi-time-scale property can lead to ill-conditioned optimization problems in typical first-principles model-based strategies \cite{yin2017distributed}, while the extensive decision-making factors create an unmanageable computational burden in centralized frameworks \cite{christofides2013distributed}. Moreover, the grid-connected IES is highly susceptible to fluctuating external conditions, such as volatile renewable energy generation, unstable local microgrid demands, and variable utility grid instructions. Consequently, a major hurdle in designing an effective scheduling and control scheme for the IES is finding an alternative to conventional first-principles dynamic models. The corresponding surrogate-based control approach must address the challenges of time-scale multiplicity and computational complexity while minimizing the negative impacts of disturbances on system dynamics. The goal is to coordinate the operating units within the IES to respond quickly to power regulation requests from the utility grid, satisfy local customers' cooling and electricity needs, and increase system profitability. To evaluate the performance of the IES, we employ the following three key indices:
	\begin{subequations} \label{E7}
		\begin{align}
			J_1 &= \|y_{1}-(1+\xi)y_e^b\|^2 \label{E7a} \\
			J_2 &= \|y_{2}-y_{sp,t}\|^2 \label{E7b} \\
			J_3 &= -\left(p_{mg}w_3 + p_{se}y_{1} + p_{cm}\xi_{as} y_e^b - p_f(u_{1}+u_{2}) - p_{pn}\|y_{1} - (1+\xi)y_e^b\|^2\right) \label{E7c}
		\end{align}
	\end{subequations}
	where $\xi$ stands for the regulation coefficient, defined as the ratio of real-time requests from the utility grid for unscheduled power to the day-ahead planned baseline power, $y_e^b$. The term $y_{sp,t}$ indicates the target building temperature within the customers' desired range. The variables $p_{mg}$, $p_{se}$, and $p_{cm}$ denote the electricity sale price in the microgrid, the wholesale electricity price in the utility grid, and the subsidy for providing unscheduled power responses, respectively. The coefficient $\xi_{as} = |\xi|$ is the absolute value of $\xi$. $p_f$ and $p_{pn}$ are fuel costs and penalty charges for failing to supply the requested power to the utility grid on time. Among these indices, $J_1$ and $J_2$ assess the power delivery deviation from the utility grid's real-time instructions and the building temperature deviation from the customers' acceptable range. $J_3$ reflects the system's economic status, being equal to the negative value of the earnings.
	
	\section{Creating \rev{physics-informed} hybrid neural networks for IESs}
	
	\begin{figure}[!ht]
		\centering
		\includegraphics[width=0.8\linewidth]{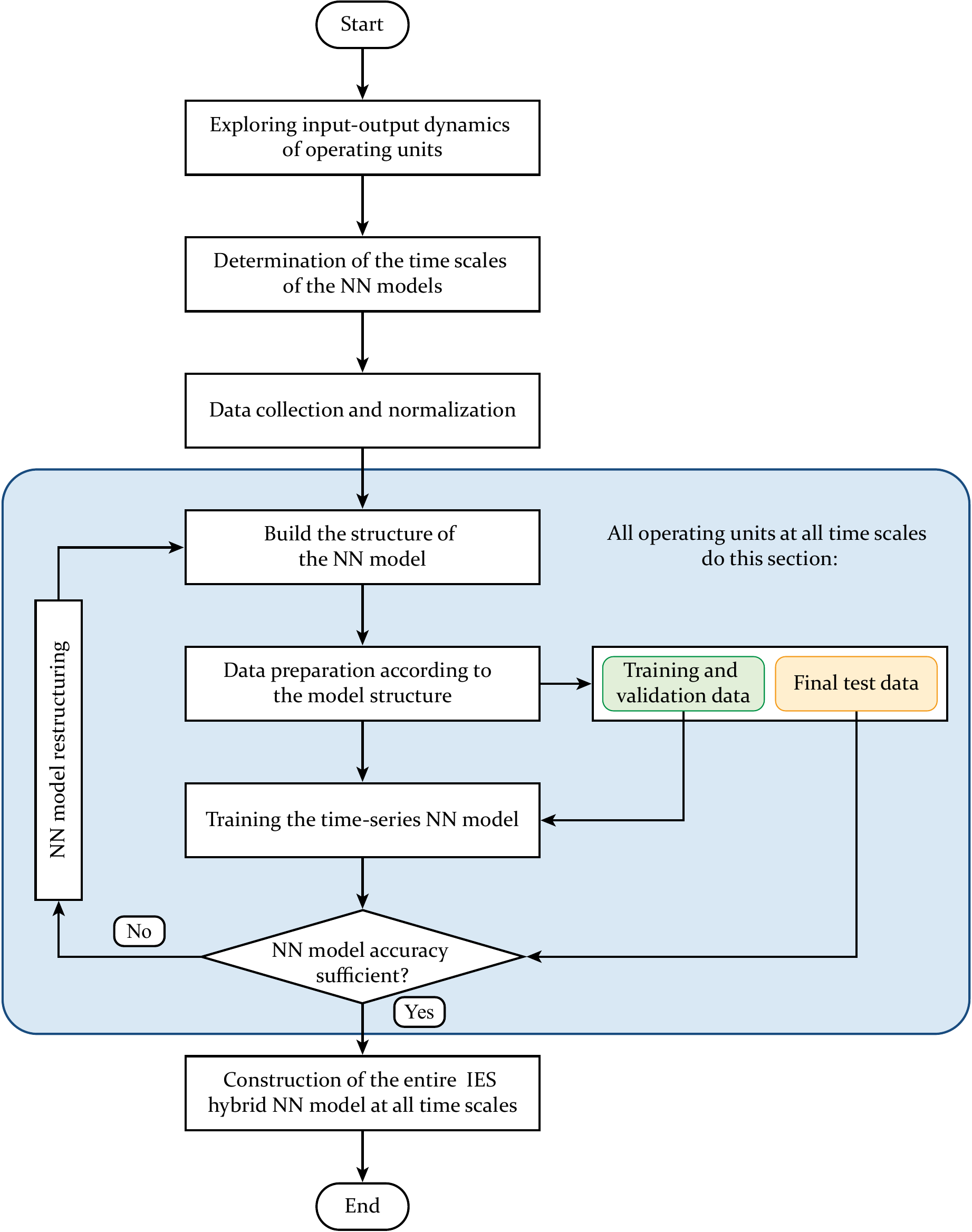}
		\caption{The workflow of establishing the hybrid neural network models for IESs.}
		\label{F2}
	\end{figure}
	
	Machine learning modeling techniques, particularly neural network tools, are increasingly seen as promising alternatives to common first-principles models. However, directly applying neural networks to IES modeling can lead to the creation of large-scale networks due to the complex dynamic interactions and various thermal-electrical processes involved \cite{gizaw2023artificial, wu2020process}. These oversized networks can be challenging to train during model development and may struggle to efficiently solve subsequent scheduling and control problems \cite{ren2022tutorial, alhajeri2022process}. Furthermore, they often fail to deal with the time-scale multiplicity present within the IES, which can ultimately compromise prediction accuracy \cite{alnajdi2023machine, abdullah2021data}. In this context, this section proposes hybrid time-series neural networks for IESs, enhanced by incorporating limited prior process knowledge. The flowchart of the proposed modeling approach is shown in Figure~\ref{F2}. First, we will investigate the dynamic response characteristics of the operating units to identify the dynamic multi-time-scale property. Next, neural networks are trained for multiple time scales based on the input-output structure of these units. Finally, the IES's multi-time-scale \rev{physics-informed} hybrid neural networks are rebuilt by integrating information on the physical interconnections between the operating units and basic dynamic formulations. The reconstructed hybrid time-series neural networks will later be used in the design of non-centralized predictive control and scheduling systems.
	
	\subsection{Dynamic response analysis of system units}
	
\begin{figure}[!ht]
		\centering
		\includegraphics[width=0.8\linewidth]{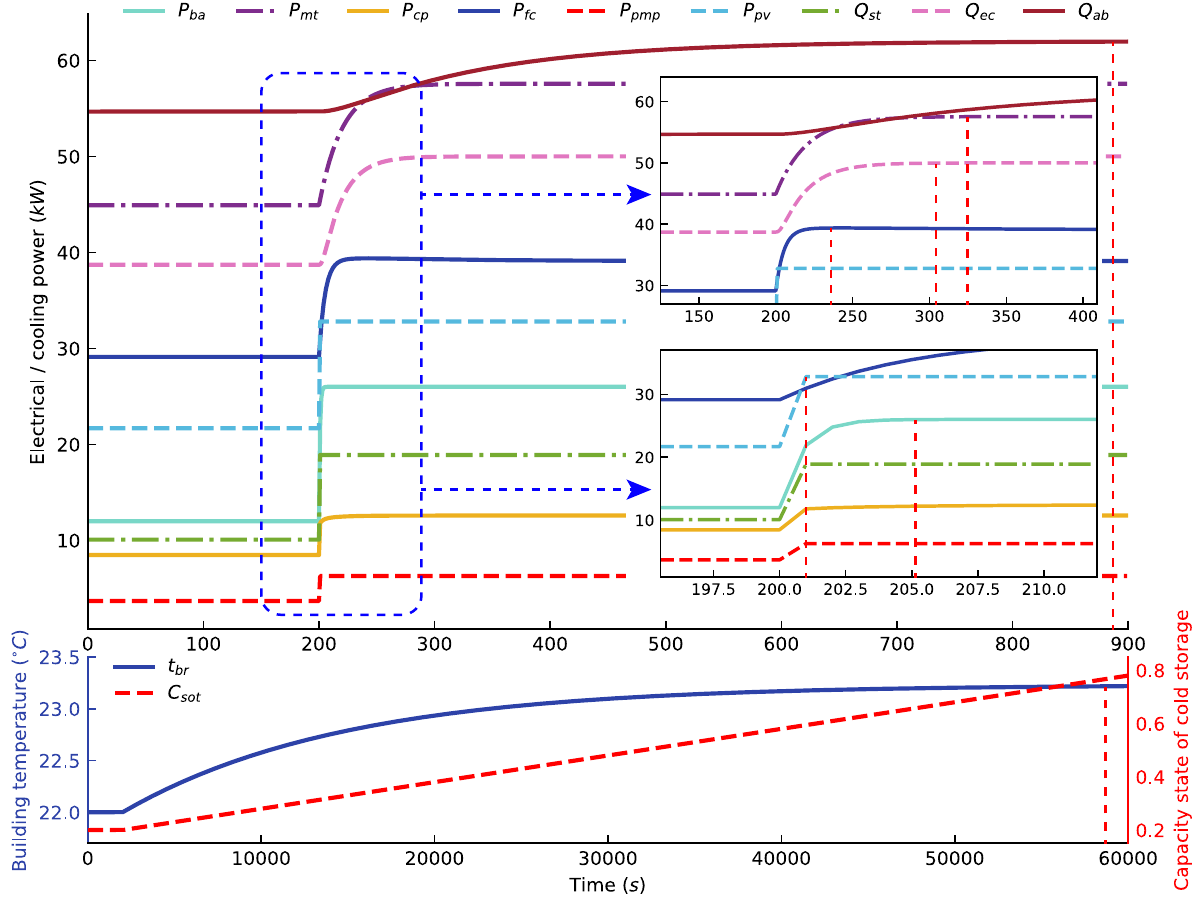}
		\caption{Step responses of operating units within the IES: the red dashed lines indicate where the approximate time constants of the outputs are located.}
		\label{F3}
	\end{figure}

	Investigating dynamic time scales using first-principles models is costly for engineers due to the difficulty in accessing highly accurate nonlinear dynamic models. However, in engineering practice, simple dynamic response tests of operating units can usually be performed to analyze their dynamic behavior and identify their dynamic time scales. Assuming we are allowed to conduct step response tests for each operating unit within the IES, this would enable us to observe the behavior of the units' outputs over time. The results, showing the units' output changes in response to step changes in their main inputs, are illustrated in Figure~\ref{F3}.
	
	Figure~\ref{F3} demonstrates that significant differences in dynamic responses exist among the various units within the IES. The battery, photovoltaics, pump, and compressor exhibit fast responses, while the fuel cell responds slightly slower, followed by the microturbine. The electric chiller is much slower than these units, and the absorption chiller is even slower than the electric chiller. Besides, the building and energy storage capacity states have the slowest responses. To confirm these observations, the approximate dominant time constants and residence times are adopted to quantify their dynamic response time scales. The dominant time constant is the time it takes for the output to reach 63.2\% of its final asymptotic value, while the residence time refers to the duration the working substance spends in a control volume \cite{skogestad2008chemical}. These values, determined from the step response tests, are listed in Table~\ref{T3}.
	
	\begin{table}[!ht] \small
		\centering
		\caption{Typical variables' approximate dominant time constants or residence times}
		\label{T3}
		\renewcommand{\arraystretch}{1.3}
		\tabcolsep 2pt
		\begin{tabular}{p{2.3cm}p{2.3cm}p{2.3cm}p{2.3cm}p{2.3cm}p{2.3cm}} \hline
			\textbf{Variable} & \textbf{Time (s)} & \textbf{Variable} & \textbf{Time (s)} & \textbf{Variable} & \textbf{Time (s)} \\ \hline
			$P_{pv}$ & 0.1 & $P_{cp}$ & 1.9 & $t_{br}$ & 12500 \\
			$P_{fc}$ & 7 & $Q_{ab}$ & 155 & $C_{sot}$ & 16000 \\
			$P_{mt}$ & 21 & $Q_{ec}$ & 25 & $C_{soc}$ & 14500 \\
			$P_{ba}$ & 0.9 & $Q_{st}$ & 1 &  &  \\
			\hline
		\end{tabular}
	\end{table}
	
	The values of the time constants or residence times of the typical variables for these operating units, shown in Table~\ref{T3}, are generally consistent with the above analysis of the response test results. These values exhibit explicit differences in magnitude, ranging from seconds to hours, indicating the presence of multiple dynamic time scales in the IES. The dynamic responses of cooling-related operating units, which have time scales of minutes to hours, are generally slower than those of electricity-related units, which have time scales from seconds to minutes. Some operating units display both fast and slow responses simultaneously, such as the compressor power and cooling generation in the electric chiller, as well as the differences between the microturbine and absorption chiller. In addition, the building's thermal inertia and the energy storage capacity states are notably slower than the responses of all other units.
	
	Given that the studied IES is a utility grid-connected system participating in the energy market, we need to consider both the time scales of electricity market operations and the multiple dynamic responses within the IES when determining the scheduling and control time intervals. The electricity market typically operates in two stages: the day-ahead market and the intraday real-time market \cite{gao2022multiscale}. In the day-ahead stage, energy systems are required to submit their hourly power generation plan (the scheduled power baseline) for the following day to the electricity market. During the intraday stage, energy systems must quickly adjust their power generation within minutes or even shorter intervals in response to changes in utility operator instructions, if they take part in real-time grid response \cite{dowling2017multi, kaundinya2009grid}.
	
	Therefore, the operational optimization of the IES should also encompass two stages: day-ahead scheduling and intraday real-time control. To tackle these optimization problems effectively and align with the time scales of the dynamic characteristics and the electricity market, we have chosen a 1-hour interval for day-ahead scheduling and 1-minute and 5-second intervals for intraday control. Certainly, the \rev{physics-informed} hybrid neural networks used for these scheduling and control tasks should be designed to accommodate these three time scales.
	
	\begin{remark}
		Note that the response of the cooling power in the cold storage, $Q_{st}$, is comparatively fast, as listed in Table~\ref{T3}. This is because two reservoir tanks are adopted in this work for cold and hot water storage. Once the water flow rates are rapidly changed, the cooling power transferred also changes rapidly. However, the states associated with storage capacity, such as $C_{sot}$, still change very sluggishly.
	\end{remark}
	
	\subsection{Dynamic time-series multi-layer perceptron for IES operating units}
	
	As previously discussed, neural network technologies are powerful modeling tools that enable the accurate prediction of intricate nonlinear dynamic systems without requiring a deep understanding of their underlying mechanisms. This trait frees engineers from extensive modeling work required by first-principles models, allowing them to focus on designing optimal operation strategies for the system. This, in turn, reduces the cost of implementing advanced control methods for complex large-scale energy systems. Hence, this subsection first introduces a generic time-series neural network model, which will be used to predict the dynamic behavior of the IES operating units on multiple time scales. Later, this model will be employed to reconstruct the hybrid neural network-based dynamic model of the entire IES.
	
	\begin{figure}[!ht]
		\centering
		\includegraphics[width=0.8\linewidth]{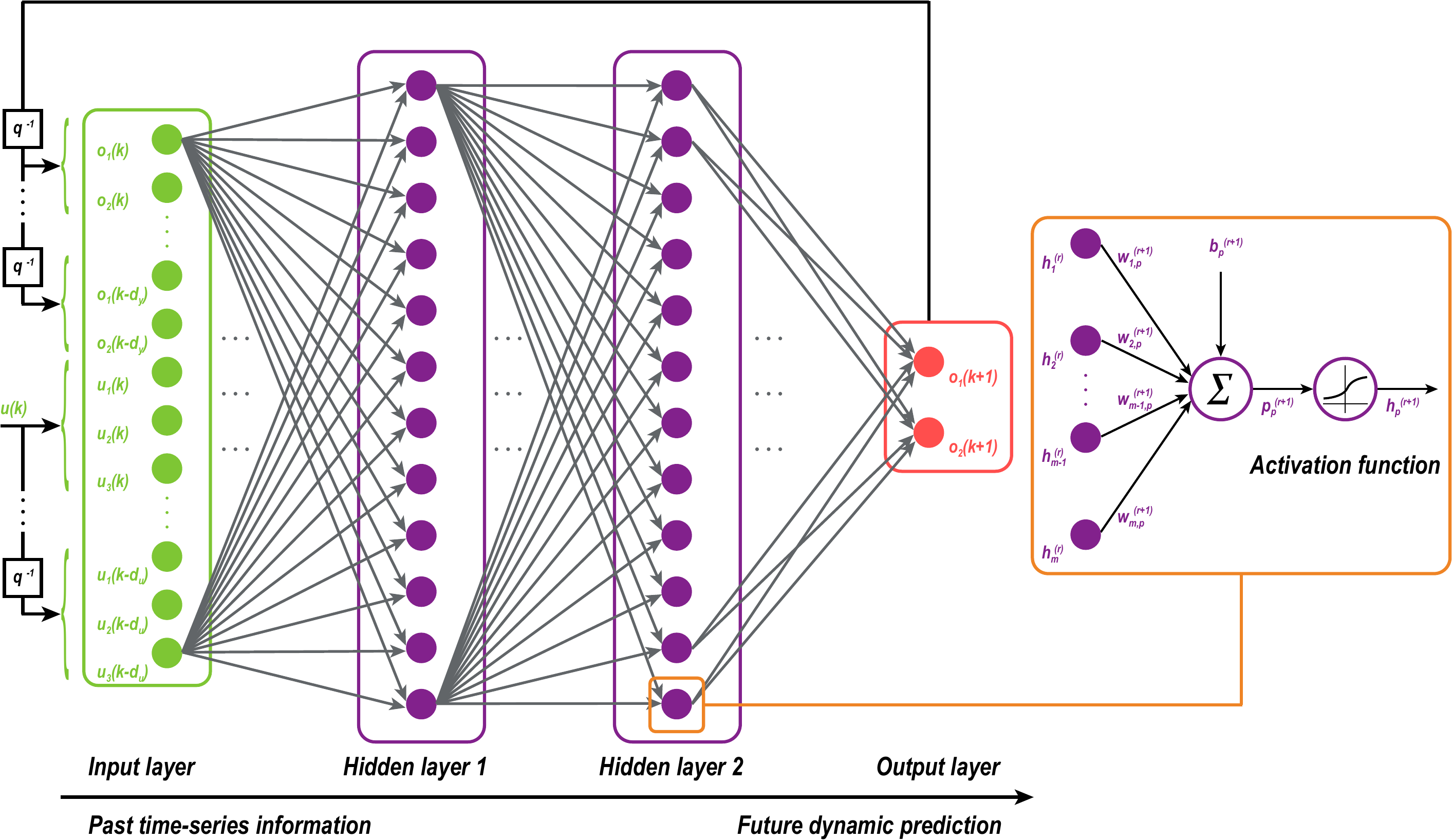}
		\caption{The dynamic time-series multi-layer perceptron employed in this work: $q^{-1}$ stands for the unit time delay.}
		\label{F4}
	\end{figure}
	
	Figure~\ref{F4} illustrates a time-series multi-layer perceptron, which is a specialized type of fully connected feedforward neural network characterized by dynamic nonlinear auto-regression with exogenous input \cite{menezes2008long, ren2022tutorial}. The key difference from a standard MLP, as depicted in Figure~\ref{F4}, is that the inputs to the time-series MLP include past information about the manipulated inputs and predicted outputs. This allows the MLP to learn the latent time-evolutionary features of the input data, thereby enabling accurate future output predictions. Specifically, if a set of time-series data pairs of a dynamic system with output vector $o$ and input vector $u$ is available at a given time instant $k$, including the output sequence of the past $d_o$ time steps $[o(k), o(k-1), \dots, o(k-d_o)]$ and the input sequence of the past $d_u$ time steps $[u(k), u(k-1), \dots, u(k-d_u)]$, then the time-series MLP with dynamic feedback attributes, as described in Figure~\ref{F4}, can be mathematically formulated in vector form as follows:
	\begin{subequations} \label{E8}
		\begin{align}
			& \mathsf{P}^{(r+1)} = \mathsf{H}^{(r)} \mathsf{W}^{(r+1)} + \mathsf{b}^{(r+1)} \label{E8a} \\
			& \mathsf{H}^{(r+1)} = \sigma_h^{(r+1)}(\mathsf{P}^{(r+1)}) \label{E8b}
		\end{align}
	\end{subequations}
	where the superscript $r$ ($r = 0, \dots, n_{hl}$) represents the $r$-th hidden layer of the neural network, in which $n_{hl}$ is the total number of hidden layers. In Eq.\eqref{E8a}, $\mathsf{P}^{(r+1)}$ denotes the pre-activation state vector of the neurons in hidden layer $r+1$; $\mathsf{H}^{(r)}$ indicates the hidden state vector of the neurons in the previous hidden layer $r$; $\mathsf{W}^{(r+1)}$ is the weight matrix for the neurons in hidden layer $r+1$, which weights the information propagation from the previous hidden layer to the current one; and $\mathsf{b}^{(r+1)}$ is the bias vector for layer $r+1$. Once $\mathsf{P}^{(r+1)}$ is obtained, the hidden state vector of the neurons in the current hidden layer $r+1$, $\mathsf{H}^{(r+1)}$, can be computed from Eq.\eqref{E8b}. In Eq.\eqref{E8b}, $\sigma_h^{(r+1)}$ denotes the nonlinear activation function for the hidden layer $r+1$ (such as Sigmoid, ReLU, or tanh), which is utilized to capture the non-linearity in the original dataset. The procedure is repeated for the next hidden layer $r+2$ using Eqs.\eqref{E8a} and \eqref{E8b} to evaluate the hidden state vector $\mathsf{H}^{(r+2)}$ until the final hidden state vector $\mathsf{H}^{(n_{hl})}$ is calculated. Notably, when $r = 0$, the zeroth layer is the input layer of the time-series MLP, meaning that $\mathsf{H}^{(0)}$ corresponds to the past input-output discrete-time sequence $[o(k), o(k-1), \dots, o(k-d_o), u(k), u(k-1), \dots, u(k-d_u)]$. Lastly, the output vector of the neural network $\mathsf{O}$ can be evaluated by:
	\begin{equation} \label{E9}
		\mathsf{O} = \mathsf{H}^{(n_{hl})} \mathsf{W}^{(out)} + \mathsf{b}^{(out)}
	\end{equation}
	where the calculated $\mathsf{O}$ equals the predicted system output at the next time instant, $o(k+1)$, while $\mathsf{W}^{(out)}$ and $\mathsf{b}^{(out)}$ are the weight matrix and bias vector of the output layer, respectively. For simplicity, the time-series MLP equations (Eqs.\eqref{E8} and \eqref{E9}) at time instant $k$ can be rewritten as the concise discrete-time prediction model expressed below:
	\begin{equation} \label{E10}
		o(k+1) = G({\mathbf{o}}(k), \mathbf{u}(k))
	\end{equation}
	where $o(k+1)$ is the predicted output at time instant $k+1$, and $G$ symbolizes the introduced time-series MLP. The vector $\mathbf{o}(k) = [o(k), o(k-1), \dots, o(k-d_o)]$ denotes the past output discrete-time sequence, while $\mathbf{u}(k) = [u(k), u(k-1), \dots, u(k-d_u)]$ represents the input discrete-time sequence. The parameters $d_o$ and $d_u$ indicate the number of past output and input values, respectively, and are tunable during MLP training.
	
	Considering the operating units expressed by Eq.\eqref{E5}, we can further formulate the specialized time-series MLP models for these units, adapted from Eq.\eqref{E10}, as follows:
	\begin{equation} \label{E11}
		o^{j}(k+1) = G^{j}(\mathbf{o}^{j}(k), \mathbf{u}^{j}(k), \mathbf{z}^{j}(k), \mathbf{w}^{j}(k), \mathbf{v}^{j}(k))
	\end{equation}
	where the superscript $j$ indicates the operating unit and $j \in \{PV, FC, BA, MA, EC, PMP, CS, BD\}$; $\mathbf{o}^{j}(k) = [o^{j}(k), o^{j}(k-1), \dots, o^{j}(k-d_o)]$ means the output time sequence of unit $j$; $\mathbf{u}^{j}(k) = [u^{j}(k), u^{j}(k-1), \dots, u^{j}(k-d_u)]$ is the manipulated continuous input time sequence of unit $j$; similarly, $\mathbf{z}^{j}(k) = [z^{j}(k), z^{j}(k-1), \dots, z^{j}(k-d_z)]$, $\mathbf{w}^{j}(k) = [w^{j}(k), w^{j}(k-1), \dots, w^{j}(k-d_u)]$, and $\mathbf{v}^{j}(k) = [v^{j}(k), v^{j}(k-1), \dots, v^{j}(k-d_u)]$ define the time sequences of the manipulated binary inputs, disturbances, and linked inputs, respectively. Using Eq.\eqref{E11}, we will later build time-series MLPs for the operating units within the IES to describe their dynamic characteristics on 1-hour, 1-minute, and 5-second time scales.
	
	\subsection{Developing hybrid time-series MLPs for IES}
	
	Upon investigating dynamic response times and introducing the time-series MLP for operating units, we will establish a hybrid time-series MLP on each time scale for the entire IES. This approach will integrate known physical information to simplify the development of machine learning models for complex energy systems and improve their prediction accuracy, while taking into account day-ahead scheduling and intraday control problems that need to be solved.
	
	\subsubsection{Streamlining 1-hour input-output configurations of operating units}
	
	Before training time-series MLPs, let's review the control problem formulation of the IES. As mentioned earlier, the operational optimization of grid-connected IESs typically consists of two stages: day-ahead scheduling and intraday real-time control. Generally, day-ahead scheduling involves decision-making about the hourly operational status of systems for the next day based on approximate forecasts of external conditions, such as environmental factors and customer demands. This includes determining the on/off status of operating units to ensure energy flow balance, managing the charging/discharging of energy storage for long-term load shifting, and planning power delivery to the utility grid.
	
	However, due to the uncertainty of long-term forecasts and the volatile nature of demands from both the microgrid and utility grid, day-ahead scheduling can only optimize the system's operational status approximately. More precise regulation is achieved through intraday real-time control, which adjusts operations based on real-time conditions. In this context, a simplified model that accurately reflects energy balance is sufficient for day-ahead scheduling \cite{qin2022multi, wu2022economic}. Moreover, day-ahead scheduling usually involves integer decision variables, leading to a mixed-integer optimization problem. Using a highly complicated model can result in computational issues and make the problem difficult to solve \cite{urbanucci2018limits}. Therefore, a simplified model is also a practical and acceptable solution in terms of computational costs.
	
	Taking these factors into account, we will slightly simplify the 1-hour input-output structures of the operating units in the IES. This will ensure that the resulting time-series MLP, which will later be converted to mixed-integer linear programming (MILP) in our proposed algorithm, is technically feasible while maintaining model accuracy for the system's energy generation and consumption under various working conditions. To be specific, on the input side of the operating units, we first neglect changes in the water flow rates and return water temperature in the chillers and building, assuming they remain at nominal values when the relevant operating units start. Thus, $G_{ab}$, $G_{ec}$, $t_{rec}$, and $t_{re}$ can be treated as constants in the 1-hour model. Based on this assumption, all variations in the supplied water temperature in the chillers' output side can be substituted by the corresponding cooling power changes, meaning that $Q_{ab}$ and $Q_{ec}$ can be viewed as the output of the chillers, ignoring $t_{ab}$ and $t_{ec}$. Meanwhile, $t_{cp}$ and $t_{hp}$ in the cold storage can be omitted, and the cooling power $Q_{st}$ is treated as its output since they depend on the chilled or return water temperature in the IES. Consequently, $C_{sth}$ and $C_{stc}$ given in Eqs.\eqref{E16} and \eqref{E17} are unnecessary. Furthermore, the building's linked inputs $t_{sl}$ and $G_{sl}$ can also be substituted by the final cooling power provided to the building, i.e., $Q_{sl}$ from Eq.\eqref{E18}, which can be approximately evaluated using a straightforward energy balance as follows:
	\begin{equation} \label{E21}
		Q_{sl} = Q_{ab} + Q_{ec} + Q_{st}
	\end{equation}
	Similarly, the linked input of the pump, the total circulated water flow rate $G_{all}$ from Eq.\eqref{E3}, is revised as:
	\begin{equation} \label{E22}
		G_{all} = z_{ma}G_{ab,0} + z_{ec}G_{ec,0} + G_{stu}
	\end{equation}
	In this case, Eqs.\eqref{E1}-\eqref{E2} are no longer needed, whereas the electrical power supplied by the IES to the utility grid, $P_{sl}$ from Eq.\eqref{E4}, is still computed by:
	\begin{equation} \label{E23}
		P_{sl} = P_{pv} + P_{fc} + P_{mt} + P_{ba} - P_{cp} - P_{pmp} - P_{d}
	\end{equation}
	The energy storage capacity states, $C_{soc}$ and $C_{sot}$ given in Eqs.\eqref{E12} and \eqref{E13}, remain as follows:
	\begin{subequations} \label{E14}
		\begin{align}
			& \dot{C}_{soc}(\tau) = - \frac{I_{ba}(\tau)}{3600 C_{eb} \eta_{pb}} \\
			& \dot{C}_{sot}(\tau) = -\frac{G_{st}(\tau)}{M_{st}}
		\end{align}
	\end{subequations}
	where $G_{st}$ is still the circulating water flow and is derived from $G_{st} = z_{st} G_{stu} + (z_{st} - 1) G_{stu}$, which is one of the outputs of the cold storage. After completing the simplification, the operating units' generic expression Eq.\eqref{E5} still holds. However, their inputs and outputs are partly modified on the 1-hour time scale, as summarized in Table~\ref{T4}. This streamlined input-output configuration will be employed to develop the 1-hour time-series MLP for the day-ahead scheduling of the IES. It is important to note that the entire simplification procedure is based on the known input-output structures and prior physical knowledge of the system, as described earlier, without requiring further information on their detailed dynamics or other aspects. Besides, both the original and adjusted outputs are measurable in reality, allowing us to collect their data.
	
	\begin{table}[!ht] \small
		\centering
		\caption{Summary of simplified input-output variables for operating units on the 1-hour time scale}
		\label{T4}
		\renewcommand{\arraystretch}{1.3}
		\tabcolsep 2pt
		\begin{tabular}{p{2.5cm}p{2.5cm}p{2.5cm}p{2.5cm}p{2.5cm}p{2.5cm}} \hline
			\textbf{Operating unit} & \textbf{Continuous input} & \textbf{Binary input} & \textbf{Disturbance} & \textbf{Linked input} & \textbf{Output} \\ \hline
			PV & / & / & $t_{a}$, $S_{ra}$ & / & $P_{pv}$ \\
			FC & $G_{ff}$ & $z_{fc}$ & / & / & $P_{fc}$ \\
			BA & $P_{bar}$ &/ & / & / & $P_{ba}$, $I_{ba}$, $C_{soc}$ \\
			MA & $G_{fm}$ & $z_{ma}$ & / & / & $P_{mt}$, $Q_{ab}$ \\
			EC & $N_{ec}$ & $z_{ec}$ & / & / & $P_{cp}$, $Q_{ec}$ \\
			PMP & / & / & / & $G_{all}$ & $P_{pmp}$ \\
			BD & / & / & $t_{a}$, $Q_{o}$ & $Q_{sl}$ & $t_{br}$ \\
			CS & $G_{stu}$ & $z_{st}$ & / & / & $Q_{st}$, $G_{st}$, $C_{sot}$ \\
			\hline
		\end{tabular}
	\end{table}
	
	\subsubsection{Data generation and preparation}
	
	Adequate and proper data are prerequisites for successful machine learning approaches. In this subsection, we will collect and preprocess the datasets to complete the preparation for time-series MLP training.
	
	In this study, pseudo-random multi-level signals (PRMSs) \cite{braun1999multi} are utilized to excite the operating units for data generation. Specifically, to simulate various daily operational data, we first employ ten-level PRMSs with randomly repeated sampling intervals as inputs to the operating units across multiple time scales, generating the respective outputs. For the 1-hour time scale, PRMSs with a 1-hour sampling time are employed to excite each unit listed in Table~\ref{T4} to collect outputs data, except for $C_{soc}$ and $C_{sot}$, which can be directly constructed via Eq.\eqref{E14} in establishing the final 1-hour hybrid MLP for the entire IES. For the 1-minute and 5-second time scales, PRMSs with 1-minute and 5-second sampling times are used to excite each unit listed in Table~\ref{T2}, excluding the cold storage and $C_{soc}$, which can be straightly built using prior physical knowledge (Eqs.\eqref{E12}-\eqref{E13}) when reconstructing the hybrid MLPs for the IES. Here, data for the cold storage is collected only on a 1-hour time scale, since it will be trained in the 1-hour MLP but not in the 1-minute and 5-second MLPs. This is because, in our proposed optimization method, the final 1-hour MLP and the resulting day-ahead scheduling problem will be exactly recast to MILP, requiring all nonlinear units to be represented by the MLP. Additionally, datasets can also be generated and gathered through real engineering practice, experiments, or other rigorous models or leading commercial process simulator platforms such as Apros, Dymola, Aspen, and PowerFactory \cite{chen2020model}.
	
\begin{figure}[!ht]
		\centering
		\includegraphics[width=0.7\linewidth]{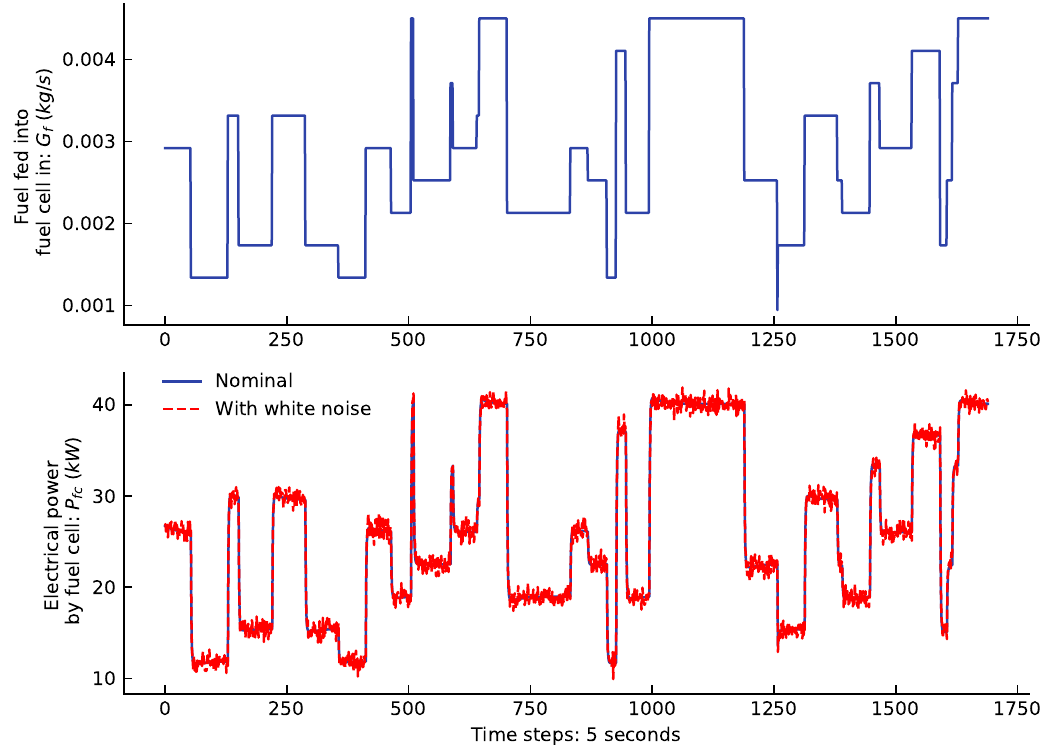}
		\caption{A segment of the fuel cell training datasets on the 5-second time scale: (i) the upper subfigure shows the training input signal, which includes ten-level PRMSs with a 5-second sampling interval, where the duration is randomly repeated; (ii) the lower subfigure depicts the corresponding system output, where the blue line represents the nominal output and the red dashed line represents the processed output with white noise.}
		\label{F5}
	\end{figure}

	Once the original training datasets on multiple time scales have been obtained, they will undergo further processing. To enhance the generalization ability of the MLPs, prevent overfitting, and mimic real operational conditions \cite{huang2023model, liu2023state}, a set of Gaussian white noise, with a zero mean and a standard deviation of 2\% of the outputs' varying amplitude, is added to the nominal training datasets for each output trajectory across all time scales. Figure~\ref{F5} presents a segment of the fuel cell training datasets on the 5-second time scale.
	
	Since the IES is a complex, large-scale system with various cross-disciplinary processes and parameters of differing numerical magnitudes, all processed datasets are subsequently normalized to mitigate the negative effects of these differences on MLPs' prediction accuracy during training. Here, we resort a min-max scaler to normalize the training datasets, scaling the data values within a specified feature range. The min-max scaler is defined as follows:
	\begin{equation} \label{E19}
		a_i^{scaled} = \frac{a_i - a_{min}}{a_{max} - a_{min}} (e_{max} - e_{min}) + e_{min}
	\end{equation}
	where $a^{scaled}$ represents the data point after normalization; $a_i$ is the sample point to be scaled; $a_{min}$ and $a_{max}$ are the corresponding minimum and maximum values within the dataset; $e_{max}$ and $e_{min}$ are our defined maximum and minimum feature values, set to 1 and 0 respectively, resulting in $a_{scaled}$ falling within the range of 0 to 1. We then apply the scaler of Eq.\eqref{E19} to all datasets to achieve normalization. 
	
	Afterwards, the final step in dataset preprocessing is to split the data into three parts: training data, validation data, and testing data, with ratios of 70\%, 20\%, and 10\%, respectively. The training and validation data will be used for training and validating the MLPs, while the testing data will be used to evaluate the prediction performance of the trained MLPs. Following this split, we will proceed to train time-series MLPs for the operating units across multiple time scales and rebuild hybrid MLPs for the entire IES.
	
	\begin{remark}
		Establishing input-output time-series neural networks for individual operating units with prior process knowledge offers significant advantages over constructing them for the entire IES. Aside from reducing the structural size and computational complexity of the networks, leading to concise and efficient data-driven representations, this approach provides better access to training datasets and ensures compatibility within the neural network. First, training data directly influence the prediction accuracy of neural networks. Collecting appropriate data for an entire IES is more challenging as against a single operating unit due to the IES’s complexity, which includes various energy conversion and intricate dynamic processes and numerous distinct inputs. This difficulty in data collection can negatively affect the neural networks' predictions. Moreover, the IES is a large-scale system that integrates diverse operating units, which may undergo maintenance, replacement, or the addition of new units. Modeling neural networks at the modular-unit level allows for easier updates when the system undergoes changes. In such cases, only the neural network of the affected unit needs to be re-trained and substituted in the existing hybrid neural network, rather than re-training the entire system. This modular method makes the proposed \rev{physics-informed} hybrid neural network, which incorporates prior physical knowledge, more adaptable and practical in engineering applications.
	\end{remark}
	
	\subsubsection{Training and reconstructing MLPs for operating units and IES}
	
	In this subsection, we will first train the time-series MLPs for operating units on three time scales based on the prepared data. Next, we will reconstruct the hybrid MLPs on each time scale for the entire IES using these MLPs, incorporating prior knowledge of the input-output physical interconnectivity between operating units and the known straightforward dynamic formulations.
	
	As previously stated, on the 1-hour time scale, we will train time-series MLPs for all the units listed in Table~\ref{T4}, excluding $C_{soc}$ and $C_{sot}$. On the 1-minute and 5-second time scales, we will train time-series MLPs for all the units listed in Table~\ref{T2}, with the exception of cold storage and $C_{soc}$. These units or processes are not trained because introducing MLPs would complicate matters unnecessarily compared to using their straightforward first-principles models that we already know. For the 1-hour MLPs, we select ReLU as the activation function to build a ReLU network. This network, as previously mentioned, will be used to formulate a MILP optimization problem for day-ahead scheduling. For the 1-minute and 5-second MLPs, we employ Sigmoid as the activation function to create differentiable neural networks for intraday real-time coordinated control. Table~\ref{T5} provides the structural parameters of these MLPs. We prepare $25 \times 10^5$ samples for each operating unit on the 5-second time scale, $8 \times 10^5$ samples for each unit on the 1-minute time scale, and $2.5 \times 10^5$ samples on the 1-hour time scale. During the training of these MLPs, the loss function is mean squared error (MSE), and the optimizer is Adam. The batch size and number of epochs are 20 and 40, respectively. The training procedure is carried out using the TensorFlow platform (2.10.0) \cite{dillon2017tensorflow} in Python (3.10.13).
	
	\begin{table}[!ht] \small
		\centering
		\caption{The structural parameters of MLPs for operating units on three time scales}
		\label{T5}
		\renewcommand{\arraystretch}{1.3}
		\tabcolsep 2pt
		\begin{tabular}{p{3.5cm}p{1.4cm}p{1.4cm}p{1.4cm}p{1.4cm}p{1.4cm}p{1.4cm}p{1.4cm}p{1.4cm}} \hline
			\textbf{5-second} & \textbf{PV} & \textbf{FC} & \textbf{BA} & \textbf{MA} & \textbf{EC} & \textbf{PMP} & \textbf{BD} & \textbf{CS} \\ \hline
			Hidden layers & 2 & 2 & 2 & 2 & 2 & 2 & 2 & / \\
			Neurons & 15, 8 & 23, 12 & 20, 10 & 32, 18 & 25, 12 & 12, 8 & 70, 40 & / \\
			Output series delay & 0 & 3 & 1 & 3 & 3 & 0 & 15 & / \\
			Input series delay & 1 & 2 & 1 & 3 & 3 & 1 & 15 & / \\
			\hline
			\textbf{1-minute} & \textbf{PV} & \textbf{FC} & \textbf{BA} & \textbf{MA} & \textbf{EC} & \textbf{PMP} & \textbf{BD} & \textbf{CS} \\ \hline
			Hidden layers & 2 & 2 & 2 & 2 & 2 & 2 & 2 & / \\
			Neurons & 15, 8 & 18, 10 & 16, 8 & 28, 14 & 20, 10 & 12, 8 & 60, 30 & / \\
			Output series delay & 0 & 1 & 1 & 3 & 1 & 0 & 9 & / \\
			Input series delay & 1 & 2 & 1 & 3 & 2 & 1 & 9 & / \\
			\hline
			\textbf{1-hour} & \textbf{PV} & \textbf{FC} & \textbf{BA} & \textbf{MA} & \textbf{EC} & \textbf{PMP} & \textbf{BD} & \textbf{CS} \\ \hline
			Hidden layers & 1 & 1 & 1 & 1 & 1 & 1 & 1 & 1 \\
			Neurons & 4 & 3 & 4 & 4 & 5 & 5 & 6 & 3 \\
			Output series delay & 0 & 0 & 0 & 0 & 0 & 0 & 1 & 0 \\
			Input series delay & 1 & 1 & 1 & 1 & 1 & 1 & 1 & 1 \\
			\hline
		\end{tabular}
	\end{table}
	
	Consequently, we can derive the time-series MLPs for the operating units within the IES across three time scales, as described by the following expression:
	\begin{equation} \label{E20}
		o^{j}_{t}(k+1) = G^{j}_{t}(\mathbf{o}^{j}_{t}(k), \mathbf{u}^{j}_{t}(k), \mathbf{z}^{j}_{t}(k), \mathbf{w}^{j}_{t}(k), \mathbf{v}^{j}_{t}(k))
	\end{equation}
	The symbolic references are essentially the same as those in Eq.\eqref{E11}, with the modification of adding the subscript $t \in \{l, s, f\}$ to denote discrete-time intervals of 1 hour, 1 minute, and 5 seconds, respectively. Additionally, when $t = l$, $j \in \{PV, FC, BA, MA, EC, PMP, BD, CS\}$; otherwise, $j \in \{PV, FC, BA, MA, EC, PMP, BD\}$.
	
	\begin{figure}[!ht]
		\centering
		\includegraphics[width=0.95\linewidth]{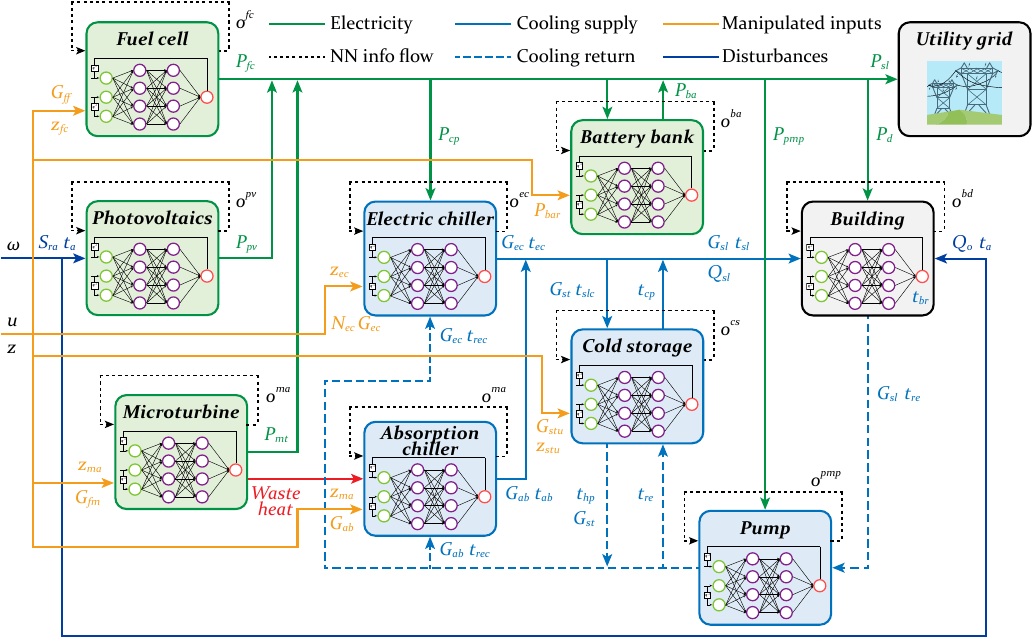}
		\caption{Restructured hybrid MLP architecture for the entire IES.}
		\label{F6}
	\end{figure}
	
	With the obtained time-series MLPs (Eq.\eqref{E20}), we can rebuild the hybrid MLPs for the entire IES by integrating them with prior physical knowledge, as shown in Figure \ref{F6}. For convenience, we name the hybrid MLP for the IES on a 1-hour time scale as the long-term MLP, and refer to the hybrid MLPs for the IES on 1-minute and 5-second time scales as slow and fast MLPs, respectively. To assemble the long-term MLP, the relevant outputs in the 1-hour MLPs (i.e., $o^{j}_{l}$ in Eq.\eqref{E20}) are connected via Eqs.\eqref{E21} and \eqref{E23}, which will serve as the linked inputs for the operating units and eventually for the outputs of the IES ($P_{sl}$ and $t_{br}$). In addition, Eqs.\eqref{E22} and \eqref{E14} are incorporated into the system to formulate the linked input for the pump and the capacity states of the energy storage. The reconstructed hybrid long-term MLP with a 1-hour discrete-time interval can be represented by the following compact formulation:
	\begin{equation} \label{E24}
		y_{l}(k+1) = F_{l}(o_{l}(k), u_{l}(k), z_{l}(k), w_{l}(k))
	\end{equation}
	where $y_{l} = [P_{sl}, t_{br}, C_{soc}, C_{sot}, o_{l}]$ is a measurement vector that includes the concerned outputs $P_{sl}$ and $t_{br}$, the capacity states of the energy storage $C_{soc}$ and $C_{sot}$, and an augmented vector containing the individual operating unit's outputs $o_{l}$. Specifically, $o_{l}$ is given by $o_{l} = [o_{l}^{PV}, o_{l}^{FC}, o_{l}^{BA}, o_{l}^{MA}, o_{l}^{EC},\\ o_{l}^{PMP}, o_{l}^{BD}, o_{l}^{CS}]$. These outputs $o_{l}$ will assist in formulating the output time sequences for the corresponding units' MLP, namely $\mathbf{o}^{j}_{l}$ in Eq.\eqref{E20}. Besides, $u_{l} = [G_{ff}, G_{fm}, N_{ec}, G_{stu}, P_{bar}]$ is the simplified 1-hour continuous manipulated input vector, $z_{l} = [z_{fc}, z_{ma}, z_{ec}, z_{st}]$ is the binary manipulated input vector, and $w_{l} = [t_a, S_{ra}, P_d, Q_o]$ is the disturbance vector. It should be mentioned that the hybrid long-term MLP of Eq.\eqref{E24} is also a hybrid ReLU network.
	
	To proceed, the reconstruction of the hybrid slow MLP is similar to that of the long-term MLP. Based on the connectivity defined by Eqs.\eqref{E1}-\eqref{E2}, Eq.\eqref{E3}, and Eq.\eqref{E4} among the operating units, and the known formulations, Eqs.\eqref{E15}-\eqref{E17}, of the cold storage, we can interconnect the pertinent input-output variables in the operating units' 1-minute MLPs (Eq.\eqref{E20}). These connected variables will either serve as linked inputs for the units or help determine the final outputs of the IES ($P_{sl}$ and $t_{br}$) and the associated measurements. Furthermore, we combine the connected network with Eqs.\eqref{E12} and \eqref{E13} to evaluate the capacity states of the energy storage ($C_{soc}$ and $C_{sot}$). Ultimately, the rebuilt hybrid slow MLP with a 1-minute discrete-time interval can be concisely expressed as follows:
	\begin{equation} \label{E25}
		y_{s}(k+1) = F_{s}(o_{s}(k), u_{s}(k), z_{s}(k), w_{s}(k))
	\end{equation}
	where $y_{s} = [P_{sl}, t_{br}, C_{soc}, C_{sot}, o_{s}]$ denotes a measurement vector that contains the outputs of the IES, the capacity states of the energy storage, and an augmented vector, $o_{s}$, comprising the operating unit's outputs. And $o_{s} = [o_{s}^{PV}, o_{s}^{FC}, o_{s}^{BA}, o_{s}^{MA}, o_{s}^{EC}, o_{s}^{PMP}, o_{s}^{BD}, o_{s}^{CS}]$, which will be used in building the output time sequences for the respective units' MLP, namely $\mathbf{o}^{j}_{s}$ in Eq.\eqref{E20}. The vectors $u_{s}$, $z_{s}$, and $w_{s}$ represent the continuous manipulated input vector, the binary manipulated input vector, and the disturbance vector, respectively. Here, $u_{s} = [G_{ff}, G_{fm}, G_{ab}, N_{ec}, G_{ec}, G_{stu}, P_{bar}]$, $z_{s} = [z_{fc}, z_{ma}, z_{ec}, z_{st}]$, and $w_{s} = [t_{a}, S_{ra}, P_{d}, Q_{o}]$.
	
	The rebuilding procedure for the hybrid fast MLP for the entire IES closely resembles that of the slow MLP and will not be repeated here. The rebuilt fast MLP with a 5-second sampling interval can be described using the following discrete-time representation:
	\begin{equation} \label{E26}
		y_{f}(k+1) = F_{f}(o_{f}(k), u_{f}(k), z_{f}(k), w_{f}(k))
	\end{equation}
	where the symbolic definitions are much the same as their counterparts in Eq.\eqref{E25} and are omitted here.
	
	\begin{remark}
		It is worth noting that when determining the MLPs' structural parameters, as provided in Table~\ref{T5}, three main factors need to be considered: the discrete time of the built MLP, the speed of the dynamic response of the trained operating unit, and the input-output type of the trained unit (i.e., either single-input single-output or multiple-input multiple-output). Generally, if the unit to be trained has slower dynamic characteristics and the MLP we want to build has a smaller discrete-time period (i.e., a faster time scale), the number of past time steps in the input-output time series (the range of time delays) will be larger. This is because we need a sufficient time horizon to cover the main dynamic response of the unit and accurately identify its current state. Additionally, longer time series or multivariate input-output structures require more hidden layers and neurons to capture the latent relationships in the data. Conversely, for univariate units or discrete frequencies with larger time scales, or units with faster dynamic characteristics, fewer time delays in the input-output time series and fewer neurons and hidden layers are usually required. This is because the transient process of the system will not dominate the entire sampling period, and a simpler neural network can adequately reflect its operational behavior.
	\end{remark}
	
	\begin{remark}
		It is noteworthy that understanding the physical input-output structure and basic dynamic information of a system is fundamental to engineering practice. Acquiring this straightforward information is not demanding, making it a reasonable assumption that this knowledge is readily available. Therefore, this does not restrict the application of our proposed \rev{physics-informed} hybrid neural network method in this work. Furthermore, during the initial system description and subsequent reconstruction, we only utilized this basic prior knowledge and physical structure without delving into deep system dynamic characteristics. This approach makes our method more generic, applicable, and scalable.
	\end{remark}
	
	\begin{remark}
		Please note that for the long-term, slow, and fast MLPs, represented by Eq.\eqref{E24}-\eqref{E26}, we simplify these expressions by directly using the relevant variables at the current moment $k$ as inputs instead of their time sequences. This is done for clarity and to reduce complexity, unlike in Eq.\eqref{E20}, where the time series of relevant variables is used as inputs. For example, consider a simple time series $\mathbf{a}(k) = [a(k), a(k-1), \dots, a(k-d_a)]$. If the past information $[a(k-1), \dots, a(k-d_a)]$ is known at time instant $k$, the only decision variable is $a(k)$. Therefore, for simplicity, in formulating Eq.\eqref{E24}-\eqref{E26}, we use the current moment $k$ variables as the inputs for the function mapping $F$, implicitly including the construction of the time series in the initial condition required by the time series MLPs of Eq.\eqref{E20}. Besides, in contrast to the MLP of a single operating unit of Eq.\eqref{E20}, the MLPs of the entire IES, expressed by Eq.\eqref{E24}-\eqref{E26}, no longer require the linked input $v$, since it is already included in the interconnected relationships within the reconstructed hybrid MLPs.
	\end{remark}
	
	\subsubsection{MLPs model validation and evaluation}
	
	This subsection validates and evaluates the dynamic prediction performance of the developed MLPs for both the operating units and the entire IES. To achieve this, multi-step ahead prediction tests are first conducted for all the MLPs outlined in Eqs.\eqref{E20}–\eqref{E26}. Multi-step ahead prediction involves auto-regressive forecasts made prior to the time evolution. At a given time instant, the MLP makes a one-step ahead prediction based on the current initial condition. At the next time instant, the MLP utilizes the previous prediction as the current initial condition. This procedure is repeated to provide multi-step ahead predictions, which are essentially identical to those used in following day-ahead scheduling and model predictive control.
	
	\begin{figure}[!ht]
		\centering
		\includegraphics[width=1\linewidth]{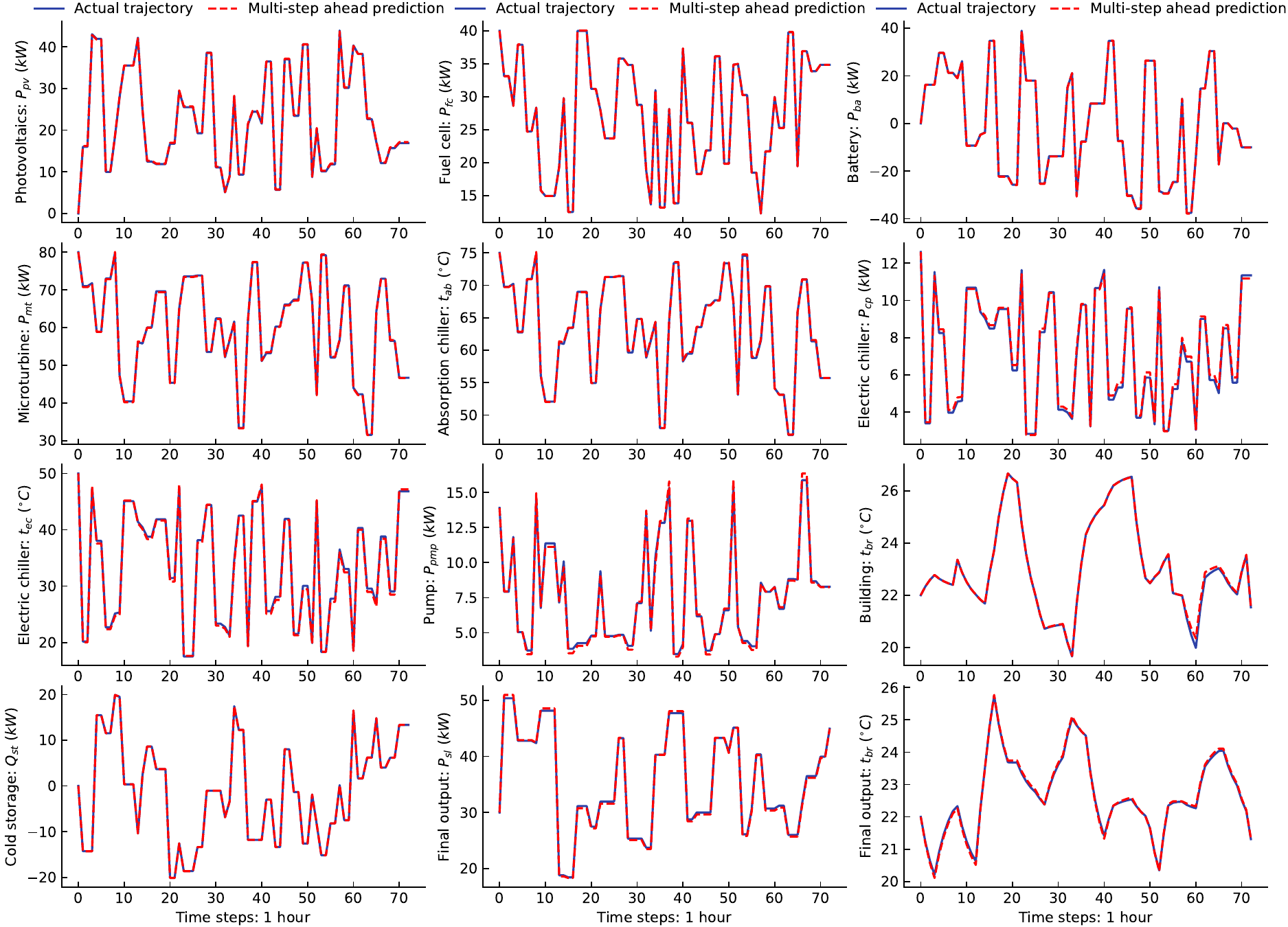}
		\caption{Multi-step ahead predictions vs. actual trajectories with 1-hour discrete-time step.}
		\label{F7}
	\end{figure}
	
	\begin{figure}[!ht]
		\centering
		\includegraphics[width=1\linewidth]{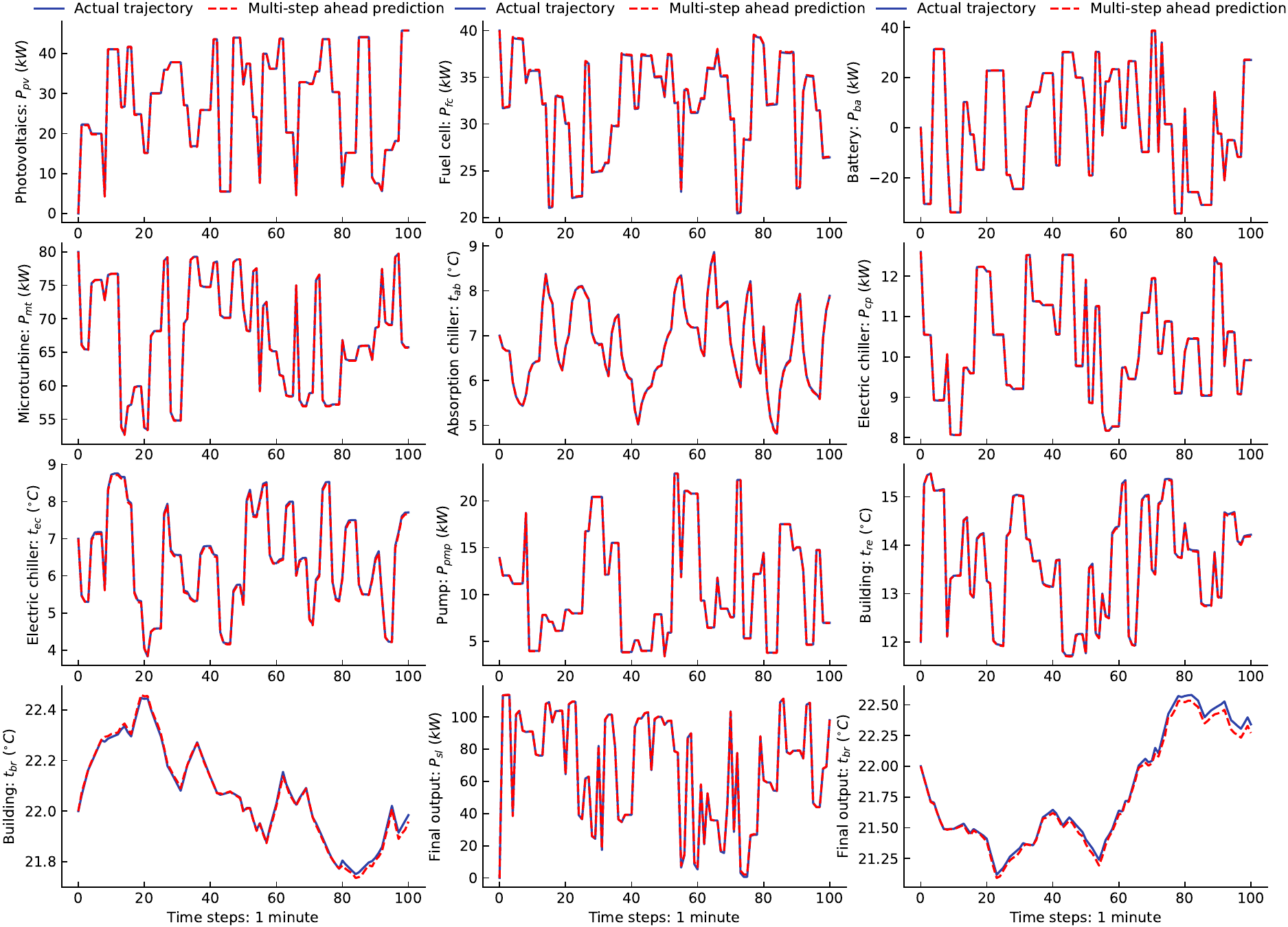}
		\caption{Multi-step ahead predictions vs. actual trajectories with 1-minute discrete-time step.}
		\label{F8}
	\end{figure}
	
	\begin{figure}[!ht]
		\centering
		\includegraphics[width=1\linewidth]{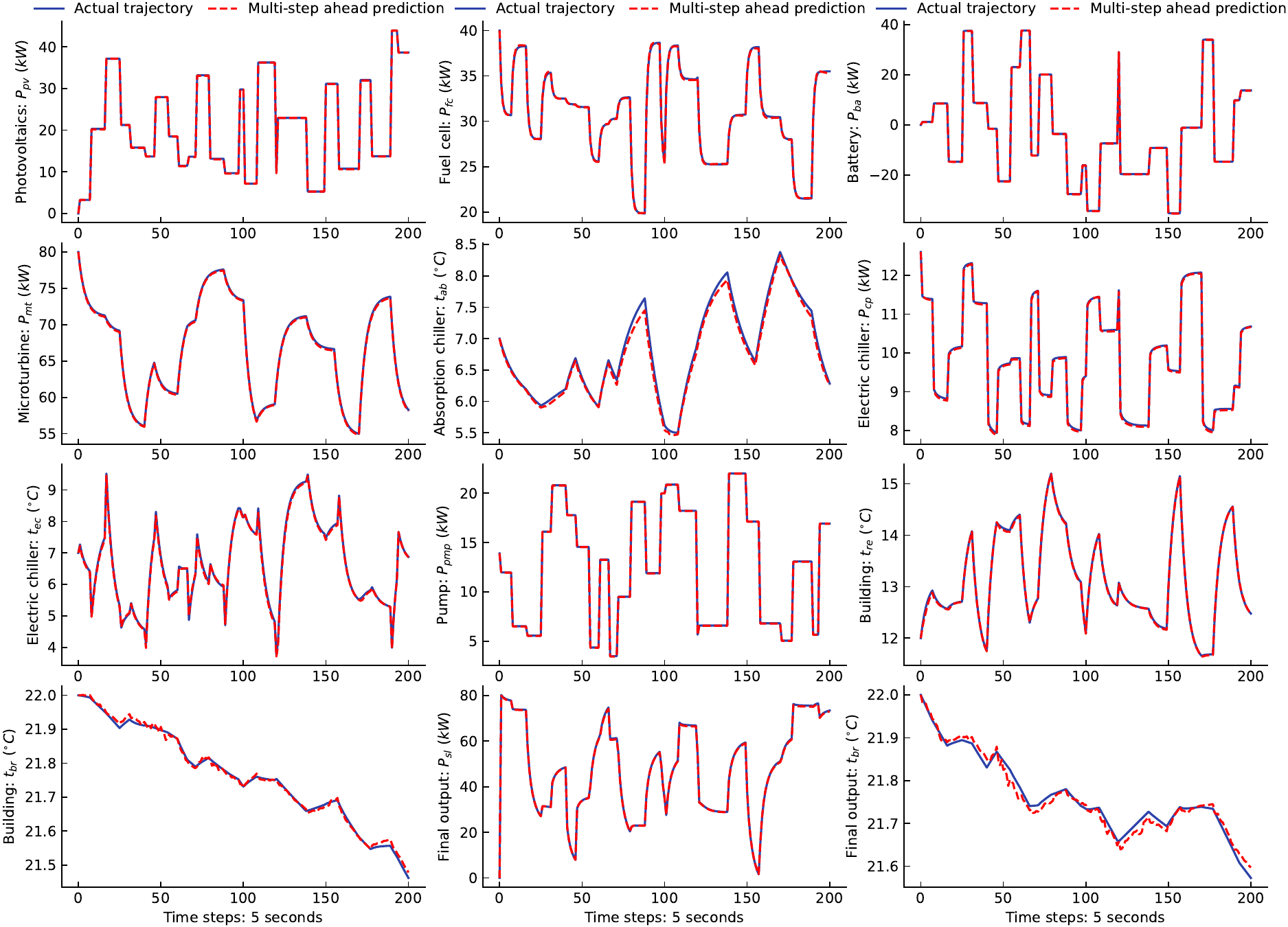}
		\caption{Multi-step ahead predictions vs. actual trajectories with 5-second discrete-time step.}
		\label{F9}
	\end{figure}
	
	Figures~\ref{F7}–\ref{F9} present the multi-step ahead open-loop prediction performance of the trained operating units' MLPs and the rebuilt hybrid IES's MLPs. The prediction steps cover the control horizon of interest for scheduling and control applications. Figure~\ref{F7} shows the prediction performance of the units' 1-hour MLP (Eq.\eqref{E20}) and the IES's long-term MLP (Eq.\eqref{E24}) versus the actual trajectories from the streamlined 1-hour systems. Figure~\ref{F8} depicts the prediction results of the units' 1-minute MLP (Eq.\eqref{E20}) and the IES's slow MLP (Eq.\eqref{E24}) versus the actual trajectories from the original systems. Figure~\ref{F9} portrays the prediction performance of the units' 5-second MLP (Eq.\eqref{E20}) and the IES's fast MLP (Eq.\eqref{E24}) versus the actual trajectories from the original systems. These figures illustrate that the proposed MLPs can effectively predict the dynamic behavior of operating units across the specified time scales. Since the proposed MLPs only need to identify relatively simple input-output responses in the systems, rather than detailed state-space models, they can achieve performance close to that of first-principles models. Thus, the hybrid MLPs can also predict the entire system's dynamics with high accuracy as illustrated in the figures. Additionally, the MLPs accurately reflect unit performance under a wide range of working conditions, which is of critical importance to system operational optimization. It is also noted that as the number of time steps increases, some MLP predictions slightly diverge from the actual trajectories. However, this deviation is insignificant since the prediction horizons for desired scheduling and control applications do not require such large prediction steps.
	
	To further assess the prediction performance of the MLPs, we utilize the normalized root mean square error (NRMSE) to measure the multi-step ahead predictions. The NRMSE is defined as follows:
	\begin{equation}
		NRMSE = \frac{\sqrt{\frac{1}{n_a} \sum_{i=1}^{n_a}(a_i^{act} - a_i^{pred})^2}}{a_{max} - a_{min}}
	\end{equation}
	where $n_a$ is the number of measured data points; $a_i^{act}$ is the actual value of the $i$-th data point; $a_i^{pred}$ is the predicted value of the $i$-th data point; and $a_{max}$ and $a_{min}$ are the maximum and minimum values within the measured data points. Tables~\ref{T6}-\ref{T8} summarize the NRMSE of the multi-step ahead predictions on the 1-hour, 1-minute, and 5-second time scales, respectively. From these tables, we observe that all the NRMSE values are comparatively low, demonstrating that the MLPs' multi-step ahead predictions are sufficiently accurate and comparable to those of the first-principles models. The capability of neural networks to predict dynamics, approaching that of first-principles models, has also been confirmed by other research \cite{wu2019machine, ren2022tutorial}. However, the NRMSE of the hybrid MLPs for the entire IES is generally slightly larger than those for a single operating unit due to error accumulation in the rebuilt hybrid MLPs. Despite the increased prediction error following the MLP reconstruction, the long-term, slow, and fast MLPs maintain excellent prediction accuracy, supporting the development of the day-ahead scheduling and real-time coordinated control proposed later.
	
	\begin{table}[!ht] \small
		\centering
		\caption{Evaluation of MLPs on 1-hour time scale: NRMSE of multi-step ahead prediction}
		\label{T6}
		\renewcommand{\arraystretch}{1.3}
		\tabcolsep 2pt
		\begin{tabular}{p{1.9cm}<{\centering}p{1.9cm}<{\centering}p{1.9cm}<{\centering}p{1.9cm}<{\centering}p{1.9cm}<{\centering}p{1.9cm}<{\centering}p{1.9cm}<{\centering}p{1.9cm}<{\centering}} \hline
			\textbf{Variables} & $P_{pv}$ & $P_{fc}$ & $P_{ba}$ & $I_{ba}$ & $P_{mt}$ & $Q_{ab}$ & $P_{cp}$ \\ \hline
			\textbf{NRMSE} & 0.00367 & 0.00113 & 0.00168 & 0.00654 & 0.00404 & 0.00424 & 0.0180 \\ \hline
			\textbf{Variables} & $Q_{ec}$ & $P_{pmp}$ & $t_{br}$ & $Q_{st}$ & $G_{st}$ & $P_{sl}$ (IES) & $t_{br}$ (IES) \\ \hline
			\textbf{NRMSE} & 0.0118 & 0.0167 & 0.0106 & 0.000281 & 0.000296 & 0.00945 & 0.00993 \\ \hline
		\end{tabular}
	\end{table}
	
	\begin{table}[!ht] \small
		\centering
		\caption{Evaluation of MLPs on 1-minute time scale: NRMSE of multi-step ahead prediction}
		\label{T7}
		\renewcommand{\arraystretch}{1.3}
		\tabcolsep 2pt
		\begin{tabular}{p{1.9cm}<{\centering}p{1.9cm}<{\centering}p{1.9cm}<{\centering}p{1.9cm}<{\centering}p{1.9cm}<{\centering}p{1.9cm}<{\centering}p{1.9cm}<{\centering}p{1.9cm}<{\centering}} \hline
			\textbf{Variables} & $P_{pv}$ & $P_{fc}$ & $P_{ba}$ & $I_{ba}$ & $P_{mt}$ & $t_{ab}$ & $P_{cp}$ \\ \hline
			\textbf{NRMSE} & 0.00157 & 0.00459 &0.000761 & 0.00172 & 0.00192 & 0.00349 & 0.00215 \\ \hline
			\textbf{Variables} & $t_{ec}$ & $P_{pmp}$ & $t_{re}$ & $t_{br}$ & $P_{sl}$ (IES) & $t_{br}$ (IES) & \\ \hline
			\textbf{NRMSE} & 0.00792 & 0.00129 & 0.00481 & 0.0159 & 0.00350 & 0.0258 & \\ \hline
		\end{tabular}
	\end{table}
	
	\begin{table}[!ht] \small
		\centering
		\caption{Evaluation of MLPs on 5-second time scale: NRMSE of multi-step ahead prediction}
		\label{T8}
		\renewcommand{\arraystretch}{1.3}
		\tabcolsep 2pt
		\begin{tabular}{p{1.9cm}<{\centering}p{1.9cm}<{\centering}p{1.9cm}<{\centering}p{1.9cm}<{\centering}p{1.9cm}<{\centering}p{1.9cm}<{\centering}p{1.9cm}<{\centering}p{1.9cm}<{\centering}} \hline
			\textbf{Variables} & $P_{pv}$ & $P_{fc}$ & $P_{ba}$ & $I_{ba}$ & $P_{mt}$ & $t_{ab}$ & $P_{cp}$ \\ \hline
			\textbf{NRMSE} & 0.00182 & 0.00514 & 0.00114 & 0.00138 & 0.00638 & 0.0273 & 0.00745 \\ \hline
			\textbf{Variables} & $t_{ec}$ & $P_{pmp}$ & $t_{re}$ & $t_{br}$ & $P_{sl}$ (IES) & $t_{br}$ (IES) & \\ \hline
			\textbf{NRMSE} & 0.00997 & 0.00125 & 0.00721 & 0.0146 & 0.00349 & 0.0315 & \\ \hline
		\end{tabular}
	\end{table}
	
	The construction of the long-term, slow, and fast MLPs is now complete. The following section will illustrate how to exploit them for day-ahead scheduling and intraday real-time control of the IES.
	
	\section{Neural network-based day-ahead scheduling and distributed EMPC}
	
	\begin{figure}[!ht]
		\centering
		\includegraphics[width=1\linewidth]{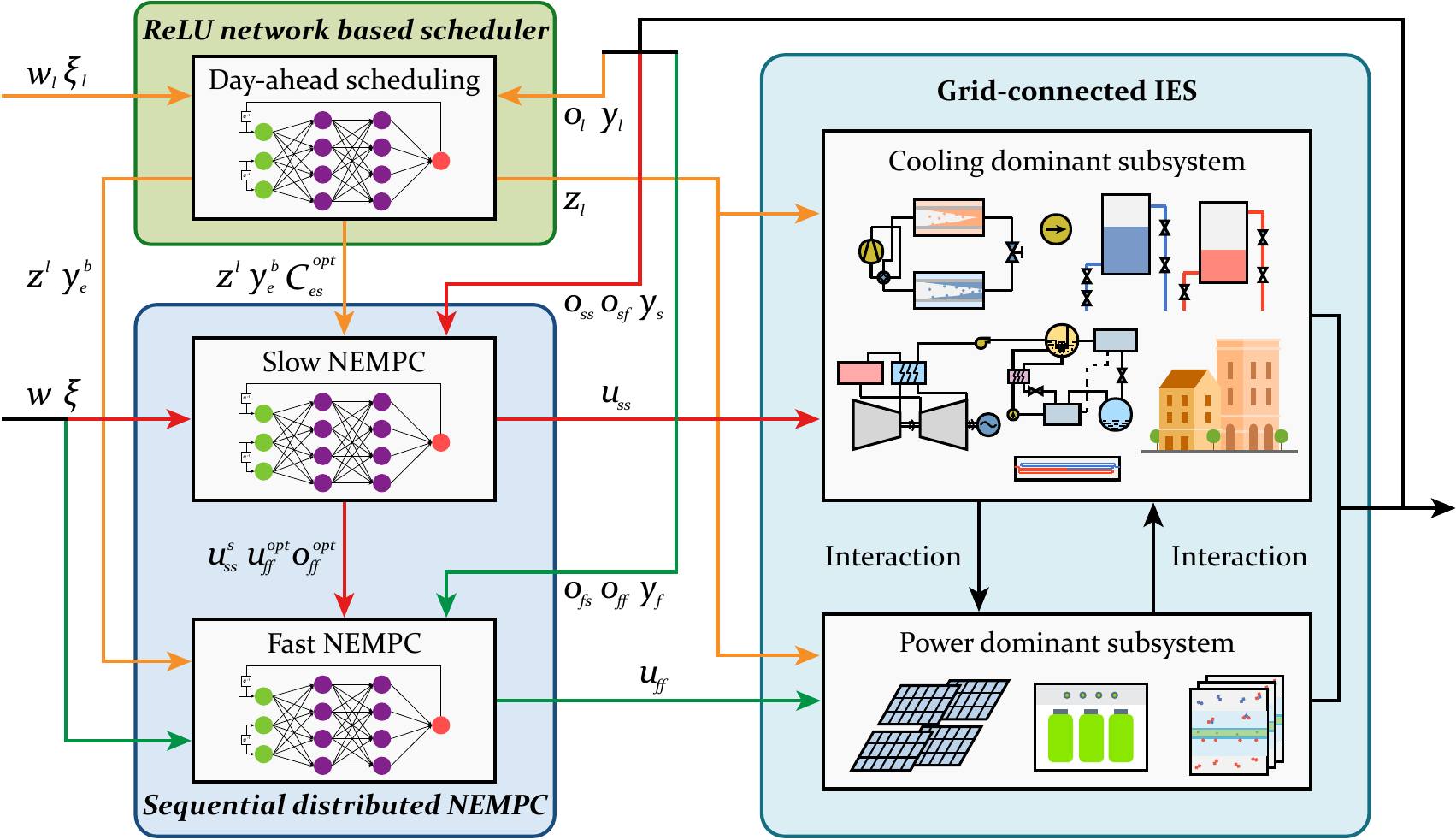}
		\caption{The diagram of the proposed machine learning-based day-ahead scheduling and distributed NEMPC.}
		\label{F10}
	\end{figure}
	
	Leveraging the developed long-term, slow, and fast MLPs, this section proposes a smart energy management strategy for grid-connected IESs. Taking into account the requirements of day-ahead and real-time optimization, our scheme is composed of: a ReLU network-based optimization for day-ahead scheduling, and a neural network-based economic model predictive control in a sequential distributed fashion for intraday real-time coordination. The proposed solution copes with challenges arising from dynamic time-scale multiplicity, varying external conditions, and potential computational complexity. It ensures that the operating units within the IES dynamically cooperate to respond swiftly to utility grid instructions, fulfill local electricity demands, maintain building temperatures within acceptable ranges, and enhance operational income. Figure~\ref{F10} displays the architecture of the ReLU network-based scheduler and the sequential distributed NEMPC.
	
	First, the day-ahead scheduling, based on the developed long-term MLP (i.e., the ReLU network), is designed to optimize the hourly operational behavior of the operating units within the IES to maintain its basic energy balance and profitability for the following day. This ReLU network-based scheduler will take into account the global objectives $J_1$, $J_2$, and $J_3$, and then be equivalently converted into a MILP optimization problem and solved. As shown in Figure~\ref{F10}, the decision-making process for day-ahead scheduling requires the external environmental conditions forecasts $w_l$, the hourly power supply regulation coefficient $\xi_l$, and the outputs of the operating units $o_l$ and the entire system $y_l$. Based on this information, the scheduling process determines the hourly baseline power $y_e^b$ to be submitted to the utility grid, the on-off status of the units and the charging/discharging of the cold storage (binary variables $z_l$), as well as the optimal hourly trajectories of the energy storage capacity states, denoted by $C_{es}^{opt}$, for long-term load shifting. Among these optimized variables, $z_l$ is simultaneously sent to the IES for switching units on/off and to the sequential distributed NEMPC as a known fact $z^l$, where $z^l = z_l$. Both $C_{es}^{opt}$ and $y_e^b$ are transmitted to NEMPC. $C_{es}^{opt}$ serves as an optimal reference for further energy storage management, while $y_e^b$ is a necessary parameter in real-time NEMPC to coordinate the operating units for more precise regulation.
	
	Subsequently, to cope with the negative impact of the IES's multiple dynamic time-scale properties in real-time coordinated control, the IES is first divided into two subsystems based on dynamic response analysis: a slow cooling-dominant subsystem and a fast power-dominant subsystem. Employing the established slow MLP and fast MLP, we propose a sequential distributed NEMPC with asynchronous control frequency. This scheme enables dynamic collaboration between the slow cooling-dominant and fast power-dominant subsystems through cooperative control between the slow and fast agents. Furthermore, during the cooperation between these two control agents, one-way information exchange from the slow agent to the fast agent assures the consistency of their decision optimization. As a result, they achieve dynamic synergy among all units with diverse dynamics in the IES, accurately respond to power requests from the utility grid, meet power supply and cooling demands within the microgrid, and enhance system economics.
	
	Specifically, the slow NEMPC, based on the slow MLP with a 1-minute sampling time, optimizes the system's dynamic performance over the next ten to fifteen minutes. To reach this goal, the slow NEMPC must consider the current status of the operating units and energy storage $z^l$, the baseline power committed to the utility grid $y_e^b$, and the energy storage operation plan, $C_{es}^{opt}$. Moreover, the slow NEMPC will collect real-time data on the actual external disturbances $w$, the actual power supply regulation coefficient $\xi$, the outputs of the operating units $o_{ss}$ and $o_{ff}$, as well as the system output feedback $y_s$. Consequently, by minimizing a global objective function that includes $J_1$, $J_2$, and $J_3$, the slow NEMPC determines the optimal control action $u_{ss}$ for the cooling-dominant subsystem, the optimal reference input $u_{ff}^{opt}$ for the power-dominant subsystem, and the optimal reference output trajectory $o_{ff}^{opt}$ for the relevant units in the fast subsystem. Following this, the input $u_{ss}$ is immediately applied to the cooling-dominant subsystem to control its dynamics. Both $u_{ff}^{opt}$ and $o_{ff}^{opt}$ are sent to the fast NEMPC for further decision-making and rapid adjustment of the power-dominant subsystem. Additionally, $u_{ss}$ is passed to the fast NEMPC as known information $u_{ss}^{s}$ ($u_{ss}^{s}=u_{ss}$), enabling the parameterization of the cooling-dominant subsystem-related variables in the fast NEMPC and improving its computational efficiency.
	
	Finally, the fast NEMPC, based on the fast MLP with a 5-second sampling interval, dedicates to high-frequency precise optimization of the dynamic behavior of the power-dominant subsystem within a one-minute prediction horizon. In the fast NEMPC, a joint objective function for the power-dominant subsystem, encompassing the global objectives $J_1$ and $J_3$, is minimized to determine the optimal control action $u_{ff}$. This control action is then directly applied to manage the dynamics of the power-dominant subsystem. To ensure consistency in optimization results and accuracy in fast MLP predictions, the fast NEMPC receives real-time data on the external conditions ($w$ and $\xi$), the optimized results transmitted from the day-ahead scheduler ($z_l$ and $y_e^b$), optimal reference information from the slow NEMPC ($u_{ss}^s$, $o_{ff}^{opt}$, and $u_{ff}^{opt}$), and feedback from the relevant systems ($o_{fs}$, $o_{ff}$, and $y_f$). The remainder of this section will provide a mathematical elaboration of the proposed neural network-based day-ahead scheduler and distributed EMPC.
	
	\subsection{ReLU network-based day-ahead scheduler}
	
	This subsection proposes a ReLU network-based day-ahead scheduler designed to optimize the IES's long-term operational performance. Before formulating the detailed optimization problem, we first need to establish the scheduling objective function. All three global objectives, outlined in Eq.\eqref{E7}, must be considered for the IES's long-term operation. However, since the day-ahead scheduler operates in discrete time with a one-hour interval, the objectives in Eqs.\eqref{E7a} and \eqref{E7b} will be transformed into constraints. This conversion assures a balance of energy flow between production and consumption over two discrete-time intervals. Thus, for the proposed day-ahead scheduling based on the ReLU network Eq.\eqref{E24}, the following objectives are formulated at the initial time instant $k$:
	\begin{subequations} \label{E27}
		\begin{align}
			J_1^l(k) =& -\alpha_1^l (p_{mg}(k)w_{l,3}(k) + p_{se}(k)y_{l,1}(k+1) + p_{cm}(k)\xi_{as}^l(k) y_e^b(k+1) \notag \\ &- p_f(k)(u_{l,1}(k)+u_{l,2}(k))) \label{E27a} \\
			J_2^l(k) =& \alpha_{2}^l \epsilon_1^l(k+1) + \alpha_{3}^l \epsilon_2^l(k+1) + \alpha_{4}^l \epsilon_3^l(k+1)+ \alpha_{5}^l \epsilon_4^l(k+1) \label{E27b}
		\end{align}
	\end{subequations}
	where the superscript $l$ indicates day-ahead scheduling; $\alpha^l_\omega$ ($\omega=1,\dots,5$) denotes the weighting coefficients, while $\epsilon^l_\omega$ ($\omega=1,\dots,4$) represent the slack variables added to the constraints of the optimization problem. The objective function of Eq.\eqref{E27a} represents the system's operational economic objective, derived from the discretized global objective of Eq.\eqref{E7c}. The penalty term for failing to deliver requested electricity to the utility grid, $p_{pn}|y_{1} - (1+\xi)y_e^b|^2$ in Eq.\eqref{E7c}, is ignored here since it will be included as one of the constraints to balance the energy flow, thereby approaching zero and can be neglected. Eq.\eqref{E27b} is adopted to ensure the feasibility of the scheduling optimization problem by minimizing the slack variables.
	
	Taking advantage of the built ReLU network Eq.\eqref{E24} and the above optimization objectives Eq.\eqref{E27}, the day-ahead scheduler for the IES at the initial time instant $k$ is presented as follows:
	\begin{subequations} \label{E28}
		\begin{align}
			\min_{u_{l}, z_{l}, y_e^b, y_l} & \sum_{i=0}^{N_p^l-1}
			J_1^l(k+i) + J_2^l(k+i) \label{E28a} \\
			s.t.\ &y_{l}(k+1+i) = F_{l}(o_{l}(k+i), u_{l}(k+i), z_{l}(k+i), w_{l}(k+i)) \label{E28b} \\
			&u_l^{min} z^u_l(k+i) \leq u_l(k+i)\leq u_l^{max} z^u_l(k+i) \label{E28c} \\
			&\Delta u^{min}_{l} \leq \Delta u_{l}(k+i) \leq \Delta u^{max}_{l} \label{E28d} \\
			&o_{l}^{min}\leq o_{l}(k+1+i)\leq o_{l}^{max} \label{E28e} \\
			&y_{l,1}(k+1+i) = (1+\xi_l)y_e^b(k+1+i) + \epsilon_1^l(k+1+i) - \epsilon_2^l(k+1+i) \label{E28f} \\
			&y_{sp,t}^{min}(k+1+i) - \epsilon_4^l(k+1+i) \leq y_{l,2}(k+1+i) \leq y_{sp,t}^{max}(k+1+i) + \epsilon_3^l(k+1+i) \label{E28g} \\
			&\epsilon_1^l(k+1+i),\ \epsilon_2^l(k+1+i),\ \epsilon_3^l(k+1+i),\ \epsilon_4^l(k+1+i) \geq 0 \label{E28h} 
		\end{align}
	\end{subequations}
	where $N_p^l$ stands for the prediction horizon; the superscripts $min$ and $max$ indicate the lower and upper bounds of the respective variables; the term $z^u_l(k) = {\rm diag}([z_{l,1}(k), z_{l,2}(k), z_{l,3}(k), 1, 1])$; $\Delta u_{l}(k)$ accounts for the increment of $u_{l}$ between time instants $k$ and $k-1$, defined as $\Delta u_{l}(k) = u_{l}(k) - u_{l}(k-1)$; $y_{sp,t}^{min}$ and $y_{sp,t}^{max}$ denote the customer's desired building temperature range.
	
	In the ReLU network-based scheduler of Eq.\eqref{E28}, the optimization objective is to minimize a weighted sum of the objectives $J_1^l$ and $J_2^l$ to maximize the system's profitability over the long term. The constraint of Eq.\eqref{E28b} guarantees that the scheduling optimization problem adheres to the hybrid long-term MLP described in Eq.\eqref{E24}. Eqs.\eqref{E28c}-\eqref{E28e} impose physical constraints on the inputs of the entire system and the outputs of the respective operating units included in the hybrid long-term MLP of Eq.\eqref{E24}. The constraint of Eq.\eqref{E28f}, derived from the global objective Eq.\eqref{E7a}, maintains electrical power balance and tracks the utility grid's power change instructions. Similarly, Eq.\eqref{E28g}, derived from the global objective Eq.\eqref{E7b}, assures the indoor temperature remains within the specified range. Together, Eqs.\eqref{E28f} and \eqref{E28g} keep the dynamic balance of energy production and consumption, with slack variables slightly loosening these constraints to improve optimization feasibility. The constraint of Eq.\eqref{E28} guarantees that these slack variables are positive definite. It is worth noting that, at the day-ahead stage, the acceptable range of building temperature, $[y_{sp,t}^{min}, y_{sp,t}^{max}]$ in Eq.\eqref{E28g}, could be tightened slightly from the customer's prescribed range, enhancing the robustness of the subsequent real-time control system against uncertainty \cite{wu2022economic}.
	
	However, solving the proposed day-ahead scheduler is mathematically and technically challenging. The scheduling problem of Eq.\eqref{E28} involves both integer decision variables and nonlinearity, resulting in a mixed-integer nonlinear programming problem (MINLP). The integer decision variables are intrinsic to the scheduling problem, while the nonlinearity arises from the ReLU activation function within the ReLU network expressed by the constraint of Eq.\eqref{E28b}. Although the ReLU network can describe a wide range of the IES's working performance, it also creates barriers to solving the problem. To this end, we will equivalently convert the optimization problem in Eq.\eqref{E28} into a mixed-integer linear programming problem via rewriting the ReLU activation function. Let us review a time-series MLP described in Eq.\eqref{E8} within the hybrid long-term MLP Eq.\eqref{E24}, where the activation function is ReLU. As mentioned, the only nonlinear component of the optimization problem of Eq.\eqref{E28} is the ReLU function $\sigma_h^{(r+1)}$ in Eq.\eqref{E8b}. Define the element-wise lower and upper bounds on the pre-activation state vector $\mathsf{P}^{(r+1)}$ as $\mathsf{P}^{(r+1)}_{min}$ and $\mathsf{P}^{(r+1)}_{max}$, respectively, i.e., $\mathsf{P}^{(r+1)} \in [\mathsf{P}^{(r+1)}_{min}, \mathsf{P}^{(r+1)}_{max}]$. Then, the ReLU activation function in Eq.\eqref{E8}, $\mathsf{H}^{(r+1)} = \sigma_h^{(r+1)}(\mathsf{P}^{(r+1)})$, can be equivalently expressed as a set of mixed-integer linear constraints as follows:
	\begin{equation} \label{E29}
		\begin{aligned}
			\mathsf{H}^{(r+1)} = \max(\mathsf{P}^{(r+1)}, 0) \Longleftrightarrow \left\{ 
			\begin{array}{ccc}
				\mathsf{H}^{(r+1)} & \geq & \mathsf{P}^{(r+1)} \\
				\mathsf{H}^{(r+1)} & \geq & 0 \\
				\mathsf{H}^{(r+1)} & \leq & \mathsf{P}^{(r+1)} - (\mathbf{I} - \gamma_{r+1}) {\rm diag}[(\mathsf{P}^{(r+1)}_{min})] \\
				\mathsf{H}^{(r+1)} & \leq & \gamma_{r+1} {\rm diag}[(\mathsf{P}^{(r+1)}_{max})]
			\end{array}\right.
		\end{aligned}
	\end{equation}
	where $\max(\mathsf{P}^{(r+1)}, 0) = \sigma_h^{(r+1)}(\mathsf{P}^{(r+1)})$ stands for the ReLU activation function; $\mathbf{I}$ is a vector with all elements equal to 1; $\gamma_{r+1} \in \{0,1\}^{n_{nr}}$ is an auxiliary vector with binary variables for the $(r+1)$-th activation layer, where $n_{nr}$ denotes the number of neurons in the $(r+1)$-th hidden layer. By assembling the mixed-integer linear constraints in Eq.\eqref{E29} for each layer in the ReLU network of Eq.\eqref{E24} and adding the corresponding $\gamma_{i+1}$ as decision variables to the optimization problem of Eq.\eqref{E28}, we can obtain the mixed-integer linear representation of the ReLU network in constraint Eq.\eqref{E28b}. Consequently, the reformulated optimization problem in Eq.\eqref{E28} becomes a mixed-integer linear programming problem. This MILP can be solved using powerful commercial solvers. It is important to note that, since the input to the ReLU network, $\mathsf{P}^{(r+1)}$, is bounded in our scheduling problem, diverse approaches, such as the interval bound propagation technique \cite{chen2021learning}, can be utilized to find the element-wise lower and upper bounds on the pre-activation state vector, so that $\mathsf{P}^{(r+1)} \in [\mathsf{P}^{(r+1)}_{min}, \mathsf{P}^{(r+1)}_{max}]$ holds. 
	
	At the initial time instant $k$, the ReLU network-based day-ahead scheduler is performed by solving the optimization problem in Eq.\eqref{E28}, yielding the optimal sequences over the prediction horizon, including the binary input ${\bf z}_l^*(k)=[z_{l}^*(k), \dots, z_{l}^*(k+N_p^l-1)]$ and the baseline power delivered to the utility grid ${\bf y}_e^{b*}(k)=[y_e^{b*}(k), \dots, y_e^{b*}(k+N_p^l-1)]$. The day-ahead scheduler also optimizes the input and output sequences ${\bf u}_l^*(k)$ and ${\bf y}_l^*(k)$. Among these optimized sequences, ${\bf z}_l^*(k)$ will be applied to the IES in a timely manner to adjust the relevant units' on-off status and energy storage charging-discharging status, while ${\bf y}_e^{b*}(k)$ is submitted to the utility grid as the committed generation plan for the next day. And then, the continuous input ${\bf u}_l^*(k)$ will be discarded after scheduling and re-optimized in the real-time coordinated control system with high-resolution MLPs according to the actual external conditions, achieving precise dynamic collaboration between the operating units. Moreover, the optimal sequences of the energy storage's capacity states are extracted from ${\bf y}_l^*(k)$, denoted as ${\bf C}_{es}^*(k) = [C_{es}^*(k+1),\dots,C_{es}^*(k+N_p^l)]$ (where $C_{es} = [y_{l,3}, y_{l,4}] = [C_{soc}, C_{sot}]$). ${\bf C}_{es}^*(k)$ will then be passed to the real-time NEMPC for further optimization to accomplish long-term load shifting. To facilitate the description of the following distributed NEMPC, let us define: $z^{l}(k):=z_{l}^*(k)$, $C_{es}^{opt}(k+1) = C_{es}^*(k+1)$, and $y_e^{b}(k) = y_e^{b*}(k)$.
	
	\subsection{Neural network based sequential distributed EMPC}
	
	In this subsection, we introduce the partition of slow cooling-dominant and fast power-dominant subsystems, then develop a sequential distributed NEMPC with asynchronous cooperative control agents based on slow and fast MLPs to realize dynamic synergy between the operating units across the IES in real time.
	
	\subsubsection{Partitioning of cooling- and power-dominant subsystems}
	
	At the outset of designing a distributed control framework, it is essential to divide the entire system into several subsystems. As observed in Section 3.1, the operating units across the IES exhibit a dynamic multi-time-scale property. The dynamic responses of cooling-related units are typically slower than those of power-related units. Cooling-related units generally respond on the scale of several minutes or more, while electricity generation units respond within several seconds or even faster. Based on this observation, the units in the IES are partitioned into two subsystems: the cooling-dominant subsystem with slow responses, which consists of the microturbine with the absorption chiller, the electric chiller, the cold storage, the pump, and the building; and the power-dominant subsystem with fast responses, which comprises the photovoltaic module, the fuel cell, and the battery bank. The microturbine and pump are divided into the cooling-dominant subsystem despite their association with power generation and consumption because the former is an upstream unit of the absorption chiller with a relatively slower time constant than other power units, and the latter directly depends on the cooling production process. Furthermore, unlike other units, energy storage plays a role in load shifting; therefore, their capacity states, with much slower dynamic responses, should be included and optimized in the slow subsystem.
	
	In this context, the slow MLP of Eq.\eqref{E25} with a larger discrete time is well-suited for managing the cooling-dominant subsystem in real-time coordinated control, while the fast MLP of Eq.\eqref{E26} with a smaller discrete time is ideal for the power-dominant subsystem, enabling precise handling of its subtle transient behaviors. Let us rewrite the slow and fast MLPs from Eqs.\eqref{E25} and \eqref{E26} by separating the outputs and manipulated inputs of their operating units into two parts—slow and fast—based on whether they pertain straightly to the slow cooling-dominant subsystem or the fast power-dominant subsystem. Corresponding to the cooling-dominant subsystem, the slow MLP in Eq.\eqref{E25} is rewritten as follows:
	\begin{equation} \label{E30}
		y_{s}(k+1) = F_{s,col}(o_{ss}(k), o_{sf}(k), u_{ss}(k), u_{sf}(k), z_{s}(k), w_{s}(k))
	\end{equation}
	where $y_s$ represents the output of the slow MLP for the cooling-dominant subsystem, defined as $y_s = [P_{sl}, t_{br}, C_{soc}, C_{sot}, o_{ss}, o_{sf}]$. The terms $o_{ss}$ and $o_{sf}$ correspond to the slow and fast outputs of the relevant units within the cooling- and power-dominant subsystems in the slow MLP, respectively. These are defined as $o_{ss} = [o_s^{MA}, o_s^{EC}, o_s^{PMP}, o_s^{BD}, o_s^{CS}]$ for the slow cooling-dominant units and $o_{sf} = [o_s^{PV}, o_s^{FC}, o_s^{BA}]$ for the fast power-dominant units. $u_{ss}$ and $u_{sf}$ are the slow and fast continuous inputs of the cooling- and power-dominant units in the slow MLP and defined as $u_{ss} = [G_{fm}, G_{ab}, N_{ec}, G_{ec}, G_{stu}]$ and $u_{sf} = [G_{ff}, P_{bar}]$, respectively. The binary input and disturbance in the slow MLP are denoted by $z_s$ and $w_s$, which are the same as those given in Eq.\eqref{E25} and will not be repeated.
	
	Similarly, the fast MLP associated with the power-dominant subsystem can also be derived from Eq.\eqref{E26} as follows:
	\begin{equation} \label{E31}
		y_{f}(k+1) = F_{f,pow}(o_{fs}(k), o_{ff}(k), u_{fs}(k), u_{ff}(k), z_{f}(k), w_{f}(k))
	\end{equation}
	where $y_f$ denotes the output, defined as $y_f = [P_{sl}, t_{br}, C_{soc}, C_{sot}, o_{fs}, o_{ff}]$. The slow and fast outputs $o_{fs}$ and $o_{ff}$ correspond to the relevant units within the cooling- and power-dominant subsystems, respectively, in the fast MLP. These are defined as $o_{fs} = [o_f^{MA}, o_f^{EC}, o_f^{PMP}, o_f^{BD}, o_f^{CS}]$ for the slow cooling-dominant units and $o_{ff} = [o_f^{PV}, o_f^{FC}, o_f^{BA}]$ for the fast power-dominant units. $u_{fs}$ and $u_{ff}$ are the slow and fast continuous inputs of the the cooling- and power-dominant units in the fast MLP and defined as $u_{fs} = [G_{fm}, G_{ab}, N_{ec}, G_{ec}, G_{stu}]$ and $u_{ff} = [G_{ff}, P_{bar}]$, respectively. $z_f$ and $w_f$ denote the binary input and disturbance in the fast MLP, which are the same as those given in Eq.\eqref{E26}.
	
	Next, we will exploit the rewritten slow and fast MLPs to design a sequential distributed NEMPC, featuring two asynchronous cooperative NEMPCs for real-time coordinated control of the IES.
	
	\begin{remark}
		It is important to note that while the rewritten slow and fast MLPs in Eqs.\eqref{E30} and \eqref{E31} share symbols for input and output elements, \rev{they operate on different MLPs with their respective discrete times, i.e., sampling intervals.} The slow MLP in Eq.\eqref{E30}, derived from Eq.\eqref{E25}, has a discrete time of 1 minute, whereas the fast MLP in Eq.\eqref{E31}, derived from Eq.\eqref{E26}, operates with a 5-second interval. Additionally, in this work, we assume that only the basic input-output structures and dynamic responses of the operating units are known, without detailed knowledge of their internal dynamics. Consequently, we cannot fully decompose the subsystems based on their internal dynamics. Hence, the slow and fast MLPs retain aspects of each other's expressions within their mathematical representations. This inclusion is crucial because, when optimizing the slow cooling-dominant subsystem over a longer time interval, it is necessary to account for the fast power-dominant subsystem's behavior to ensure the validity of the solution. Conversely, although the slow subsystem is not completely decoupled from the fast MLP's expression, the decision variables related to the slow subsystem, $u_{ss}$, are determined using the slow MLP. These variables then become constants $u_{fs}$ in the fast MLP, allowing $o_{fs}$ to be straightly calculated as known time-varying parameters in the fast MLP, which benefits from its discrete-time formulation. As a result, transient power changes associated with the slow cooling-dominant subsystem are still updated and predicted in a timely manner with high resolution in the fast MLP, without increasing the computational burden.
	\end{remark}
	
	\subsubsection{Design of sequential distributed slow and fast NEMPCs}
	
	This subsection presents a sequential distributed NEMPC scheme for real-time coordination of the transient behaviors of operating units, aiming to achieve dynamic synergy. The scheme comprises two control agents with different sampling times: a slow NEMPC based on the slow MLP for controlling the cooling-dominant subsystem and a fast NEMPC based on the fast MLP for optimizing the power-dominant subsystem.
	
	Considering the real-time coordinated control requirements for power, cooling, and economics of the IES, as defined by the global objectives in Eq.\eqref{E7}, the slow NEMPC based on the slow MLP in Eq.\eqref{E30} should tackle the following control objectives at time instant $k$:
	\begin{subequations} \label{E32}
		\begin{align}
			J_1^s(k) = &\ \alpha_1^s\|\hat{y}_{s,1}(k+1)-(1+\xi(k))y_e^b(k+1)\|^2 \label{E32a} \\
			J_2^s(k) = &\ \alpha_2^s\|\hat{y}_{s,2}(k+1)-y_{sp,t}(k+1)\|^2 \label{E32b} \\
			J_3^s(k) = & -\alpha_3^s(p_{mg}(k)w_{s,3}(k) + p_{se}(k)\hat{y}_{s,1}(k+1) + p_{cm}(k)\xi_{as}(k) y_e^b(k+1) \notag \\
			&- p_f(k)(u_{sf,1}(k)+u_{ss,1}(k)) - p_{pn}(k)\|\hat{y}_{s,1}(k+1) - (1+\xi(k))y_e^b(k+1)\|^2) \label{E32c} \\
			J_4^s(k) = &\ \alpha_4^s\| C_{es}(k+1) - C_{es}^{opt}(k+1)\| \label{E32d} 
		\end{align}
	\end{subequations}
	where $\alpha^s_\omega$ ($\omega=1,\dots,4$) are the weighting coefficients, and $\hat{y}_{s}$ is the corrected output of the slow MLP, details of which will be provided later. The vector $C_{es} = [\hat{y}_{s,3}, \hat{y}_{s,4}]$ stands for the corrected capacity state of the energy storage. The objectives in Eqs.\eqref{E32a}-\eqref{E32c} are derived from the discretization of the global objectives in Eqs.\eqref{E7a}-\eqref{E7c}, reflecting the IES's performance in tracking real-time instructions to adjust power supply to the utility grid, maintaining indoor temperature within the specified set-point range, and generating operational revenue. The objective in Eq.\eqref{E32d} measures the deviation of the energy storage’s capacity states from the scheduler-prescribed optimal reference $C_{es}^{opt}$ for load shifting.
	
	Taking into account the control objectives in Eq.\eqref{E32} and the slow MLP in Eq.\eqref{E30}, the slow NEMPC at a time instant $k$, focused on managing the cooling-dominant subsystem, is expressed as follows:
	\begin{subequations} \label{E33}
		\begin{align}
			\min_{u_{ss}, u_{sf}, y_{sp,t}, o_{ss}, o_{sf}} & \sum_{i=0}^{N^s_p-1} J_1^s(k+i) + J_2^s(k+i) + J_3^s(k+i) + J_4^s(k+i) \label{E33a} \\ 
			s.t. \ 
			& y_{s}(k+1+i) = F_{s,col}(o_{ss}(k+i), o_{sf}(k+i), u_{ss}(k+i), \notag \\
			&u_{sf}(k+i), z_{s}(k+i), w_{s}(k+i)) \label{E33b} \\
			&u_{ss}^{min} z^u_{ss}(k+i) \leq u_{ss}(k+i)\leq u_{ss}^{max} z^u_{ss}(k+i) \label{E33c} \\
			&u_{sf}^{min} z^u_{sf}(k+i) \leq u_{sf}(k+i)\leq u_{sf}^{max} z^u_{sf}(k+i) \label{E33d} \\
			& \Delta u^{min}_{ss} - V_{ss} \leq \Delta u_{ss}(k+i) \leq \Delta u^{max}_{ss} + V_{ss} \label{E33e} \\
			& \Delta u^{min}_{sf} - V_{sf} \leq \Delta u_{sf}(k+i) \leq \Delta u^{max}_{sf} + V_{sf} \label{E33f} \\
			&o_{ss}^{min}\leq o_{ss}(k+1+i)\leq o_{ss}^{max} \label{E33g} \\
			&o_{sf}^{min}\leq o_{sf}(k+1+i)\leq o_{sf}^{max} \label{E33h} \\
			&y_{sp,t}^{min}(k+1+i) \leq y_{sp,t}(k+1+i) \leq y_{sp,t}^{max}(k+1+i) \label{E33i} \\
			&\hat{y}_{s}(k+1+i) = y_{s}(k+1+i) + \beta_s^i(y_{s}^{act}(k) - y_{s}(k|k-1)) \label{E33j} \\
			& z_{s}(k+i) = z^{l}(k+i) \label{E33k} 
		\end{align}
	\end{subequations}
	where $N^s_p$ is the prediction horizon of the slow NEMPC; the superscripts $min$ and $max$ denote the lower and upper bounds of the corresponding decision variables; $z^u_{ss}$ and $z^u_{sf}$ are defined as $z^u_{ss} = {\rm diag} ([z_{s,2}, z_{s,2}, z_{s,3}, z_{s,3}, 1])$ and $z^u_{sf} = {\rm diag} ([z_{s,1}, 1])$; the term $\Delta u_{ss}(k+i) = u_{ss}(k+i) - u_{ss}(k+i-1)$ indicates the increment of the slow input related to the slow subsystem between two time instants; the expression $V_{ss} = |z^u_{ss}(k+i) - z^u_{ss}(k+i-1)| L_c$, with $L_c$ being a sufficiently large constant, is used to bypass the constraint in Eq.\eqref{E33e} when the on-off status $z^u_{ss}$ changes; the terms $\Delta u_{sf}(k+i)$ and $V_{sf}$ are defined similarly; $y_{sp,t}$ is the building temperature set-point, treated as a decision variable here; the variable $\hat{y}_{s}$ refers to the corrected output of the slow MLP; and $\beta_s^i$ is a tunable correction factor used in the output feedback procedure, where a larger $\beta_s^i$ indicates more aggressive model correction; $y_{s}^{act}(k)$ stands for the actual measured value of $y_{s}$ at the current time instant $k$, while $y_{s}(k|k-1)$ accounts for the predicted value of $y_{s}(k)$ made at the previous time instant $k-1$.
	
	In the slow NEMPC formulation in Eq.\eqref{E33}, the final control objective is to minimize a weighted objective function, as defined in Eq.\eqref{E33a}, which incorporates the mentioned global objectives. The constraint in Eq.\eqref{E33b} specifies that this optimization problem is based on the slow MLP, which is used as the prediction model. Eqs.\eqref{E33c}-\eqref{E33f} impose physical constraints on the manipulated inputs, while Eqs.\eqref{E33g}-\eqref{E33h} set physical constraints on the relevant units' outputs. Eq.\eqref{E33i} ensures that the building temperature set-point remains within the specified temperature range. The constraint in Eq.\eqref{E33j} represents the output feedback mechanism, utilizing both predicted and actual output values to compensate for model-plant mismatches and complete model correction. Lastly, Eq.\eqref{E33k} indicates that the on-off status of the relevant units $z_{s}$ is predetermined and corresponds to the optimized value $z^{l}$ from the scheduler.
	
	At time instant $k$, the optimization problem in Eq.\eqref{E33} is solved to execute the slow NEMPC based on the currently available information. The slow NEMPC determines the optimum slow input sequence within the prediction window, ${\bf u}_{ss}^*(k)=[u_{ss}^*(k), \dots, u_{ss} ^*(k+N^s_p-1)]$, and optimizes the optimum future references for the input and output sequences associated with the fast subsystem, ${\bf u}_{sf}^*(k)$ and ${\bf o}_{sf}^*(k)$. And then, the first optimal element in ${\bf u}_{ss}^*(k)$, i.e., $u_{ss}^*(k)$, immediately enters the slow cooling-dominant subsystem to adjust its transient behavior and is also dispatched to the fast NEMPC as a settled variable. Simultaneously, the first optimal elements of ${\bf u}_{sf}^*(k)$ and ${\bf o}_{sf}^*(k)$, $u_{sf}^*(k)$ and $o_{sf}^*(k+1)$, are forwarded to the fast NEMPC as references in decision-making. The remaining elements of ${\bf u}_{ss}^*(k)$, ${\bf u}_{sf}^*(k)$, and ${\bf o}_{sf}^*(k)$ are discarded and will be re-optimized at the next sampling time $k+1$ based on the latest environmental information and system conditions. This process is commonly referred to as receding horizon implementation in model predictive control \cite{rawlings2017model}. For the convenience of the subsequent fast NEMPC, we define $u_{ss}^s(k) := u_s^*(k)$, $u_{ff}^{opt}(k) := u_{sf}^*(k)$, and $o_{ff}^{opt}(k+1) := o_{sf}^*(k+1)$.
	
	Regarding the fast NEMPC, which targets precise control of the power-dominant subsystem, the relevant global objectives from Eqs.\eqref{E7a} and \eqref{E7c} should be considered, along with the prescribed optimal references from the slow NEMPC. The objective in Eq.\eqref{E7b}, which pertains to the cooling-dominant subsystem, can be disregarded. The control objectives of the fast NEMPC, based on the fast MLP of Eq.\eqref{E31} at time instant $k$, are therefore developed as follows:
	\begin{subequations}\label{E34}
		\begin{align}
			J_1^f(k) = &\ \alpha_1^f\|\hat{y}_{f,1}(k+1)-(1+\xi(k))y_e^b(k+1)\|^2 \label{E34a} \\
			J_2^f(k) = & -\alpha_2^f(p_{mg}(k)w_{f,3}(k) + p_{fe}(k)\hat{y}_{f,1}(k+1) + p_{cm}(k)\xi_{as}(k) y_e^b(k+1) \notag \\
			&- p^f(k)(u_{ff,1}(k)+u_{fs,1}(k)) - p_{pn}(k)\|\hat{y}_{f,1}(k+1) - (1+\xi(k))y_e^b(k+1)\|^2) \label{E34b} \\
			J_3^f(k) = &\ \alpha_3^f\| u_{ff}(k) - u^{opt}_{ff}(k)\|^2 \label{E34c} \\ 
			J_4^f(k) = &\ \alpha_4^f\| o_{ff}(k+1) - o^{opt}_{ff}(k+1)\|^2 \label{E34d}
		\end{align}
	\end{subequations}
	where $\alpha^f_\omega$ ($\omega=1,\dots,4$) represents the weighting coefficients, while $\hat{y}_{s}$ refers to the corrected output of the fast MLP, which will be detailed later. Since the decision-making for the cooling-dominant subsystem is handled by the slow NEMPC, the fast NEMPC focuses solely on minimizing deviations in the power response to the utility grid and maximizing the system's income, as defined in Eqs.\eqref{E34a} and \eqref{E34b}, which are discretized from Eqs.\eqref{E7a} and \eqref{E7c}. The input and output tracking objectives in Eqs.\eqref{E34c} and \eqref{E34d} are designed to guarantee that the optimization of the fast NEMPC remains in the vicinity of the optimal results obtained from the slow NEMPC.
	
	Utilizing the designed objectives in Eq.\eqref{E34} and the fast MLP in Eq.\eqref{E31}, the fast NEMPC for controlling the power-dominant subsystem at time instant $k$ is established as follows:
	\begin{subequations} \label{E35}
		\begin{align}
			\min_{u_{ff}, o_{ff}} & \sum_{i=0}^{N^f_p-1} J_1^f(k+i) + J_2^f(k+i) + J_3^f(k+i) + J_4^f(k+i) \label{E35a}\\ 
			s.t. \ 
			&y_{f}(k+1+i) = F_{f,pow}(o_{fs}(k+i), o_{ff}(k+i), u_{fs}(k+i), \notag \\
			&u_{ff}(k+i), z_{f}(k+i), w_{f}(k+i)) \label{E35b}\\
			&u_{ff}^{min} z^u_{ff}(k+i) \leq u_{ff}(k+i)\leq u_{ff}^{max} z^u_{ff}(k+i) \label{E35c}\\
			& \Delta u^{min}_{ff} - V_{ff} \leq \Delta u_{ff}(k+i) \leq \Delta u^{max}_{ff} + V_{ff} \label{E35d}\\
			&o_{ff}^{min}\leq o_{ff}(k+1+i)\leq o_{ff}^{max} \label{E35e}\\
			&\hat{y}_{f}(k+1+i) = y_{f}(k+1+i) + \beta_f^i(y_{f}^{act}(k) - y_{f}(k|k-1)) \label{E35f}\\
			&z_{f}(k+i) = z^{l}(k+i) \label{E35g}\\
			&u_{fs}(k+i) = u_{ss}^s(k+i) \label{E35h}
		\end{align}
	\end{subequations}
	where $N^f_p$ denotes the prediction horizon of the fast NEMPC; the superscripts $min$ and $max$ serve as the corresponding lower and upper bounds, while $z^u_{ff} = {\rm diag}([z_{f,1},1])$; the terms $\Delta u_{ff}(k+i)$ and $V_{ff}$ are defined similarly to their counterparts in the slow NEMPC of Eq.\eqref{E33}, in which the former indicates the increment of the fast input between two consecutive time instants, while the latter deactivates the constraint in Eq.\eqref{E35d} when the relevant $z_f$ changes; $\hat{y}_{f}$ refers to the corrected output of the fast MLP, and $\beta_f^i$ is the tunable correction factor for the lumped output feedback; $y_{f}^{act}(k)$ is the actual value of $y_{f}$ at the current time instant $k$, and $y_{f}(k|k-1)$ is the predicted value of $y_{f}(k)$ made at the previous time instant $k-1$.
	
	In the fast MLP Eq.\eqref{E35}, Eq.\eqref{E35a} presents the optimization objective, which is to minimize a joint objective function—the weighted sum of the objectives defined in Eq.\eqref{E34}. The first constraint, Eq.\eqref{E35b}, specifies that the optimization problem is formulated by taking the fast MLP as the prediction model. Eqs.\eqref{E35c} through \eqref{E35e} impose physical constraints on the manipulated fast input and the controlled output of the operating units in the power-dominant subsystem. Eq.\eqref{E35f} establishes an output feedback mechanism for model prediction correction using the measurements and predictions of the outputs. The final two constraints, Eqs.\eqref{E35g} and \eqref{E35h}, indicate that $z_f$ and $u_{fs}$ are equal to the predetermined $z^l$ from the scheduler and $u_{ss}^s$ from the slow NEMPC, respectively, and are treated as constants in this context.
	
	At each sampling time instant $k$, the fast NEMPC is implemented by solving the optimization problem in Eq.\eqref{E35}. This process evaluates the optimal fast input sequence over the next prediction horizon, ${\bf u}_{ff}^*(k)=[u_{ff}^*(k), \dots, u_{ff}^*(k+N^f_p-1)]$, based on the current system conditions. The first element in the optimal sequence, $u_{ff}^*(k)$, is then straightly applied to the power-dominant subsystem to optimize its dynamic performance, enabling rapid and precise adjustments in power generation while ensuring profitability. The other elements in ${\bf u}_{ff}^*(k)$ are discarded and will be re-evaluated at the next sampling time $k+1$ using updated system conditions, thus completing the receding horizon optimization.
	
	This completes the proposed sequential distributed NEMPC for achieving real-time dynamic synergy among the operating units within the IES. Notably, communication between the slow and fast agents is incorporated into the cooperative control scheme by sharing optimization results and formulating additional control objectives, decision variables, and references. Furthermore, the multiple shooting method is employed to solve the nonlinear optimization problems associated with the sequential distributed NEMPC. For more details on this method, please refer to \cite{biegler2010nonlinear}.
	
	\begin{remark}
		\rev{As previously discussed, to exactly recast the nonlinear, neural network-based scheduler with integer variables into a computationally efficient MILP, we employ ReLU-based time-series MLPs as prediction models for long-term scheduling. To maintain consistency with the scheduler and avoid unnecessary complexity, we also use time-series MLPs as prediction models in the real-time, neural network-based EMPCs. However, it is important to note that the proposed framework—including the development of the physics-informed, multi-time-scale neural network and distributed NEMPC for complex energy systems—is not limited to MLPs. It is equally compatible with other time-series recurrent neural networks, such as RNNs or LSTMs.}
	\end{remark}
	
	\section{Simulation and discussion}
	
	In this section, we apply the developed hybrid neural network-based day-ahead scheduling and sequential distributed NEMPC to the considered IES and compare its control performance with classical control strategies. All simulations are conducted in Python (3.10.13) on a machine equipped with 16 GB of RAM and 2.60 GHz Intel Core i7-10750H processors. The proposed day-ahead scheduler is implemented and solved using the commercial solver Gurobi (11.0.0) \cite{pedroso2011optimization}, while the distributed NEMPC is performed on the open-source CasADi platform (3.6.3) \cite{andersson2019casadi} and solved using the built-in solver IPOPT (3.14.11).
	
	\subsection{Comparison and scenario settings}
	
	\begin{figure}[!ht]
		\centering
		\includegraphics[width=0.35\hsize]{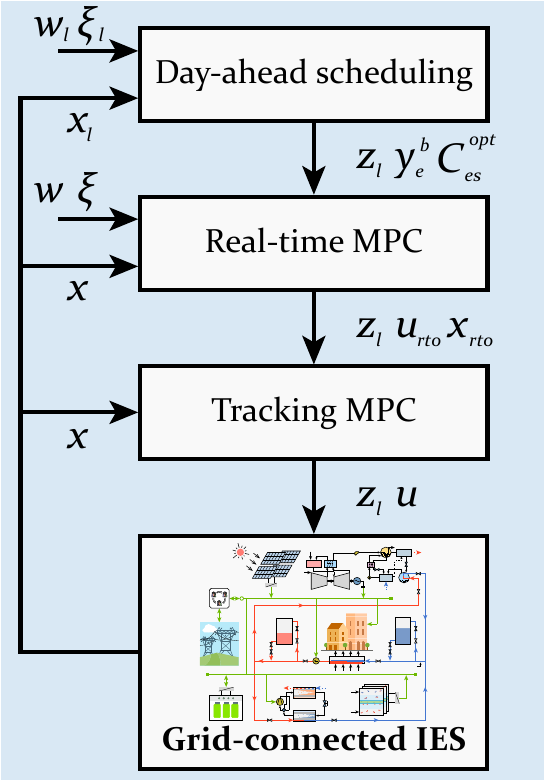}
		\caption{The standard hierarchical operational optimization paradigm.}
		\label{F11}
	\end{figure}
	
	To demonstrate the effectiveness of the proposed machine learning-based optimal scheduling and distributed control for the IES, we introduce a conventional hierarchical paradigm for comparison, as illustrated in Figure~\ref{F11}. This framework is a typical optimization structure with multiple optimal control levels across time scales, widely used in energy and process systems for operational management \cite{baldea2014integrated, ellis2017economic, ren2022tutorial}. Generally, the highest level of hierarchical control is the scheduling layer, which relies on a linearized economic model of the entire IES. This scheduling layer utilizes forecasts of external conditions, $w_l$ and $\xi_l$, to plan the approximate operational trajectories of the operating units for the next day. These plans include the on-off status $z_l$, the scheduled baseline power supplied to the utility grid $y_e^b$, and the energy storage references $C_{es}^{opt}$. Following the scheduling layer is the real-time control level, commonly employing a real-time MPC based on a rigorous \rev{nonlinear} steady-state model. This level optimizes the global performance of the entire system by evaluating the optimal references for the manipulated input $u_{rto}$ and the dynamic states $x_{rto}$, considering the real-time external and internal conditions $w$, $\xi$, and $x$. Finally, the lowest level, often a tracking MPC, operates based on the current states $x$ and the detailed \rev{nonlinear dynamic state-space} model of the entire system. This tracking level tightly follows the references prescribed by the real-time MPC to determine the final optimal inputs $u$, which are then immediately used to control the entire system. Please note that the models used at each control level are typically derived from first-principles models.
	
	Based on the proposed control approach and the classical control method, we formulate the following four control schemes for extensive comparison:
	
	(i) the proposed hybrid neural network-based day-ahead scheduler integrated with sequential distributed NEMPC, as illustrated in Figure~\ref{F10}, is referred to as Problem 1 (P1).
	
	(ii) the standard day-ahead scheduler, based on a linearized economic model within the classical hierarchical control strategy, substitutes for the proposed ReLU network-based day-ahead scheduler in Problem 1, while retaining the developed distributed NEMPC. This scheme is referred to as Problem 2 (P2).
	
	(iii) the standard real-time MPC and tracking MPC, as used in the typical hierarchical control framework, replace the designed distributed NEMPC in Problem 1, while retaining the developed day-ahead scheduler. This scheme is termed Problem 3 (P3).
	
	(iv) the standard hierarchical control scheme shown in Figure~\ref{F11}, which includes the linearized model-based scheduler, real-time MPC, and tracking MPC, is referred to as Problem 4 (P4).
	
	\begin{table}[!ht] \small
		\centering
		\caption{Parameters of the proposed scheduler and NEMPCs}
		\label{T9}
		\renewcommand{\arraystretch}{1.3}
		\tabcolsep 2pt
		\begin{tabular}{p{4.3cm}<{\centering}p{3.6cm}<{\centering}p{3.6cm}<{\centering}p{3.6cm}<{\centering}} \hline
			& \textbf{NN-based scheduler} & \textbf{Slow NEMPC} & \textbf{Fast NEMPC} \\ \hline
			\textbf{Discrete/Sampling time} & 3600 s & 60 s & 5 s \\ \hline
			\textbf{Prediction horizon steps} & 24 & 12 & 10 \\ \hline
		\end{tabular}
	\end{table}
	
	Regarding the selection of controller parameters, as listed in Table~\ref{T9}, the prediction horizon for the ReLU network-based scheduler is set to 24 hours (i.e., 24 discrete-time steps), covering the next operating day, with a discrete time interval of 1 hour, aligning with the long-term MLP. The prediction horizons for the slow NEMPC and fast NEMPC are set to 12 minutes (i.e., 12 discrete-time steps) and 50 seconds (i.e., 10 discrete-time steps), respectively, with corresponding sampling times of 1 minute and 5 seconds, consistent with the slow and fast MLPs. For the employed hierarchical controllers, the parameters of the standard day-ahead scheduler are chosen to be the same as those of the proposed ReLU network-based scheduler, while the parameters for the real-time MPC and tracking MPC are identical to those in the slow and fast NEMPC, respectively.
	
	\begin{figure}[!ht]
		\centering
		\includegraphics[width=1\hsize]{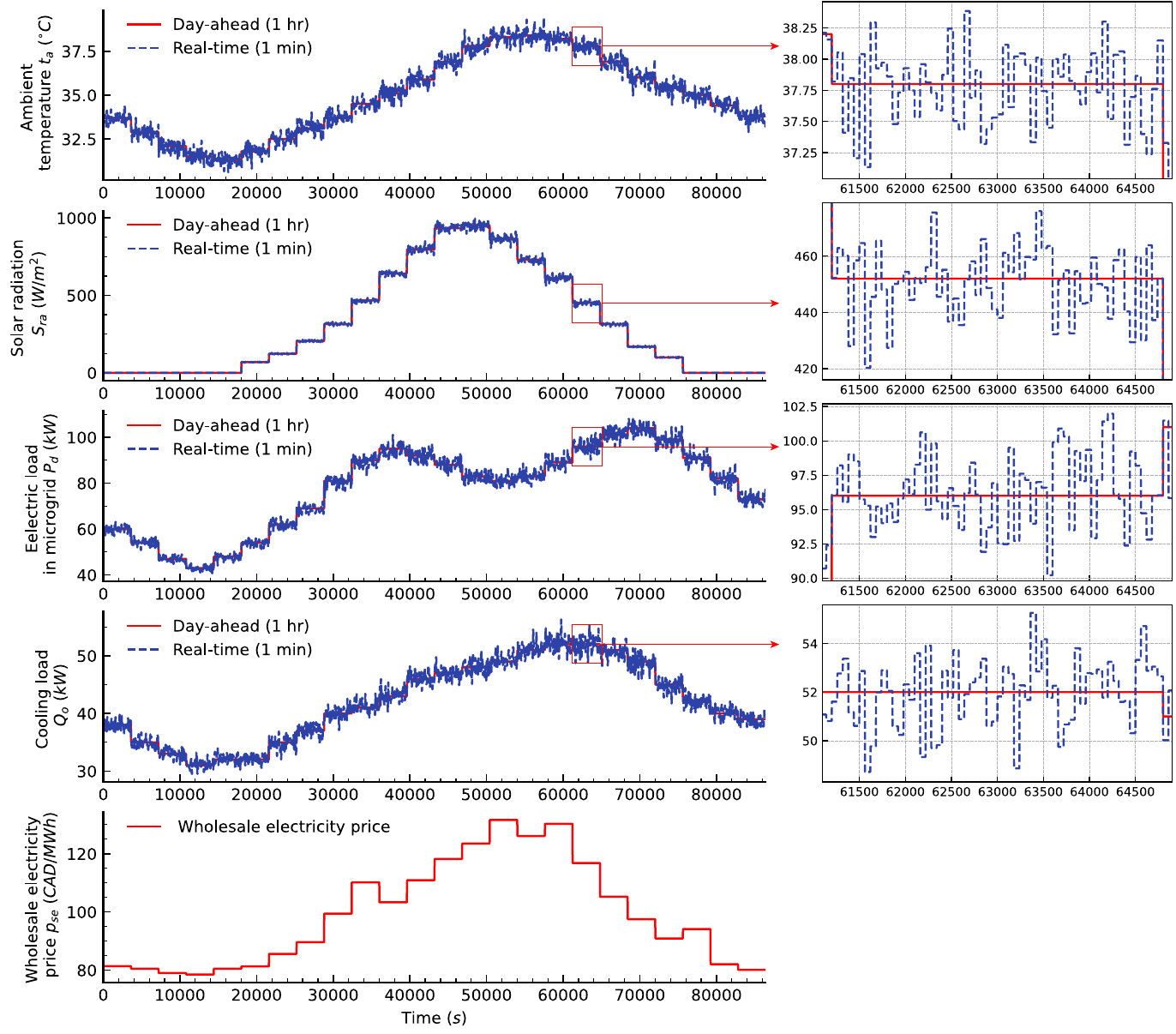}
		\caption{The evolution of external scenarios over a typical operating day: the top four subfigures display disturbances, with red solid lines representing day-ahead forecasts and blue dashed lines representing actual real-time data—ambient temperature ($t_a$), solar radiation ($S_{ra}$), electric demand in the microgrid ($P_d$), and cooling demand in the microgrid ($Q_o$). The bottom subfigure illustrates the wholesale electricity price of the utility grid ($p_{se}$).}
		\label{F12}
	\end{figure}
	
	Figure~\ref{F12} illustrates the typical evolution of external scenarios during an operating day. In this setup, the actual real-time disturbance data are randomly distributed around the respective day-ahead forecasts, following a Gaussian distribution with an amplitude of $\pm$8\%. The day-ahead forecasts, with a time resolution of 1 hour, will be used by the day-ahead schedulers, while the real-time data, with a time resolution of 1 minute, will be utilized by the intraday real-time coordinated controllers. Moreover, the price of natural gas is set at 0.2 CAD/kg, and electricity for local microgrid customers is priced at 80 CAD/MWh. These prices, along with the wholesale electricity price of the utility grid, are based on relevant policies in Ontario, Canada, during the summer of 2021 \cite{ieso2021ieso, oeb2021oeb}. We also assume that the subsidy for participating in grid response ($p_{cm}$) and the penalty for failing to supply the promised power to the utility grid ($p_{pn}$) are tied to the wholesale electricity price ($p_{se}$), with $p_{cm} = 1.5 p_{se}$ and $p_{pn} = 1.5 p_{se}$.
	
	\subsection{Case 1: Provision of 20\% response capacity of baseline power}
	
	\begin{figure}[!ht]
		\centering
		\includegraphics[width=1\linewidth]{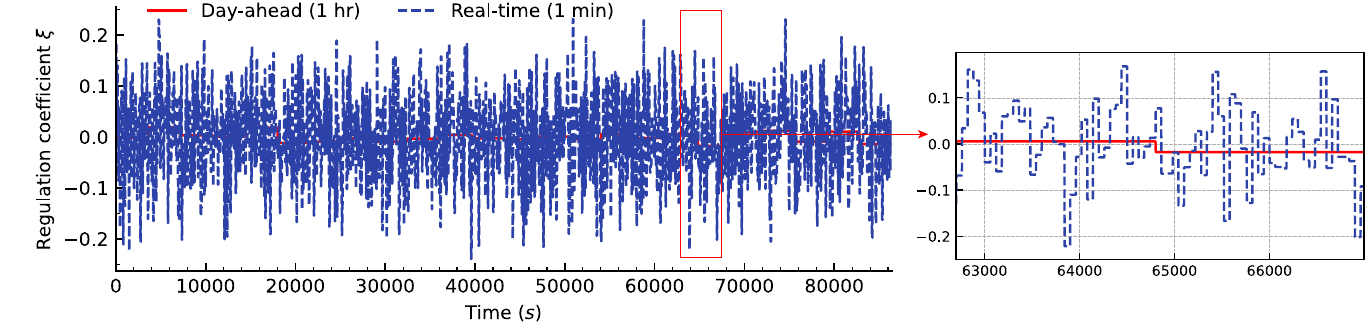}
		\caption{The evolution of the regulation coefficient $\xi$: the red solid line represents the day-ahead curve of $\xi_l$ with 1-hour temporal resolution used in the day-ahead schedulers, while the blue dashed line shows the actual real-time curve of $\xi$ with 1-minute temporal resolution used in the real-time coordinated controllers.}
		\label{F13}
	\end{figure}
	
	First, we investigate the performance of Problems 1-4 under a common working condition for the grid-connected IES that participates in the grid response as a dispatchable generation unit. In this case study, the IES is permitted to provide 20\% of the planned baseline power as available response capacity (i.e., the regulation capacity) to the utility grid. This allows the delivered power to the utility grid to be adjusted within $\pm 20\%$ of baseline power at the grid's request, meaning that the regulation coefficient $\xi$ basically satisfies $\xi \in [-20\%, 20\%]$. The specific evolution of $\xi$ is shown in Figure~\ref{F13}, where the actual real-time $\xi$ is randomly distributed around the day-ahead curve and follows a Gaussian distribution. In this context, the IES must quickly and accurately adjust the power delivered to the utility grid, based on the scheduled baseline power $y_e^b$ and the regulation coefficient $\xi$, in response to real-time instructions from the utility grid operator for regulating unscheduled power generation.
	
	\begin{figure}[!ht]
		\centering
		\includegraphics[width=1\linewidth]{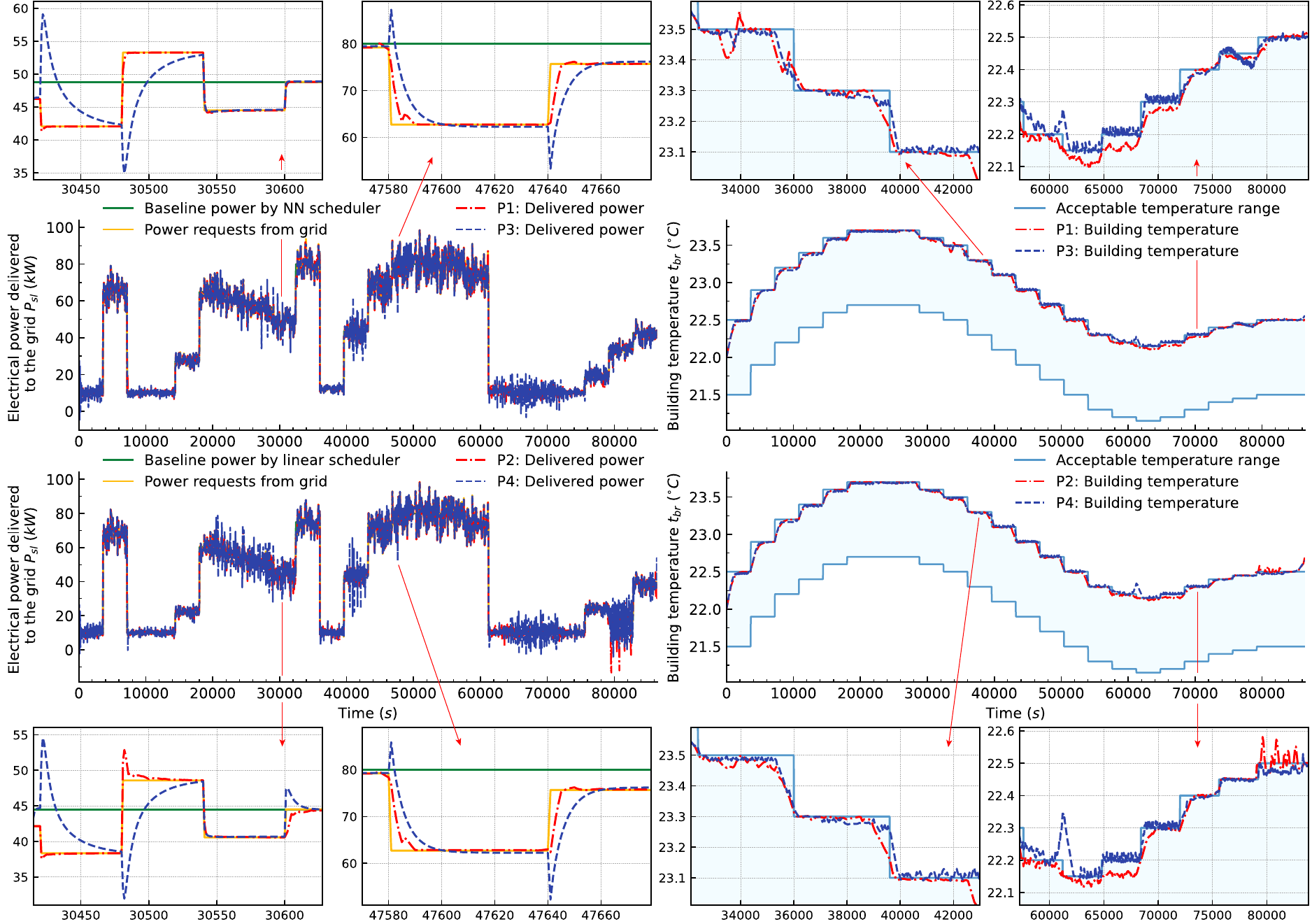}
		\caption{The outputs of the IES under P1-P4 in Case 1: The left column shows power sent to the utility grid, and the right column shows building temperature; the top row presents performance under P1 and P3 with $y_e^b$ from the ReLU network-based scheduler; the bottom row depicts performance under P2 and P4 with $y_e^b$ from the linearization model-based scheduler.}
		\label{F14}
	\end{figure}
	
	\begin{figure}[!ht]
		\centering
		\includegraphics[width=1\linewidth]{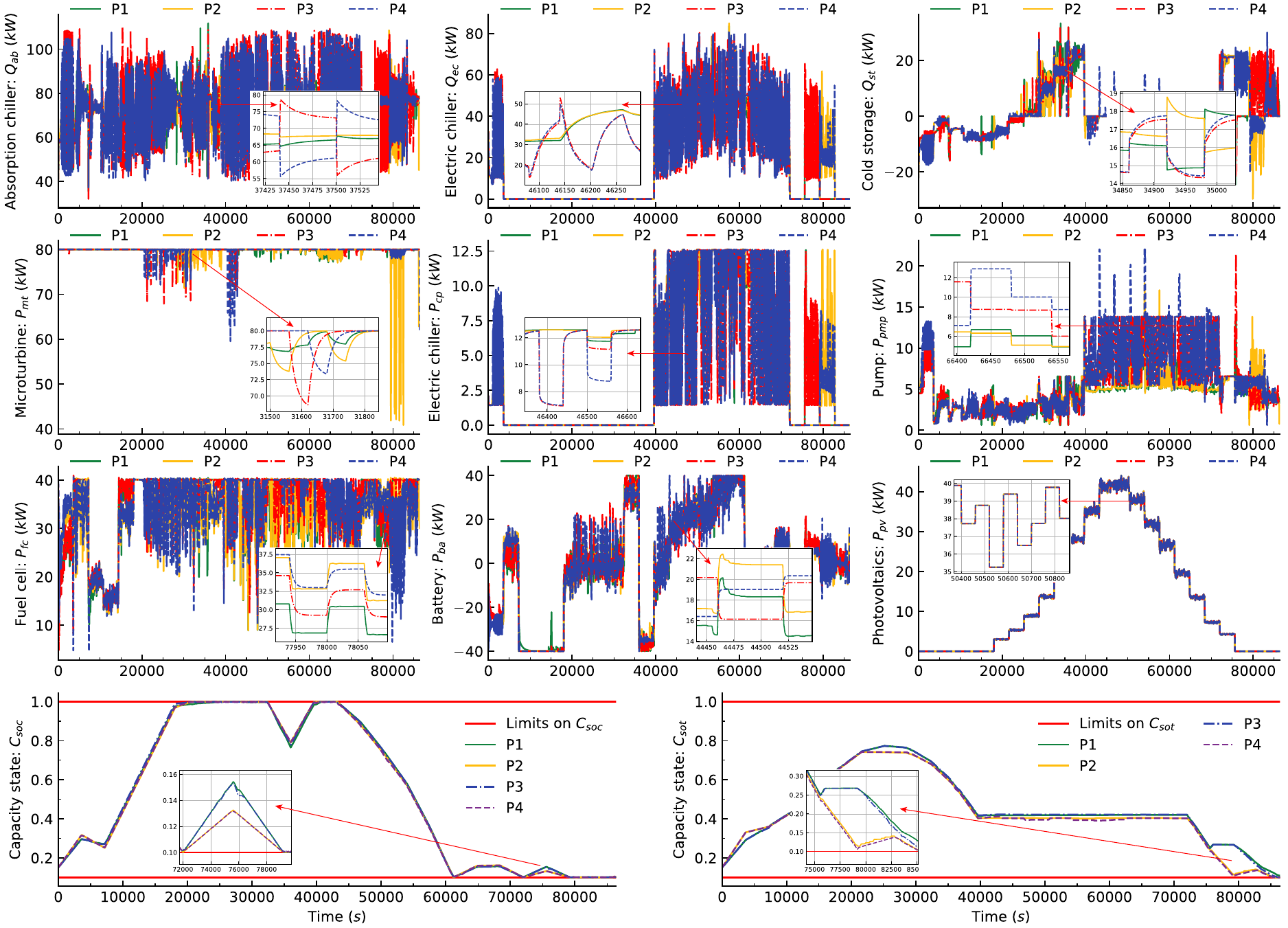}
		\caption{The outputs of the operating units under P1-P4 in Case 1: The first three rows show the electrical/cooling power production and consumption of each operating unit, while the last row displays the capacity states of energy storage.}
		\label{F15}
	\end{figure}
	
	\begin{figure}[!ht]
		\centering
		\includegraphics[width=1\linewidth]{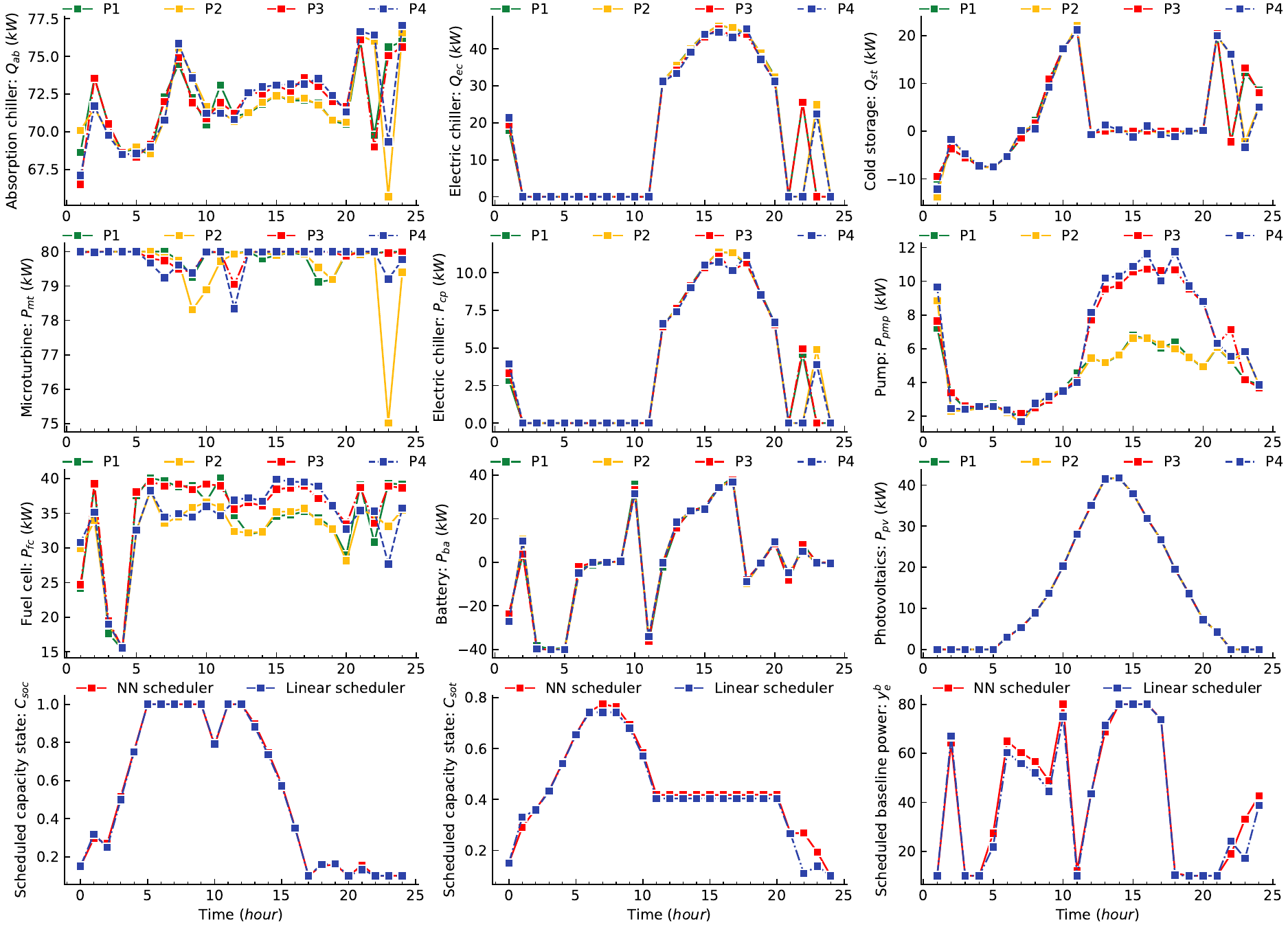}
		\caption{The hourly average performance of the systems in Case 1: The first three rows show the hourly averages of electrical/cooling power production and consumption for each operating unit; the last row displays the hourly planned capacity states of energy storage and the scheduled baseline power, from left to right, as determined by the ReLU network-based and linear model-based schedulers.}
		\label{F16}
	\end{figure}
	
	Figures~\ref{F14}-\ref{F16} illustrate the operational performance of the IES under the management of Problems 1-4. Figures~\ref{F14} and \ref{F15} depict the dynamic behavior of the entire system and the respective operating units, while Figure~\ref{F16} presents their hourly average performance. As shown in Figure~\ref{F14}, Problems 1-4 generally succeed in meeting the operational requirements of the IES, which quickly responds to the utility grid's power regulation requests and maintains indoor temperatures within the customer-specified range. Meanwhile, the indoor temperature is kept close to the upper limit to reduce cooling power loads, similar instances have been observed in other studies of building climate control \cite{halvgaard2012economic}. Among these, the IES under the proposed Problem 1 is more effective in closely tracking time-varying instructions and promptly adjusting its power supply to the utility grid, with fewer instances of the building temperature exceeding the specified range. Problem 2, compared to Problems 1, 3, and 4, provides the second-best dynamic conduct in precisely responding to electrical power regulation requests and maintaining indoor temperature. In contrast, Problem 3 is inferior to Problems 1 and 2 in controlling power transmission to the grid and maintaining indoor temperature, sometimes failing to reach the specified power target during a regulating period. Nevertheless, it is comparable to, and slightly outperforms, Problem 4, which further diminishes the IES's capability to rapidly adjust electricity generation and cooling.
	
	Moreover, as seen in Figure~\ref{F15}, under the control of Problems 1 and 2, the operating units associated with power generation—particularly those within the power-dominant subsystem managed by the fast NEMPC—exhibit superior power responsiveness, with precise, swift, and complementary transient behavior, reaching their steady state more quickly during the regulation period, in contrast to the sluggish responses observed under Problems 3 and 4. Besides, the cooling production in the units of the cooling-dominant subsystem is smoother under Problems 1 and 2 as against to Problems 3 and 4. The features of the power- and cooling-related operating units controlled by the MLP-based NEMPC in Problems 1 and 2 facilitate a fast and precise response to unscheduled power regulation requests from the grid while keeping a comparatively steady indoor temperature with minimal fluctuations. These common characteristics of Problems 1 and 2, which are attributed to the same NEMPC, therefore contribute to better meeting power and cooling requirements during transient periods, dealing with the control issues of dynamic time-scale multiplicity exhibited in the IES, and optimizing its transient behavior. Meanwhile, energy storage operates similarly across Problems 1-4, effectively achieving long-term load shifting. Taking into account the formulation of Problems 1–4, these results preliminarily demonstrate that, during the transient process, the proposed sequential distributed NEMPC, which exploits the developed slow and fast MLPs, can significantly improve system dynamics and realize dynamic synergy among the operating units, even with different scheduling strategies. Additionally, the findings also suggest that the designed ReLU network-based scheduler helps reinforce overall dynamic performance, since real-time coordinated control is always implemented on the basis of a long-term scheduler.
	
	Figure~\ref{F16} further confirms these findings. As we examine Figure~\ref{F16}, we observe that sometimes the real-time controller dominates the system's behavior, while at other times, the scheduler has a more significant influence. For instance, the fuel cell power under Problem 1 is similar to that under Problem 3 during the 6th to 10th hour, while the fuel cell power under Problem 2 resembles that under Problem 4. However, during the 11th to 20th hour, the fuel cell power under Problem 1 aligns more closely with that under Problem 2, and similarly, under Problem 3, it mirrors the behavior observed under Problem 4. Similar patterns are evident in the cooling produced by the absorption chiller and the power consumed by the pump. In addition, we note that the pump's power consumption under Problems 1 and 2 is greatly lower than under Problems 3 and 4 during the 11th to 20th hour, and the scheduled capacity state of the battery under the proposed scheduler is slightly larger than under the linear model-based scheduler. This allows the IES to operate with more dispatchable generation and greater optimization flexibility, thanks to the high-precision ReLU network, thereby enhancing overall performance. Finally, as depicted in the bottom right sub-figure of Figure~\ref{F16}, the proposed ReLU network-based scheduler optimizes a larger baseline power commitment to the utility grid than the linear scheduler during certain periods. To verify this, we calculated the total scheduled baseline power under both schedulers to determine the amount of planned electrical energy sent to the grid over a day. The result is 1006.46 kWh under the proposed scheduler against 965.14 kWh under the typical scheduler, indicating that our proposed solution offers the potential for greater profit through arbitrage. Combining the comparative discussions of Figures~\ref{F14}-\ref{F15}, these observations suggest that our proposed solution is capable of effectively handling dynamic characteristics and creating more optimization space.
	
	To assess the performance of the optimization schemes more thoroughly, we define four evaluation metrics derived from the global control objectives outlined in Eq.\eqref{E7} as follows:
	\begin{subequations} \label{E36}
		\begin{align}
			E_p &= \sum_{k=1}^{N_{L}} J_1^{\frac{1}{2}}(k) \\
			E_t &= \sum_{k=1}^{N_{L}} min\{\|y_2(k)-y_{sp,t}^{min}(k)\|, \|y_2(k)-y_{sp,t}^{max}(k)\|\},\ y_2(k) \notin [y_{sp,t}^{min}(k), y_{sp,t}^{max}(k)] \\
			E_e &= \sum_{k=1}^{N_{L}} - J_3(k) + \Delta C_{es} \\
			E_o &= \kappa_1 E_p + \kappa_2 E_t - \kappa_3 E_e
		\end{align}
	\end{subequations}
	where $E_p$ denotes the cumulative bias in electrical power delivered to the utility grid, while $N_{L}$ serves as the simulation length; $E_t$ indicates the cumulative values for which the indoor temperature falls outside the acceptable range $[y_{sp,t}^{min}, y_{sp,t}^{max}]$; $E_e$ is the actual revenue generated over an operating day, and $\Delta C_{es}$ accounts for the costs associated with surplus energy stored relative to the initial condition; $E_o$ conveys the overall operational performance of the entire IES, with scaling factors $\kappa_\omega$ ($\omega=1,2,3$). In these metrics, smaller values of $E_p$, $E_t$, and $E_o$ indicate better performance in regulating power generation, cooling production, and overall system operation, respectively, while a larger $E_e$ suggests enhanced profitability.
	
	\begin{table}[!ht] \small
		\centering
		\caption{Performance evaluation metrics in Case 1}
		\label{T10}
		\renewcommand{\arraystretch}{1.3}
		\tabcolsep 2pt
		\begin{tabular}{p{2.6cm}<{\centering}p{2.6cm}<{\centering}p{2.6cm}<{\centering}p{2.6cm}<{\centering}p{2.6cm}<{\centering}} \hline
			$\xi \in [-20\%,20\%]$ & $E_{p}$ & $E_{t}$ & $E_{e}$ & $E_{o}$ \\ \hline 
			\textbf{P1} & 11193.3 & 67.3 & 81.4 & 511.915 \\
			\textbf{P2} & 28517.3 & 174 & 74.2 & 1438.66 \\
			\textbf{P3} & 33348.4 & 413.6 & 74.5 & 1799.72 \\
			\textbf{P4} & 46133.8 & 505.7 & 70.7 & 2488.84 \\ \hline
		\end{tabular}
	\end{table}
	
	\begin{figure}[!ht]
		\centering
		\includegraphics[width=1\linewidth]{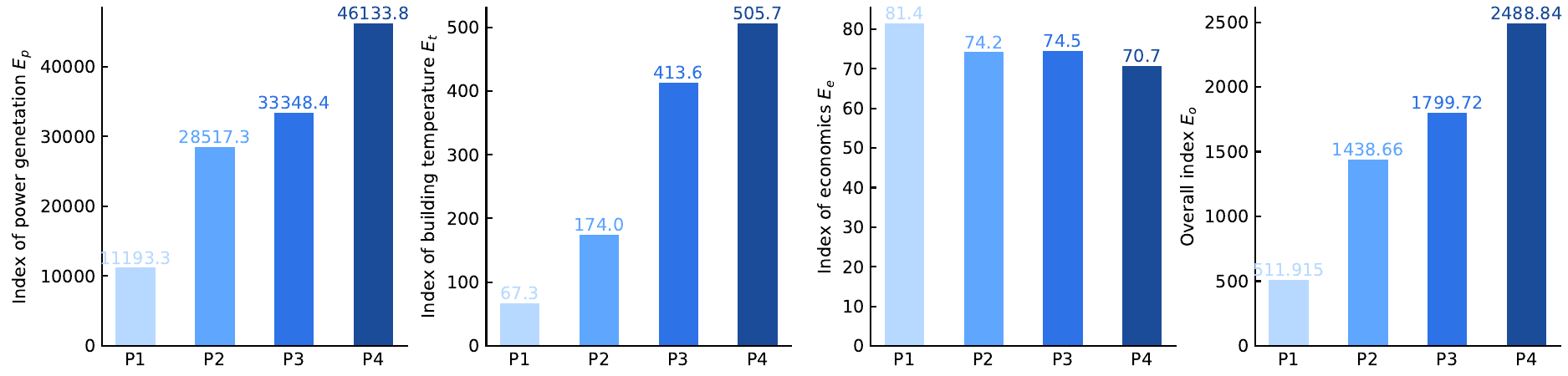}
		\caption{Visualized performance evaluation metrics in Case 1.}
		\label{F17}
	\end{figure}
	
	Table~\ref{T10} presents the evaluation metrics for Problems 1-4 in Case 1, which are also depicted in Figure~\ref{F17}. The metrics shown in these charts demonstrate that the proposed scheduling and control scheme (Problem 1), based on hybrid neural network techniques, outperforms other control schemes in terms of power generation, cooling production, and economic efficiency, significantly enhancing overall system performance. Thanks to the proposed NEMPC, Problem 1, which shares the same scheduling strategy as Problem 3, effectively reduces output deviations in electrical and cooling power, generates higher profits, and improves overall system performance in comparison with Problem 3. Similarly, the performance metrics under P2 are superior to those under P4 for the same reasons. On the other hand, due to the developed day-ahead scheduler, Problem 1 offers superior performance in power, cooling, and revenue as against Problem 3, which implements the same real-time control strategy as Problem 1. The system behavior under Problem 3 also shows improvements over Problem 4. Comparing the performance metrics of the system under Problems 2 and 3, we find that they are comparable in enhancing system revenue, with Problem 2 slightly excelling in meeting power demand. However, Problem 2 substantially surpasses Problem 3 in fulfilling cooling demand. Both Problems 2 and 3 outperform Problem 4 across all performance metrics. This indicates that both long-term scheduling and real-time control strategies greatly impact system dynamics and economics, with real-time coordinated control having a greater effect on transient processes. They complement each other, where a good scheduling strategy allows real-time coordinated control to better optimize dynamic behavior. Through these comparisons, it is evident that the enhancements in system performance under Problem 1 result from the combined efforts of the proposed day-ahead scheduler and real-time NEMPC. Ultimately, under the joint optimization of the designed scheduler and controller, the overall system performance under Problem 1 improves by over 70\% compared to the classical hierarchical control scheme (Problem 4). In relation to Problem 4, Problem 2's overall system performance improved by 42\% with the designed NEMPC, while Problem 3 achieved a 27\% enhancement due to the developed scheduler. These results demonstrate that our developed MLPs effectively capture system dynamics and performance across a wide range of operating conditions. With appropriate optimization strategies, they can handle the complex tasks of scheduling and control in energy systems. Leveraging these MLPs, the neural network-based scheduling scheme we established surpasses traditional linear model-based schemes, reaching more desirable optimization results during decision-making. Additionally, in contrast to the typical real-time and tracking MPCs, the proposed sequential distributed NEMPC based on the MLPs effectively improves system dynamic responses, economic benefits, and overall performance.
	
	\subsection{Case 2: Provision of varied response capacity of baseline power}
	
\begin{figure}[!ht]
		\centering
		\includegraphics[width=1\linewidth]{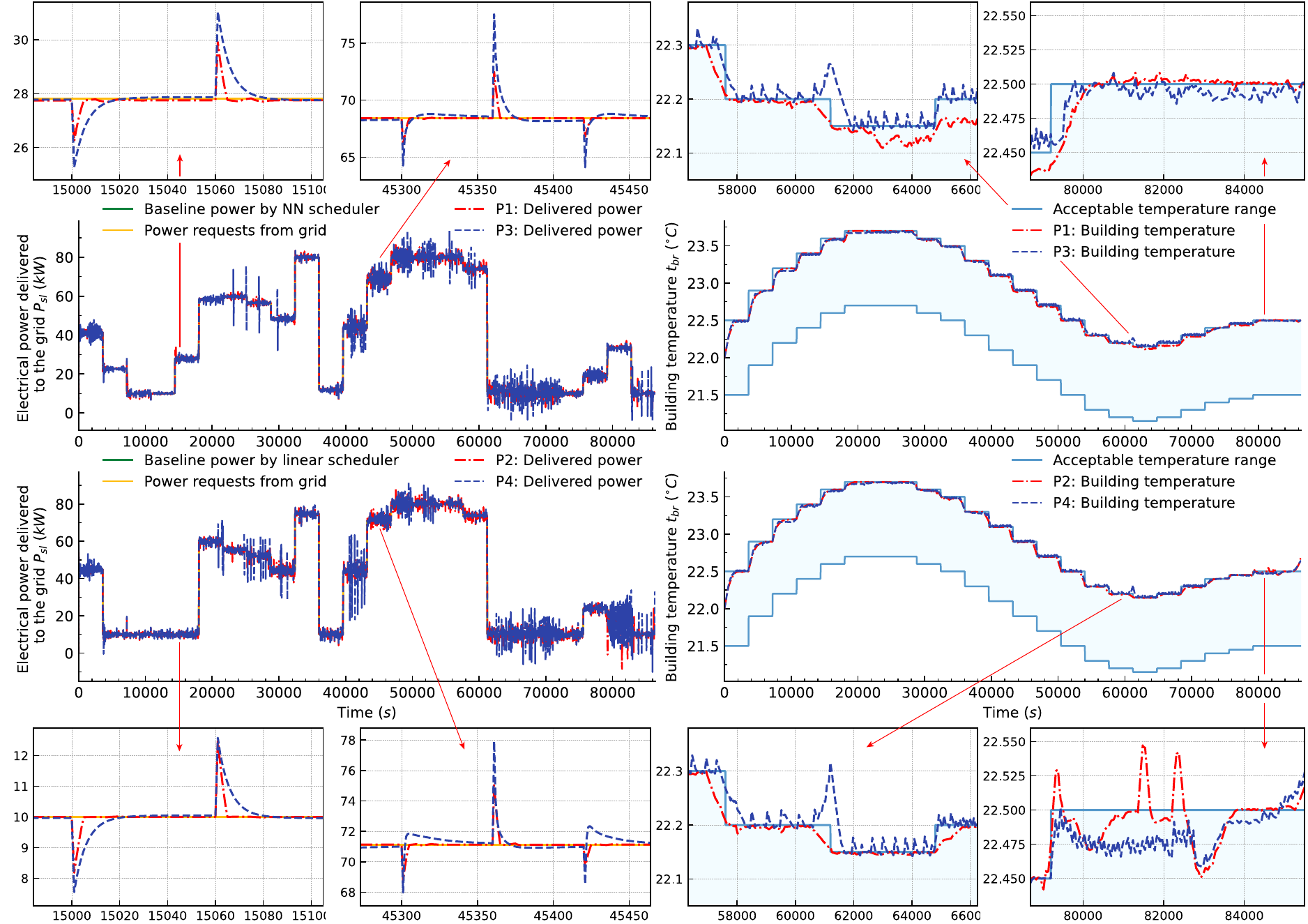}
		\caption{The outputs of the IES under P1-P4 with $\xi=0$ in Case 2.}
		\label{F18}
	\end{figure}

\begin{figure}[!ht]
		\centering
		\includegraphics[width=1\linewidth]{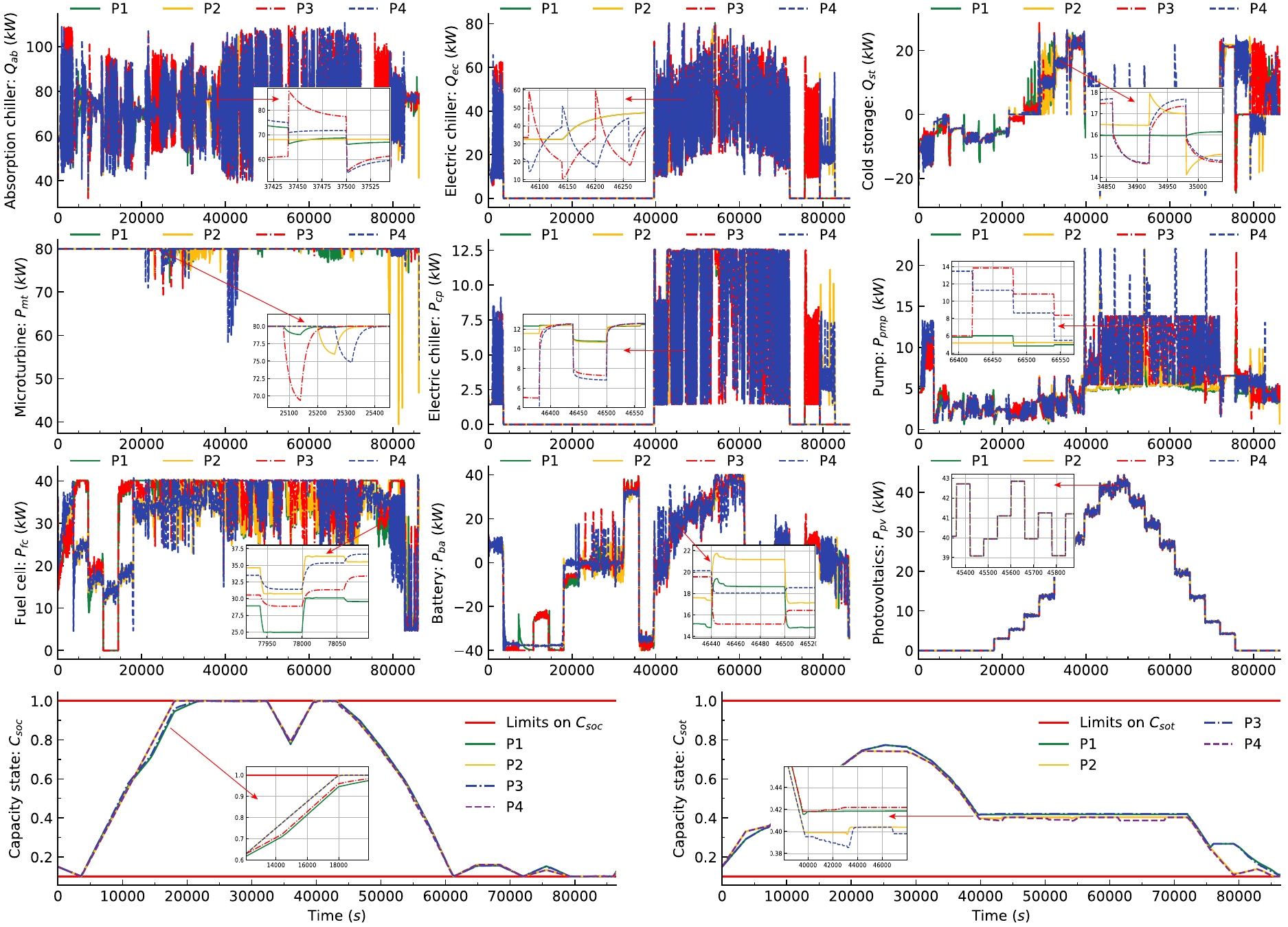}
		\caption{The outputs of the operating units under P1-P4 with $\xi=0$ in Case 2.}
		\label{F19}
	\end{figure}

\begin{figure}[!ht]
		\centering
		\includegraphics[width=1\linewidth]{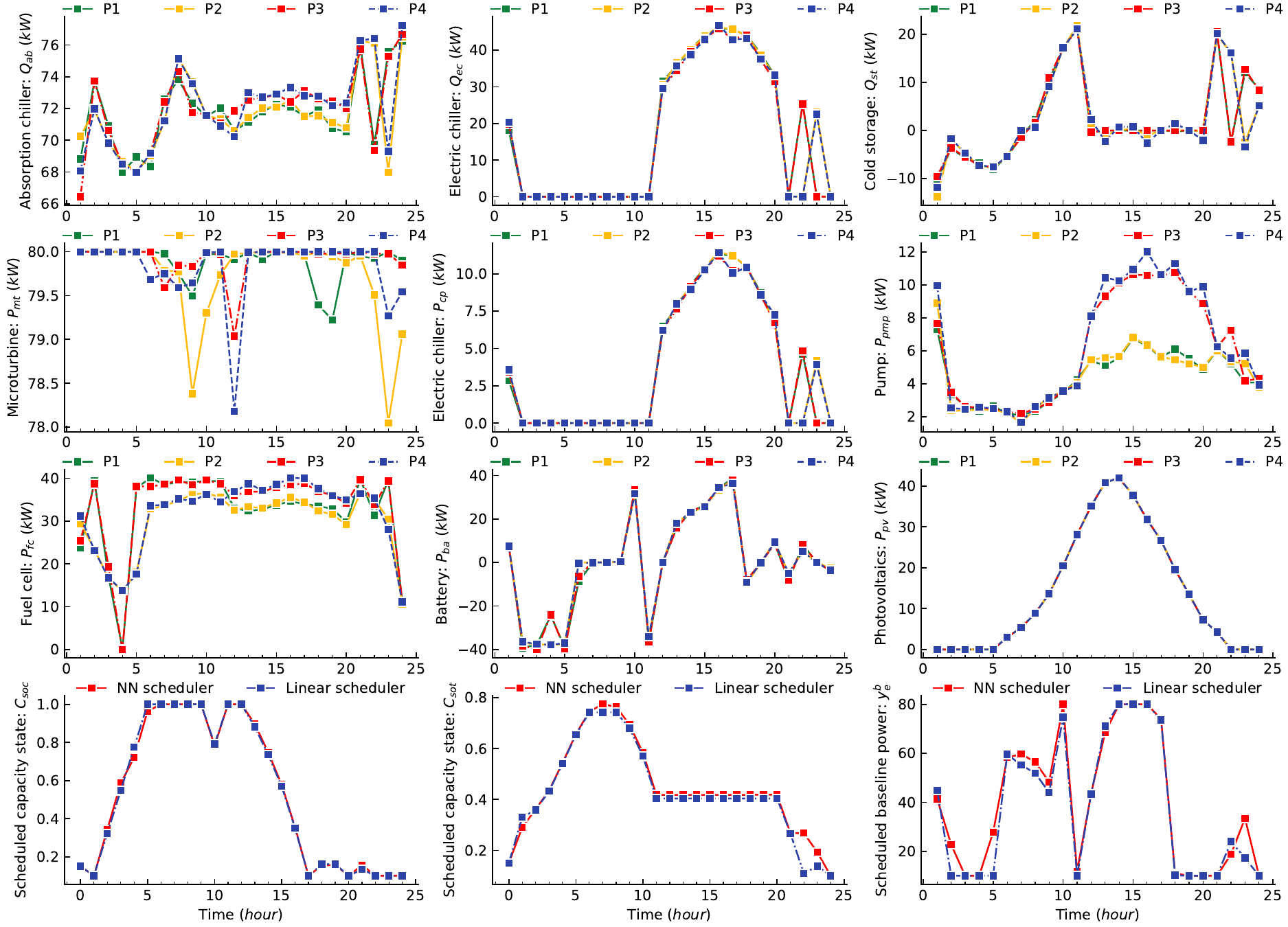}
		\caption{The hourly average performance of the systems with $\xi=0$ in Case 2.}
		\label{F20}
	\end{figure}

\begin{figure}[!ht]
		\centering
		\includegraphics[width=1\linewidth]{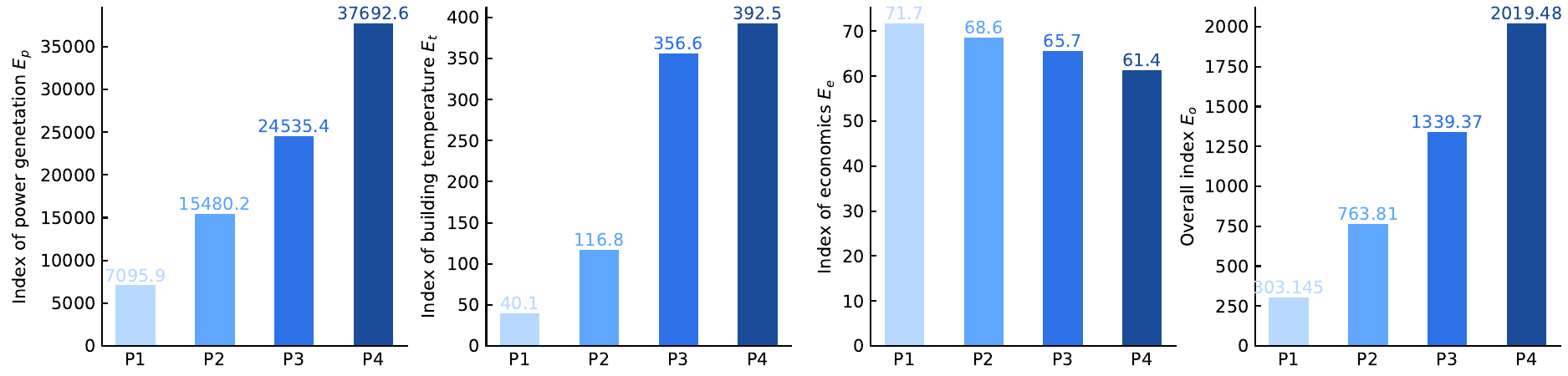}
		\caption{Visualized performance evaluation metrics for $\xi=0$ in Case 2.}
		\label{F21}
	\end{figure}

	In many cases, IES operators may choose to provide varying levels of available response capacity to the utility grid or opt not to participate in power response at all, instead delivering power according to the previously set baseline power $y_e^b$ without considering any time-varying regulation commands. Therefore, this subsection investigates the performance of Problems 1-4 when providing 0\%, 10\%, and 30\% response capacity. Due to space limitations, detailed simulation results for the case where the system does not participate in grid response—i.e., when the regulation coefficient $\xi$ is consistently zero—are presented in Figures~\ref{F18}-\ref{F21}. Simulation results for other scenarios are summarized in the performance evaluation metrics for further comparative discussion.
	
	Unlike in Case 1, where the system had to rapidly follow real-time power regulation instructions from the grid, the control system in this scenario experiences reduced uncertainty since it no longer responds to unplanned power requests. This reduction in uncertainty makes it easier for the control system to manage the IES transient processes. As shown in Figure~\ref{F18}, all control strategies demonstrate better dynamic behavior optimization, both in maintaining baseline power transmission to the grid and in regulating indoor temperature, as against their performances in Case 1. Although the system no longer needs to respond to grid requests, the inherent variability of the external environment persists. As a result, the control system must still adjust the IES operation in real time to deal with these external changes, leading to small spikes in the power delivery and indoor temperature curves. Nonetheless, under Problem 1, the IES adheres more closely to the baseline power supplied to the grid and maintains the indoor temperature within the specified range in most cases, indicating superior system operation and dynamic optimization in comparison with to other schemes.
	
	As illustrated in Figures~\ref{F19} and \ref{F20}, the dynamic behavior of operating units and their hourly performance when not participating in grid response are similar to those observed in Case 1. Under the proposed scheduler and NEMPC optimization, the power-dominant subsystem continues to exhibit more proactive transient behavior, while the cooling-dominant subsystem show comparatively smoother dynamic responses. This indicates effective coordination among operating units across different dynamic response timescales. The primary difference is that, since the system only needs to deliver power according to the promised baseline, overall dynamic fluctuations are reduced. In terms of hourly performance, the system is still influenced by both the scheduling plan and real-time control strategies. With the proposed approach, the system's self-consumption of electricity remains low, while available power storage and the planned baseline power sent to the grid are slightly higher, indicating more efficient use of the operating units. According to our calculations, the developed scheduler optimizes the total daily power supply to the grid to 955.69 kWh, in contrast to 899.96 kWh with the standard method, leading to increased revenue from the utility grid. These similarities with Case 1 validate our previous comparative analysis, suggesting that the developed MLPs accurately reflect system performance across different time scales and provide the necessary model accuracy for operational decisions. The designed scheduler and distributed NEMPC, leveraging the long-term, slow, and fast MLPs, significantly enhance the system's precise power responsiveness, cooling demand satisfaction, and overall operational efficiency.
	
	Furthermore, as shown in Figure~\ref{F21}, all performance evaluation metrics under each control scheme are lower than their values in Case 1. This occurs because maintaining power transmission to the grid at the baseline level is comparatively easier, resulting in smaller tracking errors for electrical and cooling power. However, total revenue decreases due to the loss of grid response subsidies. Despite this, our proposed solution, Problem 1, consistently outperforms other strategies in meeting power and cooling demands, enhancing system profitability, and delivering better overall performance. A similar cross-comparison reveals that the proposed ReLU network-based scheduler and sequential distributed NEMPC play a critical role in improving overall system performance by reducing power deviation, lowering cooling dissatisfaction, and increasing operational income.
	
	\begin{table}[!ht] \small
		\centering
		\caption{Performance evaluation metrics in Case 2}
		\label{T11}
		\renewcommand{\arraystretch}{1.3}
		\tabcolsep 2pt
		\begin{tabular}{p{2.6cm}<{\centering}p{2.6cm}<{\centering}p{2.6cm}<{\centering}p{2.6cm}<{\centering}p{2.6cm}<{\centering}} \hline
			$\xi=0\%$  & $E_{p}$ & $E_{t}$ & $E_{e}$ & $E_{o}$ \\ \hline 
			\textbf{P1} & 7095.9 & 40.1 & 71.7 & 303.145 \\
			\textbf{P2} & 15480.2 & 116.8 & 68.6 & 763.81 \\
			\textbf{P3} & 24535.4 & 356.6 & 65.7 & 1339.37 \\
			\textbf{P4} & 37692.6 & 392.5 & 61.4 & 2019.48 \\ \hline
			$\xi \in [-10\%,10\%]$ & $E_{p}$ & $E_{t}$ & $E_{e}$ & $E_{o}$ \\ \hline 
			\textbf{P1} & 8114.8 & 43 & 76.7 & 350.54 \\
			\textbf{P2} & 21988.3 & 161 & 70.1 & 1109.82 \\
			\textbf{P3} & 26183.8 & 345.4 & 70.6 & 1411.29 \\
			\textbf{P4} & 41933.4 & 385 & 64.9 & 2224.27 \\ \hline
			$\xi \in [-30\%,30\%]$ & $E_{p}$ & $E_{t}$ & $E_{e}$ & $E_{o}$ \\ \hline 
			\textbf{P1} & 17804 & 229.8 & 83.1 & 922 \\
			\textbf{P2} & 33447.3 & 369.3 & 77.9 & 1779.12 \\
			\textbf{P3} & 41145.2 & 509.5 & 76.2 & 2235.81 \\
			\textbf{P4} & 54941.8 & 458.9 & 70.4 & 2906.14 \\ \hline
		\end{tabular}
	\end{table}
	
	\begin{figure}[!ht]
		\centering
		\includegraphics[width=0.8\linewidth]{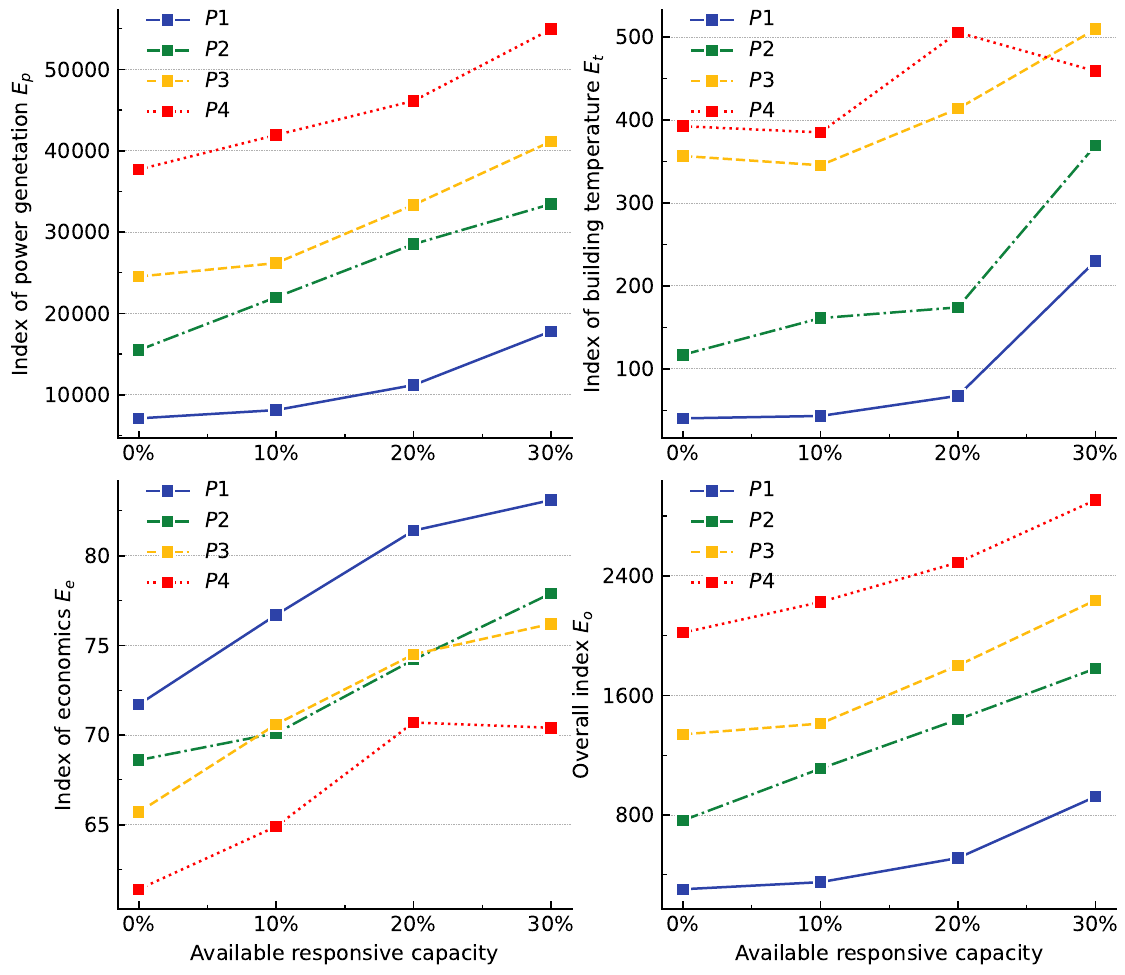}
		\caption{The performance evaluation metrics for varied response capacities: $E_p$ (upper left), $E_t$ (upper right), $E_e$ (lower left), $E_o$ (lower right).}
		\label{F22}
	\end{figure}
	
	Figure~\ref{F22} depicts the performance metrics of the IES with varying response capacities in the simulations, as listed in Tables~\ref{T10} and \ref{T11}. As shown, under all control strategies, the system's evaluation metrics generally increase as the responsive capacity provided by the IES grows. This is because a larger responsive capacity enables greater power regulation, which pushes the system closer to its regulation limits and introduces more control challenges, leading to higher tracking errors in power transmission and cooling supply. However, within a reasonable range, increased responsive capacity can enhance the system's profitability. Despite this, the overall system performance tends to decline with continued increases in nominal regulation capacity, suggesting that the actual responsive capacity the system can provide to the grid is limited—an observation supported by existing research \cite{wu2023distributed}. Notably, regardless of conditions, the proposed scheduling and control scheme in Problem 1 consistently outperforms other control strategies, offering better power regulation responses, smaller temperature deviations, higher profits, and superior overall performance. By contrast, Problem 2 with the proposed NEMPC accomplishes the next best performance, while Problem 3 with the developed scheduler still surpasses the standard Problem 4 in these metrics. These results highlight the positive impact of the proposed MLP-based scheduling and control scheme on improving the system's precise responsiveness, energy supply quality, and profitability, demonstrating its effectiveness and superiority in system operation management.
	
	\begin{figure}[!ht]
		\centering
		\includegraphics[width=1\linewidth]{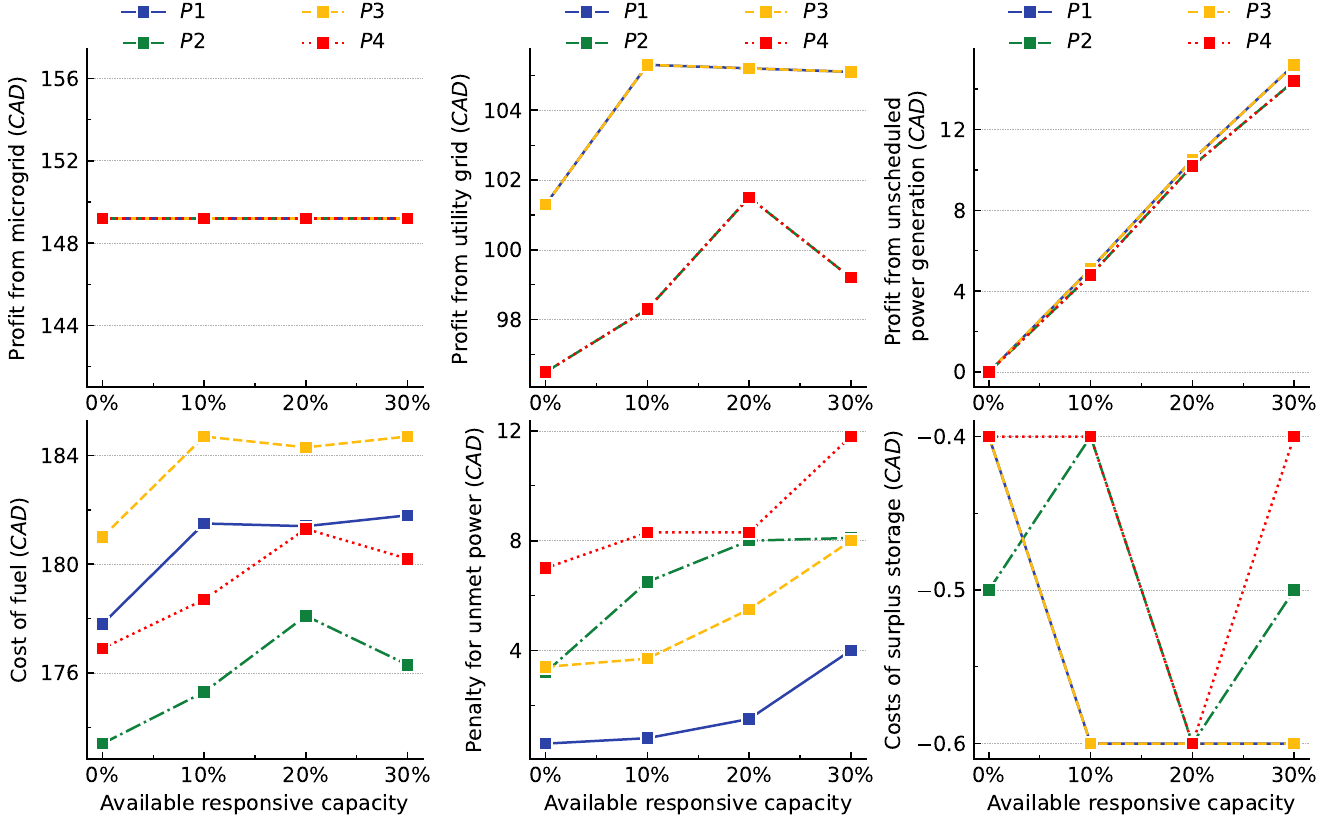}
		\caption{The multiple economic measurements for varied response capacities: Top row (left to right): profits from power supply to microgrid, utility grid, and grid response subsidy; Bottom row (left to right): fuel costs, fines for unmet power commitments to utility grid, and costs of surplus stored energy.}
		\label{F23}
	\end{figure}
	
	Moreover, Figure~\ref{F23} presents measurements relevant to system economics, which are components of the global economic objective. While these measurements do not fully reflect system performance, they help us better understand the relationship between system dynamics and economics. As shown, the revenue from the microgrid remains the same across different operational schemes due to consistent electrical loads within the microgrid. However, with the proposed scheduling approaches, the IES under Problem 1 and Problem 3 can leverage the potential of the operating units to provide more baseline power to the utility grid, thereby generating higher profits. In this case, the grid's subsidies for unscheduled generation responses are also slightly higher. Nevertheless, increased generation leads to higher fuel consumption, resulting in higher fuel costs under Problem 1 and Problem 3. Notably, fuel costs are lower under Problem 1 than Problem 2, suggesting that the proposed real-time control strategy not only improves system dynamic behavior, as previously mentioned, but also enhances energy efficiency, leading to more efficient system operation. Additionally, since the same scheduling strategies are used, the fines for failing to meet power supply requirements are lower under Problem 1 and Problem 2 in comparison with other schemes. With the developed real-time NEMPC, Problem 1 reaches relatively lower penalty levels due to the proactive dynamic tracking performance—tighter power tracking leads to fewer penalties. Ultimately, as discussed earlier, the proposed Problem 1 consistently maintains high profitability. These observations underscore that the system's transient behavior affects both energy supply quality and economic benefits. Achieving optimal economic operation and dynamic synergy requires close cooperation between effective long-term scheduling and successful real-time control.
	
	\begin{table}[!ht] \small
		\centering
		\caption{The average computational time of MPCs during simulations}
		\label{T12}
		\renewcommand{\arraystretch}{1.4}
		\tabcolsep 11pt
		\begin{tabular}{p{3.2cm}<{\centering}p{3.2cm}<{\centering}p{4.6cm}<{\centering}} \hline
			\textbf{Slow NEMPC} & \textbf{Fast NEMPC} & \textbf{Total spent time per hour} \\ \hline
			5.74 s & 0.63 s & 798 s \\ \hline
			\textbf{Real-time MPC} & \textbf{Tracking MPC} & \textbf{Total spent time per hour} \\ \hline
			0.91 s & 4.27 s & 3129 s \\ \hline
		\end{tabular}
	\end{table}
	
	Finally, we evaluated the computation time of the proposed MLP-based scheduling and control scheme to assess its practicality. The ReLU network-based scheduler had an average computation time of 6.685 seconds during simulations, making it feasible for day-ahead scheduling. For the real-time distributed NEMPC, which requires online application, the average computation times during simulations are presented in Table~\ref{T12}. These are compared with the computation times of the standard hierarchical MPC, including real-time MPC and tracking MPC. Considering the sampling intervals of 60 seconds and 5 seconds, respectively, the computation times for slow NEMPC and fast NEMPC are sufficiently fast for online implementation. In contrast to the standard hierarchical control, the developed distributed NEMPC, based on the slow and fast MLPs, significantly reduces the total computation time per hour, demonstrating improved computational efficiency. However, it is noteworthy that, owing to the use of a steady-state model, the real-time MPC's computation time is substantially shorter than that of the slow NEMPC, which is based on the complete slow MLP. On the other hand, the tracking MPC, which requires high-frequency computations and employs detailed first-principles models, has a much higher computational complexity. Consequently, its computation time is significantly longer than that of fast NEMPC, leading to a greater total computation time per hour than the proposed method. It is important to highlight that in the proposed distributed control method, during the evaluation of the fast NEMPC, the decision variables pertaining to the slow subsystem in the fast MLP have been determined and parameterized. This procedure can be viewed as a model reduction on the fast dynamic time scale, allowing the fast NEMPC, which also requires high-frequency computation, to reach a shorter computation time. In other words, in terms of model scale and complexity, control methods directly based on neural network models might face potential issues related to computational efficiency and cost, consistent with findings in existing work \cite{liu2023state, xie2015data, zhao2022machine}. However, the proposed approach addresses the optimization and control challenges arising from the time-scale multiplicity in the IES dynamics by leveraging neural networks for accurate system dynamics predictions across multiple time scales, while reducing computation time. By designing an appropriate scheduling and distributed control strategy, we mitigate potential computational burdens and enhance overall system performance, demonstrating the applicability of the proposed method in managing complex energy systems.
	
	\section{Concluding remarks}
	
	Integrated energy systems are complex nonlinear process systems \rev{which are characterized} by diverse dynamic behaviors across multiple \rev{time scales}. Establishing an accurate first-principles model for such systems is challenging due to the intricate dynamic interactions involved, which poses significant barriers to effective system management. To tackle these issues, we propose a \rev{physics-informed} hybrid time-series neural network approach that leverages machine learning and integrates limited prior process knowledge for highly accurate predictions of the IES's dynamic behavior across multiple time scales. Subsequently, using the developed hybrid neural networks, we introduce a ReLU network-based scheduler and a sequential distributed NEMPC with slow and fast MLPs for day-ahead scheduling and real-time coordination between operating units. The design of the proposed approach considers the multi-time-scale dynamics of power- and cooling-relevant units within the IES towards superior dynamic and economic performance.
	
	Extensive simulations demonstrate the effectiveness and applicability of the proposed hybrid neural network-based scheme in achieving dynamic synergy between operating units. \rev{We achieve} significant improvements in system performance, including precise and fast responsiveness to unscheduled power regulation requests, maintaining building temperature, and enhancing operational profitability. Consequently, the overall performance of the IES is substantially improved compared to classical solutions. Moreover, our investigation based on simulation results indicates that both long-term scheduling and real-time control significantly impact the system's dynamic behavior and economic efficiency. To successfully manage complex energy systems, it is essential to design an effective control method as the final decision-making tool, supported by an efficient scheduling framework. The developed hybrid neural network-based scheduler and controller outperform conventional strategies in these areas. Finally, our evaluation of computational time reveals that the proposed neural network-based distributed scheme reduces computational burden and achieves enhanced efficiency compared to the standard hierarchical control paradigm. \rev{Meanwhile, it is important to note that, in addition to challenges posed by data acquisition, directly applying neural network models to the control of large-scale energy systems may still face difficulties due to the online computational complexity during MPC evaluation.} Engineers must find appropriate solutions to reduce the scale of neural networks without sacrificing prediction accuracy for control purposes. Our proposed sequential distributed approach effectively reduces computational time while addressing dynamic time-scale multiplicity. Nonetheless, developing a systematic method for neural network model reduction in the control of large-scale complex energy systems remains an open theoretical question.
	
	\section{Acknowledgments}
	
	This work was supported by National Natural Science Foundation of China (Grant 51936003); National Key R\&D Program of China (Grant 2022YFB4100403); China Scholarship Council. The second author X. Yin would like to acknowledge the financial support from Ministry of Education, Singapore, under its Academic Research Fund Tier 1 (RS63/22), and Nanyang Technological University, Singapore (Start-Up Grant).
	

\end{document}